\documentclass[11pt,a4paper]{article}
\pdfoutput=1 %

\usepackage{jheppub} %

\usepackage[utf8]{inputenc}
\usepackage[T1]{fontenc}
\usepackage{lmodern}
\usepackage{fontawesome}
\usepackage{hyperref}
\usepackage{tikz}
\usepackage{subcaption}
\usepackage{mathtools}

\usepackage{xcolor}
\usepackage{amsmath,amssymb}

\definecolor{col1}{RGB}{100,143,255}
\definecolor{col2}{RGB}{120, 94, 240}
\definecolor{col3}{RGB}{254,97,0}
\definecolor{col4}{RGB}{220, 38, 127}
\definecolor{col5}{RGB}{255, 176, 0}

\usepackage{tikz-feynman} 

\usetikzlibrary{positioning}
\usetikzlibrary{automata,positioning,arrows,matrix,backgrounds,calc}
\usetikzlibrary{decorations.text}
\usetikzlibrary{decorations.pathmorphing}
\usetikzlibrary{shapes.geometric}
\usetikzlibrary{arrows.meta}

\usepackage{pgf}
\usepackage{relsize}

\usetikzlibrary{decorations,patterns,arrows.meta}

\usetikzlibrary{decorations.pathreplacing,decorations.markings}

\tikzset{
    ->-/.style={decoration={
  markings,
  mark=at position .5 with {\arrow{>}}},postaction={decorate}},
    -<-/.style={decoration={
  markings,
  mark=at position .5 with {\arrow{<}}},postaction={decorate}},
    ->/.style={decoration={
  markings,
  mark=at position .4 with {\arrow{>}}},postaction={decorate}},
}

\newcommand{\ep}{\varepsilon}
\newcommand{\gE}{E}
\newcommand{\gEI}{E^{\textrm{int}}}
\newcommand{\bb}{\boldsymbol}
\newcommand{\gVI}{V^{\textrm{int}}}
\newcommand{\gVE}{V^{\textrm{ext}}}
\newcommand{\gVEI}{V^{\textrm{ext}}_{\textrm{in}}}
\newcommand{\gVEO}{V^{\textrm{ext}}_{\textrm{out}}}
\newcommand{\gGE}{{\Gamma^{\textrm{ext}}}}
\newcommand{\gGI}{{\Gamma^{\textrm{int}}}}

\newcommand{\EI}{\mathcal{E}}
\newcommand{\EE}{\mathcal{R}}
\renewcommand{\AA}{a}
\newcommand{\sunny}{\text{\faSunO}}
\newcommand{\shady}{\text{\faMoonO}}

\DeclareMathOperator{\sinc}{sinc}
\newcommand{\R}{\mathbb{R}}  
\newcommand{\dd}{\textrm{d}}
\newcommand{\pp}{\textrm{p}}
\renewcommand{\tt}{\textrm{t}}
\newcommand{\cc}{\textrm{c}}
\newcommand{\rr}{\textrm{r}}
\DeclareMathOperator{\Sym}{Sym}

 \hypersetup{
   pdfauthor={Michael Borinsky, Zeno Capatti, Eric Laenen, Alexandre Salas-Bern\'ardez},
   pdftitle={Flow-oriented perturbation theory},
   pdfsubject={},
   pdfkeywords={},
 }

\preprint{\vbox{\hbox{Nikhef 2022-017}} \vbox{\hbox{IPARCOS-UCM-23-005}}}

\title{Flow-oriented perturbation theory}

\author[1]{Michael Borinsky,}
\author[2]{Zeno Capatti,}
\author[3,4,5]{Eric Laenen,}
\author[6]{Alexandre Salas-Bern\'ardez}

\affiliation[1]{Institute for Theoretical Studies, ETH Z\"urich, 8092 Z\"urich, Switzerland}
\affiliation[2]{Institute for Theoretical Physics, ETH Z\"urich, 8093 Z\"urich, Switzerland}
\affiliation[3]{IOP/ITFA, University of Amsterdam, Science Park 904, 1098 XH Amsterdam, The Netherlands}
\affiliation[4]{Nikhef, Theory Group, Science Park 105, 1098 XG, Amsterdam, The Netherlands}
\affiliation[5]{ITF, Utrecht University, Leuvenlaan 4, 3584 CE Utrecht, The Netherlands}
\affiliation[6]{Universidad Complutense de Madrid, Departamento de F\'isica Te\'orica and IPARCOS, 28040 Madrid, Spain}

\abstract{
We introduce a new diagrammatic approach to perturbative quantum field theory, which we call flow-oriented perturbation theory (FOPT). Within it, Feynman~graphs are replaced by strongly connected directed graphs (digraphs).  FOPT is a coordi\-nate space analogue of time-ordered perturbation theory and loop-tree duality, but it has the advantage of having combinatorial and canonical Feynman rules, combined with a simplified $i\ep$ dependence of the resulting integrals.  Moreover, we introduce a novel digraph-based representation for the S-matrix. The associated integrals involve the Fourier transform of the flow polytope. Due to this polytope's properties, our S-matrix representation exhibits manifest infrared singularity factorization on a per-diagram level.  Our findings reveal an interesting interplay between spurious singularities and Fourier transforms of polytopes.
}

\makeatletter
\gdef\@fpheader{}
\makeatother

\begin{document}

\maketitle

\section{Introduction}

The perturbative approach to quantum field theory  provides the basis of our understanding of the fundamental laws governing high-energy processes. Experimental observables are computed in an asymptotic formalism as a power series in a coupling whose value should be small. The coefficients of this perturbative series are represented in terms of Feynman diagrams. The practical computation of such observables is most often performed within a covariant approach in momentum space, in which each diagram is manifestly invariant under the action of the Poincaré group. Nevertheless, since the early days of quantum electrodynamics, other, non manifestly covariant approaches have been used, such as the venerable time-ordered perturbation theory (TOPT), which can be derived analogously to quantum mechanical perturbation theory (see ref.~\cite{Sterman:1993hfp,tasi_sterman} for a review). 

TOPT involves three-dimensional momentum space loop integrals, with time-ordered vertices, and is well-suited for a local singularity analysis.
Similarly, more recent formulations as loop-tree duality~\cite{Bierenbaum_2010,Catani_2008,Capatti_2019,Runkel_2019,Runkel_2020,Berghoff:2022vah} and manifestly-causal loop-tree duality~\cite{bobadilla2021lotty,capatti2020manifestly,Sborlini_2021,Aguilera_Verdugo_2020,Aguilera_Verdugo_2021,Kromin:2022txz}, have led to an advance in the understanding of the large distance singularity structure of Feynman integrals. In particular the denominator structures that arise therein are usually more directly identified with formal graph-theoretic constructs~\cite{Capatti_2021,Kreimer:2020mwn,Kreimer:2021jni,Dallaway:2022zoz}, which makes the generalization of the singularity analysis considerably easier. 

Parametric Feynman integration is still the present-day standard for the numerical evaluation of multi-loop integrals~\cite{Binoth:2000ps,Borowka:2017idc,Heinrich:2020ybq,Smirnov:2021rhf}, especially in Euclidean kinematic regimes. In this context practically all relevant Feynman diagrams can be integrated readily~\cite{Borinsky:2020rqs}. However, numerical methods based on four-dimensional~\cite{Gong:2008ww,Binoth:2005ff,Nagy:2006xy,Becker:2010ng,Becker:2012aqa,Becker:2012nk,Anastasiou:2018rib,Anastasiou:2020sdt} and especially three-dimensional representations~\cite{Soper:1999xk,Buchta:2015wna,Capatti_2020,Kromin:2022txz,Capatti_2021,Capatti:2022tit,Kermanschah:2021wbk} offer a promising alternative due to their manifest IR singularity cancellation features, and their native adaption to Minkowski space kinematics. Moreover, these three-dimensional representations allow for more efficient and robust treatment of integrals near thresholds.

In contrast to the momentum space formulation, coordinate space methods have received considerably less attention. Indeed, while field theories are almost always phrased in terms of coordinate space Lagrangians, scattering theory is naturally formulated in momentum space. 
Nevertheless, many fundamental physical principles find a beautiful and revealing formulation in coordinate space. Unitarity, for example, can be concisely encoded through the largest time equation~\cite{Veltman2,Martinus}. 
Coordinate space methods also have a prominent role in axiomatic formulations of Quantum Field Theory~\cite{Epstein:1973gw}. In the context of renormalization group invariants  at higher loop coordinate space methods are more powerful than their momentum space counterpart~\cite{Chetyrkin:1980pr,Schnetz:2013hqa,Borinsky:2021jdb,gfe,Borinsky:2022lds}.
Recently, cutting rules for eikonal diagrams~\cite{Eric1,Eric2} have also been generalized using coordinate space Green's functions, showing that these rules express and emphasize causality.
Other interesting developments include light-cone ordered perturbation theory and work on the coordinate space analysis of infrared divergences of Feynman diagrams~\cite{Sterman3,StermanErdogan,ErdoganCS,Salas-Bernardez:2022cuw}. 

In this paper we develop a new representation of coordinate space Green's functions based on the concept of energy flow: flow-oriented perturbation theory (FOPT). %
Much like TOPT, FOPT is canonical, in the sense that there is a well-defined Feynman integral for each energy-flow-oriented graph, without ambiguity. %
We derive the FOPT representation in sect.~\ref{sec:fopt} from four dimensional covariant coordinate space rules, starting with the derivation of the FOPT representation for the scalar triangle diagram, followed by the general treatment, which holds for any massless scalar diagram, independent of the number of edges incident to each internal vertex. We then summarise the FOPT representation in a concise set of Feynman rules and show that it has the UV behaviour which is expected from covariant analyses.

However, the FOPT representation comes with two inherent caveats: the external data is given in coordinate space, but, ultimately, a momentum space encoding of such data is needed for the computation of scattering cross-sections. Additionally, as we will explain in sect.~\ref{sec:singularitiesFOPT}, finite distance singularities play a quite intricate role in the FOPT representation and the presence of IR singularities is not manifest. Motivated by this, we adjust perspective in 
sect.~\ref{sec:s_matrix}: we (partially) transform the FOPT representation back to momentum space. 
We call the resulting hybrid representation, which effectively describes S-matrix elements, the $p$-$x$ S-matrix representation. 
It is hybrid in the sense that external kinematics are given in momentum space while internal integrations are performed in three-dimensional coordinate space. These internal coordinate space integrals are \emph{covariant} three-dimensional Feynman integrals, modulated by the Fourier transform of a certain polytope associated to the underlying flow-oriented Feynman diagram. 

This polytope turns out to be an instance of the well-studied \emph{flow polytope}. This type of polytope has close connections to the representation theory of arithmetic groups, diagonal harmonics, to Schubert polynomials and  other mathematical structures~\cite{gallo1978extreme,stanley2000acyclic,baldoni2004counting,meszaros2015flow,meszaros2017polytope,benedetti2019combinatorial}. 
Here, we add a new application of the flow polytope to this list: the study of Feynman integrals. This paper is hence an addition to the growing literature on the use of polytopes in the study of analyticity properties and evaluation techniques in quantum field theory~\cite{Kaneko:2009qx,Arkani-Hamed:2013jha,Brown:2015fyf,Panzer:2019yxl,Arkani-Hamed:2017tmz,Schultka:2018nrs,Arkani-Hamed:2019mrd,Borinsky:2020rqs,Ananthanarayan:2020fhl,Arkani-Hamed:2022cqe}. The inherent finiteness of such polytopes' Fourier transforms gives a concise explanation for the cancellation of spurious singularities in our new $p$-$x$ S-matrix representation. In general, the cancellation of such spurious singularities is a not a well-understood phenomenon as we will discuss in sect.~\ref{sec:spurious}. Additionally, the Fourier transform of the flow polytope expresses the oscillating behaviour of the S-matrix integrand which, especially at large distances, becomes crucial for determining the singular structure of the S-matrix itself.

Equipped with the necessary knowledge of the flow polytope's properties, we discuss IR singularities in the $p$-$x$ representation of the S-matrix in sect.~\ref{sec:IRpxsing} and identify a coordinate space analogue of collinear and soft singularities. Subsequently, we detail the factorization properties of these singular regions for diagrams contributing to the S-matrix. Our observation is that, in the $p$-$x$ S-matrix representation, IR factorization, which is expected from physical amplitudes \cite{Collins:1989gx}, is in fact already present at the \emph{diagram level}.  The general discussion is again supported by the pedagogical treatment of the triangle diagram.

Finally, we observe that the original FOPT representation has some interesting features in the context of unitarity-cut integrals and Cutkosky's theorem 
and the largest time equation~\cite{Veltman2,Cutkosky:1960sp}. For instance, virtual and real contributions to cross sections can readily be put under the same integration measure. We leave an in-depth analysis of these observations for a future work, hence we relegate these aspects to appx.~\ref{app:cutkosky}, where we sketch some of these ideas for the interested reader.

\section{Flow-oriented perturbation theory}
\label{sec:fopt}
We start by introducing flow-oriented perturbation theory (FOPT) for a massless scalar quantum field theory, and derive the FOPT Feynman rules. To motivate them, we discuss the one-loop triangle diagram in some detail. A number of useful concepts for FOPT graphs, such as their completion, cycles and routes are introduced and explained. 

\subsection{Scalar QFT in coordinate space}

\label{sec:coordinate_space_feynman_rules}
The massless coordinate-space Feynman propagator for a scalar field in $D=4$ dimensional space time (with the mostly-minus metric) reads  
\begin{align} \Delta_F(z) = \int \frac{\dd^4 p}{(2\pi)^4} e^{-i p \cdot z}\frac{i}{p^2+ i \ep} = \frac{1}{(2 \pi)^2} \frac{1}{-z^2 + i \ep}\,. \end{align}
As usual, the Feynman rules provide a recipe to translate a graph $G$ with sets of edges $\gE$, internal vertices $\gVI$ and external vertices $\gVE$ into an integral. Recall that external vertices are defined by the requirement that there is only one adjacent vertex to each of them. The usual coordinate-space Feynman rules (see for instance  \cite[Ch.~10.1]{le1991quantum} or \cite[Ch.~6.1]{itzykson2012quantum}) read
\begin{enumerate}
\item
Associate a coordinate vector to each internal or external vertex. We label the location of external vertices with $x_a$, $a\in V^{\text{ext}}$, and that of internal vertices with $y_v$, $v\in V^{\text{int}}$. 
\item
For each \emph{internal} edge $e=\{v,v'\}$ multiply by a Feynman propagator $\Delta_F(z_e) = \Delta_F(-z_e)$, where $z_e$ is the difference of the coordinates of the vertices to which the edge $e$ is incident. For example, if $e$ is an internal edge (none of the two vertices defining it is external), then $z_e=y_{v}-y_{v'}$. If $v$ is instead an external vertex, then $z_e=x_v-y_{v'}$.
\item For each interaction vertex multiply by a factor  $-ig$.
\item For each internal vertex $v\in V^{\text{int}}$ integrate over all values of the components of $y_v$, i.e. over all possible locations of the internal vertex in $4$-dimensional Minkowski space. 
\end{enumerate}
The resulting expression is a function of the external coordinates $\{x_a\}_{a\in V^{\text{ext}}}$. To be explicit, the application of the coordinate-space Feynman rules to a generic graph $G$ contributing to a Green's function of a massless scalar theory gives
\begin{align} A_G(x_1, \ldots, x_{|\gVE|})= \frac{(-ig)^{|\gVI|}}{(2\pi)^{2|\gE|}} \left[ \prod_{v \in \gVI} \int \dd^4 y_v \right] \prod_{e \in \gE}\frac{1}{-z_e^2+i\ep}\;.\label{eq:FI} \end{align}
One integrates over the four dimensional Minkowski space location of each internal vertex. Accounting for symmetry factors results in an expression for the scalar $n$-point correlation function,
\begin{align}\label{scalarcorrelator} \Gamma(x_1,...,x_{|\gVE|})=\left\langle 0 | T ( \varphi(x_1) \cdots \varphi(x_{|\gVE|}) ) | 0 \right \rangle = \sum_{G} \frac{1}{\Sym G} A_G(x_1, \ldots, x_{|\gVE|}), \end{align}
where we sum over all graphs $G$ from a given scalar QFT with the given, fixed external vertices $\gVE$.

\subsection{The triangle diagram in FOPT}%

In this section we treat the FOPT representation of the triangle. It will serve as a prototype for later derivations in this paper, because it is simple enough to show all details, while also exhibiting most of the subtleties associated with the general arguments. In coordinate space the triangle diagram can be drawn as
\vspace{0.2cm}
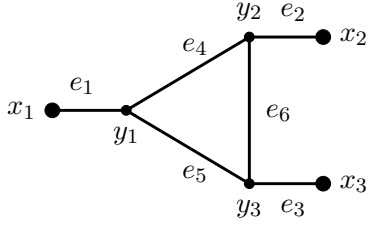
\begin{center}
\resizebox{5cm}{!}{%
\begin{tikzpicture} \begin{feynman} \vertex(1); \vertex[above right = 1.5cm and 2.5cm of 1](2); \vertex[below right = 1.5cm and 2.5cm of 1](3); \vertex[left = 1.5cm of 1](E1); \vertex[right = 1.5cm of 2](E2); \vertex[right = 1.5cm of 3](E3); \vertex[left = 0.2cm of E1](L1) {\scalebox{1.5}{$x_1$}}; \vertex[right = 0.2cm of E2](L2) {\scalebox{1.5}{$x_2$}}; \vertex[right = 0.2cm of E3](L3) {\scalebox{1.5}{$x_3$}}; \vertex[below = 0.2cm of 1](L1_I) {\scalebox{1.5}{$y_1$}}; \vertex[above = 0.2cm of 2](L2_I) {\scalebox{1.5}{$y_2$}}; \vertex[below = 0.2cm of 3](L3_i) {\scalebox{1.5}{$y_3$}}; \vertex[above left = 0.2cm and 0.5cm of 1](edge1) {\scalebox{1.5}{$e_1$}}; \vertex[above right = 0.2cm and 0.5cm of 2](edge2) {\scalebox{1.5}{$e_2$}}; \vertex[below right = 0.2cm and 0.5cm of 3](edge3) {\scalebox{1.5}{$e_3$}}; \vertex[above right = 1cm and 1cm of 1](edge4) {\scalebox{1.5}{$e_4$}}; \vertex[below right = 1cm and 1cm of 1](edge5) {\scalebox{1.5}{$e_5$}}; \vertex[below right = 1.25cm and 0.2cm of 2](edge6) {\scalebox{1.5}{$e_6$}}; \diagram*[large]{ (1) -- [-,line width=0.6mm] (2) -- [-,line width=0.6mm] (3) -- [-,line width=0.6mm] (1), (1) -- [-,line width=0.6mm] (E1), (2) -- [-,line width=0.6mm] (E2), (3) -- [-,line width=0.6mm] (E3), }; \path[draw=black, fill=black] (E1) circle[radius=0.15]; \path[draw=black, fill=black] (E2) circle[radius=0.15]; \path[draw=black, fill=black] (E3) circle[radius=0.15]; \path[draw=black, fill=black] (1) circle[radius=0.1]; \path[draw=black, fill=black] (2) circle[radius=0.1]; \path[draw=black, fill=black] (3) circle[radius=0.1]; \end{feynman} \end{tikzpicture}
}.
\end{center}
\vspace{0.2cm}
We labelled the external vertex locations with the variables $x_1$, $x_2$ and $x_3$ and their adjacent internal vertices' locations with $y_1$, $y_2$ and $y_3$ respectively. To each edge we associate a label $e_i$, $i=1,...,6$. 
Given this labelling, the coordinate-space triangle diagram, according to the Feynman rules presented in sec.~\ref{sec:coordinate_space_feynman_rules}, reads
\begin{gather} \begin{gathered} \label{eq:triangle_position_space} A_G(x_1,x_2,x_3)= \\
\frac{(-ig)^{3}}{(2\pi)^{12}} \int \left[\prod_{v\in \gVI}\dd^4 y_v\right] \frac{1}{(x_1-y_1)^2(x_2-y_2)^2(x_3-y_3)^2 (y_1-y_2)^2 (y_2-y_3)^2 (y_1-y_3)^2 }\,, \end{gathered} \end{gather}
where a negative 
$i\varepsilon$ prescription is assumed for all propagators. The expression $A_G$ is a function of the three external coordinates $x_1,x_2,x_3$.

Our first aim is to perform the time integrations $\left[ \int \mathrm{d} y_v^0 \right]$ in eq.~\eqref{eq:triangle_position_space} explicitly. We will do so by employing the \emph{residue theorem} together with a series of distributional identities, which will  allow us to cast the result in an especially elegant form. 
This approach works analogous to TOPT, and results
    in an expression that is entirely \emph{combinatorial} and \emph{canonical}, i.e.~we can read off the expression from modified, \emph{flow-oriented} Feynman graphs, and there is a unique canonical expression for each such Feynman graph. The main result of this section will then be the associated flow-oriented Feynman rules.\\

We aim for an expression for 
the following partially integrated version of $A_G$, which we denote by $\AA_G$:
\begin{gather} \begin{gathered} \label{eq:mathcal_A_triangle} \AA_G(x_1,x_2,x_3,\vec{y}_1, \vec{y}_2, \vec{y}_3)= \\
\frac{(-ig)^{3}}{(2\pi)^{12}} \int \left[\prod_{v\in \gVI}\mathrm{d} y_v^0\right] \frac{1}{(x_1-y_1)^2(x_2-y_2)^2(x_3-y_3)^2 (y_1-y_2)^2 (y_2-y_3)^2 (y_1-y_3)^2 }\,. \end{gathered} \end{gather}
Henceforth, we omit the variable dependence of $\AA_G$, but emphasize that it depends on the 4-vector coordinates of the external vertices and the 3-vector coordinates of the internal vertices.
To evaluate eq.~\eqref{eq:mathcal_A_triangle} using the residue theorem it is convenient to introduce auxiliary variables $z_e^0$, for each $e\in E$,  set equal to the time-difference between the vertices connected through $e$ via a delta function. These variables will be integrated from $-\infty$ to $\infty$. Thus we have the representation
\begin{align} \label{eq:triangle_dirac_deltas} \AA_G=& \frac{(-ig)^{3}}{(2\pi)^{12}} \int \left[\prod_{v \in \gVI}\mathrm{d} y_v^0\right]\left[\prod_{e\in\gE}\frac{\mathrm{d} z_e^0}{-z_e^2 + i\varepsilon}\right] \nonumber \\ & \times \left[\prod_{i=1}^3\delta(z_i^0-x_i^0+y_i^0)\right] \delta(z_4^0-y_{12}^0)\delta(z_5^0-y_{13}^0)\delta(z_6^0-y_{23}^0), \end{align}
where we reintroduced the $i\varepsilon$ prescription and uses the shorthand notation $y_{ab}^0 = y_{a}^0-y_b^0$. Note that we implicitly broke a symmetry of the original integrand; by fixing for instance $z_1^0 < 0$ if $y_1^0 > x_1^0$ we implicitly chose an \emph{orientation} of the edge $e_1$. The reintroduction of this symmetry will be important for the statement of the FOPT Feynman rules. 

For each edge, we can use the integral representation of the delta function, 
$\delta(z) = \int_{-\infty}^\infty \frac{\dd E}{2\pi} e^{iEz}$,
to write $\AA_G$ as integrals of oscillating exponentials
\begin{gather} \AA_G= \frac{(-ig)^{3}}{(2\pi)^{12}} \int \left[\prod_{e \in \gE}\frac{\dd z_e^0 \dd E_e/(2\pi)}{-z_e^2 + i\varepsilon}\right] \left[\prod_{i=1}^3\dd y_i^0 e^{i E_i(z_i^0-x_i^0+y_i^0)} \right] \nonumber\\ \times e^{iE_4(z_4^0-y_{12}^0)+iE_5(z_5^0-y_{13}^0)+iE_6(z_6^0-y_{23}^0)}.\label{eq:intermediatesteptriangle} \end{gather}
Eventually, we will interpret the auxiliary variables $E_e$ as the amount of energy that flows through the edge $e$ in the chosen orientation (notice the mass dimension of the $E_e$).
We are now ready to carry out the integration in the auxiliary variables $z_e^0$ of eq.~(\ref{eq:intermediatesteptriangle}). Reordering eq.~(\ref{eq:intermediatesteptriangle}) gives 
\begin{equation}
    \AA_G=
\frac{(-ig)^{3}}{(2\pi)^{12}}
\int\left[\prod_{e\in \gE} \frac{\dd E_e}{2\pi}\right] \left[\prod_{i=1}^3\dd y_i^0e^{iE_i(-x_i^0+y_i^0)}\right]  e^{-iE_4 y_{12}^0-iE_5y_{13}^0-iE_6y_{23}^0}\prod_{e\in \gE}\int\dd z_e^0\frac{ e^{i z_e^0 E_e}}{-z_e^2+i\varepsilon}.
\end{equation}
The integrals over $z_e^0$ can be performed using the residue theorem. As the denominators read $-z_e^2 + i\varepsilon = -(z_e^0)^2 + |\vec{z_e}|^2 +i\varepsilon$, the integrand has two poles, located at $z_e^0=\pm \sqrt{|\vec{z}_e|^2+i\varepsilon}$. For $E_e>0$, we close the contour of integration in the upper-half of complex plane, while for $E_e<0$, we close it in the lower half. In the former case, the pole contained in the integration contour is $z_e^0= \sqrt{|\vec{z}_e|^2+i\varepsilon}$, while in the latter case it is $z_e^0=- \sqrt{|\vec{z}_e|^2+i\varepsilon}$. Therefore, we have
\begin{equation}
\label{eq:contour}
    \int\dd z_e^0\frac{ e^{i z_e^0 E_e}}{-z_e^2+i\varepsilon} =
\frac{-2\pi i}{2|\vec{z}_e|}\left(e^{i(|\vec{z}_e|+i\ep)E_e}\Theta(E_e)+e^{-i(|\vec{z}_e|+i\ep)E_e}\Theta(-E_e)\right),
\end{equation}
where we respected the $i\ep$-prescription by adding a small imaginary part to $|\vec{z}_e|$.
The above expression means that we should treat negative and positive-energy flows through an edge differently. This is the first key step in the flow-oriented perturbation theory formalism.
We will make this aspect explicit by writing the product of the sums of two terms as a sum over $2^{|\gE|}$ terms. Each resulting term can be interpreted as an assignment of flow directions to each edge of the graph. 
We will denote such an assignment as $\bb\sigma$ that assigns  $\sigma_e =\pm 1$ to an edge $e$ to indicate a positive or negative energy flow. 
Hence, we write the product as,
\begin{equation}
    \prod_{e \in \gE}\int\dd z_e^0\frac{ e^{i z_e^0 E_e}}{-z_e^2+i\varepsilon}=\sum_{\bb\sigma\in\{\pm 1\}^6}\prod_{e \in \gE}\frac{-2\pi i}{2|\vec{z}_e|}e^{i \sigma_e( |\vec{z}_e| + i \ep)E_e}\Theta(\sigma_e E_e),
\end{equation}
where $\bb \sigma$ runs over all vectors of length $6$ with $\pm 1$ entries. Inserting this in the expression for $\AA_G$ gives
\begin{gather} \AA_G= \frac{(-ig)^{3}}{(2\pi)^{12}} \sum_{\bb\sigma\in\{\pm 1\}^6}\int \left[\prod_{e\in \gE} \frac{\dd E_e}{2i|\vec{z}_e|}e^{i\sigma_e( |\vec{z}_e| + i \ep)E_e}\Theta(\sigma_e E_e)\right] \nonumber \\
 \times \left[\prod_{i=1}^3\dd y_i^0e^{iE_i(y_i^0-x_i^0)}\right] e^{-iE_4y_{12}^0-iE_5y_{13}^0-iE_6y_{23}^0}. \end{gather}
Note that the $i\ep$ ensures convergence of the $E_e$ integrals. We can rearrange the exponentials and resolve the integration over the $y_i^0$ variables by using the integral representation of the delta function, but in reverse:
\begin{align} \AA_G= \frac{(-ig)^{3}}{i^6 (2\pi)^{9}}\sum_{\bb\sigma\in\{\pm 1\}^6} \int &\left[\prod_{e\in\gE} \frac{\dd E_e}{2|\vec{z}_e|}e^{i\sigma_e(|\vec{z}_e|+ i\ep)E_e}\Theta(\sigma_eE_e)\right]\left[\prod_{i=1}^3 e^{-iE_ix_i^0}\right] \nonumber\\ &\times \delta(E_1-E_4-E_5)\delta(E_2+E_4-E_6)\delta(E_3+E_5+E_6)\,. \end{align}
We observe that each of the delta functions is associated to an internal vertex $y_v^0$ and that they enforce \emph{energy-conservation} at each internal vertex.
Performing the change of variables $E_e\rightarrow\sigma_e E_e$, and resolving the theta functions gives
\begin{align} \label{eq:integralA123} \AA_G= \frac{(-ig)^{3}}{i^6 (2\pi)^{9}} &\sum_{\bb\sigma\in\{\pm 1\}^6}\left[ \prod_{e\in \gE} \int_{0}^\infty\frac{\dd E_e}{2|\vec{z}_e|}e^{i (|\vec{z}_e|+i\ep)E_e}\right]\left[\prod_{i=1}^3 e^{-i \sigma_i E_ix_i^0}\right] \nonumber\\
 &\times \delta(\sigma_1E_1-\sigma_4E_4-\sigma_5E_5)\delta(\sigma_2E_2+\sigma_4E_4-\sigma_6E_6)\delta(\sigma_3E_3+\sigma_5E_5+\sigma_6E_6) . \end{align}
Whenever the sign vector $\bb\sigma$ is such that the argument of a delta function is either a strictly negative or positive sum of energies, then the integral is zero, as such sums cannot vanish under the constraint $E_e > 0$. 
This implies that only a subset of the vectors $\bb\sigma$ contribute. We will make heavy use of a diagrammatic interpretation of this integral in order to correctly understand which such vectors lead to a non-zero contribution. When  we introduced in eq.~\eqref{eq:triangle_dirac_deltas} the auxiliary variables that correspond to time differences, we implicitly chose an orientation for the graph. If $z_j^0=x_j^0-y_j^0$, then we choose the orientation of the $j$-th edge such that energy flows from the vertex $y_j$ to the vertex $x_j$. Analogously, if $z_4^0=y_1^0-y_2^0$, then the $4$-th edge orientation flows from $y_2$ to $y_1$. We can depict this chosen orientation as a directed graph, or \emph{digraph},
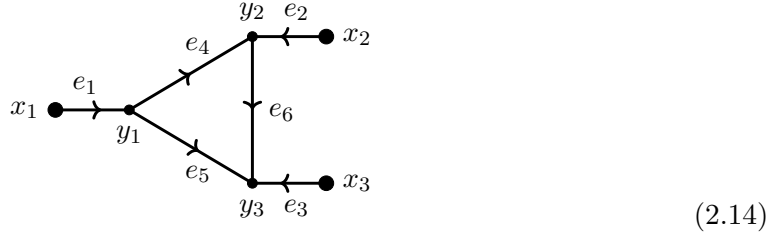
\begin{equation}
\resizebox{5cm}{!}{%
\begin{tikzpicture} \begin{feynman} \vertex(1); \vertex[above right = 1.5cm and 2.5cm of 1](2); \vertex[below right = 1.5cm and 2.5cm of 1](3); \vertex[left = 1.5cm of 1](E1); \vertex[right = 1.5cm of 2](E2); \vertex[right = 1.5cm of 3](E3); \vertex[left = 0.2cm of E1](L1) {\scalebox{1.5}{$x_1$}}; \vertex[right = 0.2cm of E2](L2) {\scalebox{1.5}{$x_2$}}; \vertex[right = 0.2cm of E3](L3) {\scalebox{1.5}{$x_3$}}; \vertex[below = 0.2cm of 1](L1_I) {\scalebox{1.5}{$y_1$}}; \vertex[above = 0.2cm of 2](L2_I) {\scalebox{1.5}{$y_2$}}; \vertex[below = 0.2cm of 3](L3_i) {\scalebox{1.5}{$y_3$}}; \vertex[above left = 0.2cm and 0.5cm of 1](edge1) {\scalebox{1.5}{$e_1$}}; \vertex[above right = 0.2cm and 0.5cm of 2](edge2) {\scalebox{1.5}{$e_2$}}; \vertex[below right = 0.2cm and 0.5cm of 3](edge3) {\scalebox{1.5}{$e_3$}}; \vertex[above right = 1cm and 1cm of 1](edge4) {\scalebox{1.5}{$e_4$}}; \vertex[below right = 1cm and 1cm of 1](edge5) {\scalebox{1.5}{$e_5$}}; \vertex[below right = 1.25cm and 0.2cm of 2](edge6) {\scalebox{1.5}{$e_6$}}; \diagram*[large]{ (1) -- [->-,line width=0.6mm] (2) -- [->-,line width=0.6mm] (3) -- [-<-,line width=0.6mm] (1), (1) -- [-<-,line width=0.6mm] (E1), (2) -- [-<-,line width=0.6mm] (E2), (3) -- [-<-,line width=0.6mm] (E3), }; \path[draw=black, fill=black] (E1) circle[radius=0.15]; \path[draw=black, fill=black] (E2) circle[radius=0.15]; \path[draw=black, fill=black] (E3) circle[radius=0.15]; \path[draw=black, fill=black] (1) circle[radius=0.1]; \path[draw=black, fill=black] (2) circle[radius=0.1]; \path[draw=black, fill=black] (3) circle[radius=0.1]; \end{feynman} \end{tikzpicture}
}
\label{eq:triangle_original_orientation}
\end{equation}
The sign vector $\bb \sigma$ can now be interpreted as flipping the orientations of edges in this digraph. For example, if $\bb\sigma=(-1,...,-1)$, then the corresponding digraph would have all edge orientations flipped. If instead $\bb\sigma=(1,-1,-1,1,1,-1)$, then the corresponding digraph is
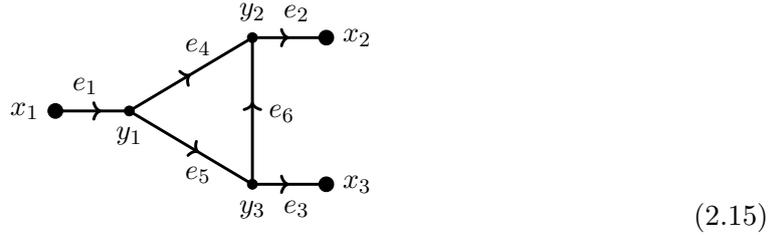
\begin{equation}
\resizebox{5cm}{!}{%
\begin{tikzpicture} \begin{feynman} \vertex(1); \vertex[above right = 1.5cm and 2.5cm of 1](2); \vertex[below right = 1.5cm and 2.5cm of 1](3); \vertex[left = 1.5cm of 1](E1); \vertex[right = 1.5cm of 2](E2); \vertex[right = 1.5cm of 3](E3); \vertex[left = 0.2cm of E1](L1) {\scalebox{1.5}{$x_1$}}; \vertex[right = 0.2cm of E2](L2) {\scalebox{1.5}{$x_2$}}; \vertex[right = 0.2cm of E3](L3) {\scalebox{1.5}{$x_3$}}; \vertex[below = 0.2cm of 1](L1_I) {\scalebox{1.5}{$y_1$}}; \vertex[above = 0.2cm of 2](L2_I) {\scalebox{1.5}{$y_2$}}; \vertex[below = 0.2cm of 3](L3_i) {\scalebox{1.5}{$y_3$}}; \vertex[above left = 0.2cm and 0.5cm of 1](edge1) {\scalebox{1.5}{$e_1$}}; \vertex[above right = 0.2cm and 0.5cm of 2](edge2) {\scalebox{1.5}{$e_2$}}; \vertex[below right = 0.2cm and 0.5cm of 3](edge3) {\scalebox{1.5}{$e_3$}}; \vertex[above right = 1cm and 1cm of 1](edge4) {\scalebox{1.5}{$e_4$}}; \vertex[below right = 1cm and 1cm of 1](edge5) {\scalebox{1.5}{$e_5$}}; \vertex[below right = 1.25cm and 0.2cm of 2](edge6) {\scalebox{1.5}{$e_6$}}; \diagram*[large]{ (1) -- [->-,line width=0.6mm] (2) -- [-<-,line width=0.6mm] (3) -- [-<-,line width=0.6mm] (1), (1) -- [-<-,line width=0.6mm] (E1), (2) -- [->-,line width=0.6mm] (E2), (3) -- [->-,line width=0.6mm] (E3), }; \path[draw=black, fill=black] (E1) circle[radius=0.15]; \path[draw=black, fill=black] (E2) circle[radius=0.15]; \path[draw=black, fill=black] (E3) circle[radius=0.15]; \path[draw=black, fill=black] (1) circle[radius=0.1]; \path[draw=black, fill=black] (2) circle[radius=0.1]; \path[draw=black, fill=black] (3) circle[radius=0.1]; \end{feynman} \end{tikzpicture}
}
\label{eq:triangle_right_orientation}
\end{equation}
As we can see from this example, the sum over all vectors $\bb\sigma$ is actually equal to the sum over all possible orientations of the triangle graph. The energy-conservation conditions imposed by the delta functions in eq.~\eqref{eq:integralA123} can then be interpreted as enforcing the conservation of energies at any internal vertex. As an example, let us look at the contribution of the sign vector $\bb\sigma=(1,\ldots,1)$ orientation to eq.~\eqref{eq:triangle_original_orientation},
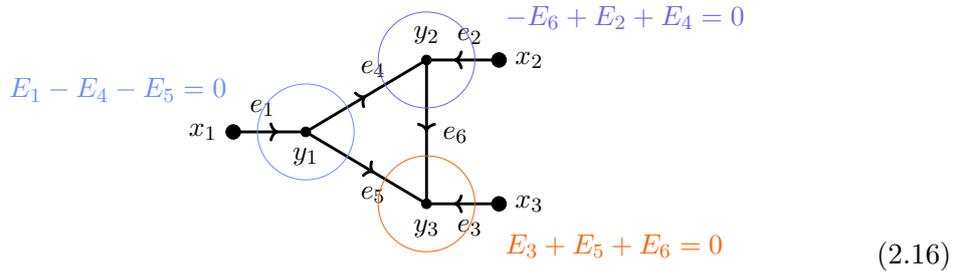
\begin{align} \resizebox{10cm}{!}{ \begin{tikzpicture} \begin{feynman} \vertex(1); \vertex[above right = 1.5cm and 2.5cm of 1](2); \vertex[below right = 1.5cm and 2.5cm of 1](3); \vertex[left = 1.5cm of 1](E1); \vertex[right = 1.5cm of 2](E2); \vertex[right = 1.5cm of 3](E3); \vertex[above left = 0.5cm and 0.cm of E1](mom1) {\scalebox{1.5}{$\color{col1}E_1-E_4-E_5=0$}}; \vertex[above right = 0.5cm and 0.cm of E2](mom2) {\scalebox{1.5}{$\color{col2} -E_6+E_2+E_4=0$}}; \vertex[below right = 0.5cm and 0.cm of E3](mom3) {\scalebox{1.5}{$\color{col3} E_3+E_5+E_6=0$}}; \vertex[left = 0.2cm of E1](L1) {\scalebox{1.5}{$x_1$}}; \vertex[right = 0.2cm of E2](L2) {\scalebox{1.5}{$x_2$}}; \vertex[right = 0.2cm of E3](L3) {\scalebox{1.5}{$x_3$}}; \vertex[below = 0.2cm of 1](L1_I) {\scalebox{1.5}{$y_1$}}; \vertex[above = 0.2cm of 2](L2_I) {\scalebox{1.5}{$y_2$}}; \vertex[below = 0.2cm of 3](L3_i) {\scalebox{1.5}{$y_3$}}; \vertex[above left = 0.2cm and 0.5cm of 1](edge1) {\scalebox{1.5}{$e_1$}}; \vertex[above right = 0.2cm and 0.5cm of 2](edge2) {\scalebox{1.5}{$e_2$}}; \vertex[below right = 0.2cm and 0.5cm of 3](edge3) {\scalebox{1.5}{$e_3$}}; \vertex[above right = 1cm and 1cm of 1](edge4) {\scalebox{1.5}{$e_4$}}; \vertex[below right = 1cm and 1cm of 1](edge5) {\scalebox{1.5}{$e_5$}}; \vertex[below right = 1.25cm and 0.2cm of 2](edge6) {\scalebox{1.5}{$e_6$}}; \diagram*[large]{ (1) -- [->-,line width=0.6mm] (2) -- [->-,line width=0.6mm] (3) -- [-<-,line width=0.6mm] (1), (1) -- [-<-,line width=0.6mm] (E1), (2) -- [-<-,line width=0.6mm] (E2), (3) -- [-<-,line width=0.6mm] (E3), }; \path[draw=black, fill=black] (E1) circle[radius=0.15]; \path[draw=black, fill=black] (E2) circle[radius=0.15]; \path[draw=black, fill=black] (E3) circle[radius=0.15]; \path[draw=black, fill=black] (1) circle[radius=0.1]; \path[draw=black, fill=black] (2) circle[radius=0.1]; \path[draw=black, fill=black] (3) circle[radius=0.1]; \path[draw=col1] (1) circle[radius=1.]; \path[draw=col2] (2) circle[radius=1.]; \path[draw=col3] (3) circle[radius=1.]; \end{feynman} \end{tikzpicture} } \label{eq:triangle_zero_vertex} \end{align}
where we included the conditions imposed by the delta functions into the graphical representation.
The energy conservation condition in orange of the bottom right vertex is a sum of positive energies. Such an energy conservation condition can never be satisfied. It follows that this orientation gives no contribution to the FOPT representation of the triangle. 
It is actually quite easy to see that the solution to the delta function constraint coincides with the physically intuitive picture 
of a realizable energy flow through the diagram along the indicated directions.
A digraph gives a non-zero contribution if
edges are followed in the positive orientation, and two conditions are fulfilled: (i) we must be able to reach each vertex by starting from some external vertex, and (ii) we must be able to reach some external vertex if we start form any vertex. Even though the $y_3$ vertex in the digraph in eq.~\eqref{eq:triangle_zero_vertex} can be reached from many external vertices by following a positive route, we cannot 
reach any external vertex if we start from it. Hence, there is no proper energy-flow possible with the assigned orientation.
The orientation of eq.~\eqref{eq:triangle_right_orientation} can be depicted with its associated delta function arguments as
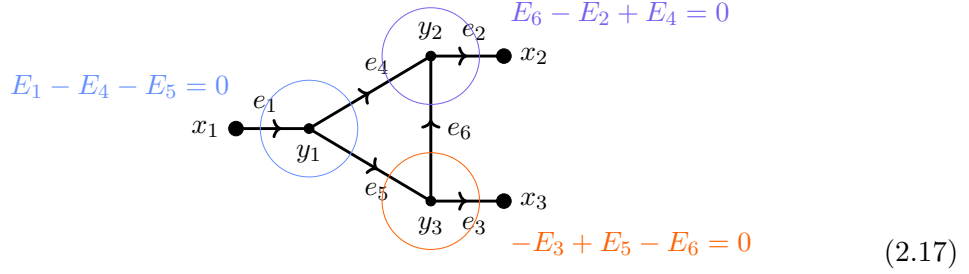
\begin{equation}
\resizebox{10cm}{!}{%
\begin{tikzpicture} \begin{feynman} \vertex(1); \vertex[above right = 1.5cm and 2.5cm of 1](2); \vertex[below right = 1.5cm and 2.5cm of 1](3); \vertex[left = 1.5cm of 1](E1); \vertex[right = 1.5cm of 2](E2); \vertex[right = 1.5cm of 3](E3); \vertex[above left = 0.5cm and 0.cm of E1](mom1) {\scalebox{1.5}{$\color{col1}E_1-E_4-E_5=0$}}; \vertex[above right = 0.5cm and 0.cm of E2](mom2) {\scalebox{1.5}{$\color{col2} E_6-E_2+E_4=0$}}; \vertex[below right = 0.5cm and 0.cm of E3](mom3) {\scalebox{1.5}{$\color{col3} -E_3+E_5-E_6=0$}}; \vertex[left = 0.2cm of E1](L1) {\scalebox{1.5}{$x_1$}}; \vertex[right = 0.2cm of E2](L2) {\scalebox{1.5}{$x_2$}}; \vertex[right = 0.2cm of E3](L3) {\scalebox{1.5}{$x_3$}}; \vertex[below = 0.2cm of 1](L1_I) {\scalebox{1.5}{$y_1$}}; \vertex[above = 0.2cm of 2](L2_I) {\scalebox{1.5}{$y_2$}}; \vertex[below = 0.2cm of 3](L3_i) {\scalebox{1.5}{$y_3$}}; \vertex[above left = 0.2cm and 0.5cm of 1](edge1) {\scalebox{1.5}{$e_1$}}; \vertex[above right = 0.2cm and 0.5cm of 2](edge2) {\scalebox{1.5}{$e_2$}}; \vertex[below right = 0.2cm and 0.5cm of 3](edge3) {\scalebox{1.5}{$e_3$}}; \vertex[above right = 1cm and 1cm of 1](edge4) {\scalebox{1.5}{$e_4$}}; \vertex[below right = 1cm and 1cm of 1](edge5) {\scalebox{1.5}{$e_5$}}; \vertex[below right = 1.25cm and 0.2cm of 2](edge6) {\scalebox{1.5}{$e_6$}}; \diagram*[large]{ (1) -- [->-,line width=0.6mm] (2) -- [-<-,line width=0.6mm] (3) -- [-<-,line width=0.6mm] (1), (1) -- [-<-,line width=0.6mm] (E1), (2) -- [->-,line width=0.6mm] (E2), (3) -- [->-,line width=0.6mm] (E3), }; \path[draw=black, fill=black] (E1) circle[radius=0.15]; \path[draw=black, fill=black] (E2) circle[radius=0.15]; \path[draw=black, fill=black] (E3) circle[radius=0.15]; \path[draw=black, fill=black] (1) circle[radius=0.1]; \path[draw=black, fill=black] (2) circle[radius=0.1]; \path[draw=black, fill=black] (3) circle[radius=0.1]; \path[draw=col1] (1) circle[radius=1.]; \path[draw=col2] (2) circle[radius=1.]; \path[draw=col3] (3) circle[radius=1.]; \end{feynman} \end{tikzpicture}
}
\label{eq:triangle_flipped_deltas}
\end{equation}
In this case, the combinatorial condition is fulfilled and we can 
comply with the previously problematic condition imposed by the delta function from the bottom right vertex. 
Hence, this orientation provides a non-zero contribution. 

This utility of the graphical representation suggests a definition of the Feynman integral $\AA_{G,\bb \sigma}$ associated to a single digraph (i.e.~a graph $G$ with an orientation $\bb \sigma$). We define
\begin{gather} \begin{gathered} \AA_{G,\boldsymbol{\sigma}}(x_1,x_2,x_3,\vec{y}_1,\vec{y}_2,\vec{y}_3)= \frac{(-ig)^{3}}{i^6 (2\pi)^{9}} \left[ \prod_{e\in \gE} \int_{0}^\infty\frac{\dd E_e}{2|\vec{z}_e|}e^{i (|\vec{z}_e|+ i\ep)E_e}\right]\left[\prod_{i=1}^3 e^{-i \sigma_i E_ix_i^0}\right] \\  \times\, \delta(\sigma_1E_1-\sigma_4E_4-\sigma_5E_5)\delta(\sigma_2E_2+\sigma_4E_4-\sigma_6E_6)\delta(\sigma_3E_3+\sigma_5E_5+\sigma_6E_6)\,, \end{gathered} \end{gather}
such that 
\begin{align} \AA_{G}(x_1,x_2,x_3,\vec{y}_1,\vec{y}_2,\vec{y}_3) = \sum_{\bb\sigma} \AA_{G,\bb \sigma} (x_1,x_2,x_3,\vec{y}_1,\vec{y}_2,\vec{y}_3)\,, \end{align}
where  the sum runs over all possible orientations of the graph $G$. Each such orientation gives rise to a digraph $(G,\bb \sigma)$.

We will  derive a compact representation for the functions $\AA_{G,\bb \sigma} (x_1,x_2,x_3,\vec{y}_1,\vec{y}_2,\vec{y}_3)$,
and illustrate the derivation for this compact representation with an example. Take the orientation from eq.~\eqref{eq:triangle_right_orientation} and set $\bb\sigma=(1,-1,-1,1,1,-1)$. %
In this case we have
\begin{align} \begin{aligned} \label{eq:trangle_ori123}  \AA_{G,\bb \sigma} = \frac{(-ig)^{3}}{i^6 (2\pi)^{9}} &\left[\prod_{e\in\gE} \int_0^\infty \frac{\dd E_e}{2|\vec{z}_e|}e^{i (|\vec{z}_e|+i\ep)E_e}\right]\left[\prod_{i=1}^3 e^{-i \sigma_i E_i x_i^0}\right] \\
 &\times\delta(E_1-E_4-E_5)\delta(-E_2+E_4+E_6)\delta(-E_3+E_5-E_6). \end{aligned} \end{align}
To resolve the delta functions we need to choose a set of linearly independent energies. We choose $E_3, \, E_4, \, E_6$, which gives as linearly dependent energies
\begin{align} E_1=E_3+E_4+E_6, \quad E_2=E_4+E_6, \quad E_5=E_3+E_6. \label{eq:basis_choice_triangle} \end{align}
We see that our choice has the property of expressing the dependent energies $E_1,E_2,E_5$ as strictly \emph{positive sums} of independent energies $E_3,E_4,E_6$. This is an important property that shall feature prominently in our derivation. 
In order to achieve a diagrammatic understanding of this property, let us look at the \emph{completed graph} for this orientation. 
Graph completion plays an important role within the theory of (dual)-conformal four-point Feynman integrals and graphical functions (see e.g.~\cite{Borinsky:2022lds}).
The completed graph is obtained from the original one by gathering all external vertices into one special vertex $\circ$. For a graph $G$ with an orientation $\bb \sigma$, i.e. $(G,\bb \sigma)$, we denote the associated completed digraph as $(G,\bb \sigma)^\circ$.
For our present triangle example from eq.~(\ref{eq:triangle_flipped_deltas}), the completed graph looks as follows: 
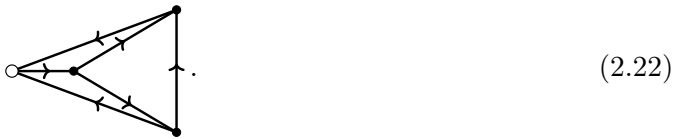
\begin{align} \raisebox{-0.8cm}{\resizebox{2.5cm}{!}{ \begin{tikzpicture} \begin{feynman} \vertex(1); \vertex[above right = 1.5cm and 2.5cm of 1](2); \vertex[below right = 1.5cm and 2.5cm of 1](3); \vertex[left = 1.5cm of 1](E1); \vertex[right = 1.5cm of 2](E2); \vertex[right = 1.5cm of 3](E3); \diagram*[large]{ (1) -- [->-,line width=0.6mm] (2) -- [-<-,line width=0.6mm] (3) -- [-<-,line width=0.6mm] (1), (1) -- [-<-,line width=0.6mm] (E1), (2) -- [->-,line width=0.6mm] (E1), (3) -- [->-,line width=0.6mm] (E1), }; \path[draw=black, fill=white,thick] (E1) circle[radius=0.15]; \path[draw=black, fill=black] (1) circle[radius=0.1]; \path[draw=black, fill=black] (2) circle[radius=0.1]; \path[draw=black, fill=black] (3) circle[radius=0.1]; \end{feynman} \end{tikzpicture} }}. \end{align}
The condition that a proper energy-conserving flow exists on the original graph translates to a graph-theoretical property of the 
completed graph, viz.~the requirement that the completed graph is \emph{strongly connected} (see e.g.~\cite[Ch.~10]{bondy1976graph}). 
A digraph is strongly connected if we can reach each vertex from any other vertex by taking some positively oriented route. 
The contribution of a digraph in FOPT will only be non-zero if the associated completed graph is strongly connected. Furthermore, a strongly connected digraph has a unique set of \emph{cycles}. A cycle is defined as a subset of edges of an oriented graph that compose a positive-energy oriented path starting at a vertex and coming back to that same vertex.

In the running example of the triangle, this graph has exactly three oriented cycles $\{\pp_1,\pp_2,\pp_3\}$ (depicted with coloured edges): 
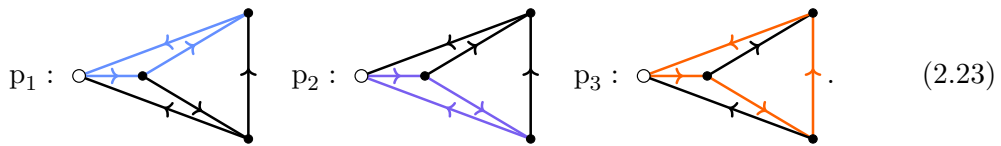
\begin{equation}
     \pp_1: \  
    \raisebox{-0.8cm}{\resizebox{2.5cm}{!}{%
\begin{tikzpicture} \begin{feynman} \vertex(1); \vertex[above right = 1.5cm and 2.5cm of 1](2); \vertex[below right = 1.5cm and 2.5cm of 1](3); \vertex[left = 1.5cm of 1](E1); \vertex[right = 1.5cm of 2](E2); \vertex[right = 1.5cm of 3](E3); \diagram*[large]{ (1) -- [->-,line width=0.6mm, col1] (2) -- [-<-,line width=0.6mm] (3) -- [-<-,line width=0.6mm] (1), (1) -- [-<-,line width=0.6mm, col1] (E1), (2) -- [->-,line width=0.6mm, col1] (E1), (3) -- [->-,line width=0.6mm] (E1), }; \path[draw=black, fill=white,thick] (E1) circle[radius=0.15]; \path[draw=black, fill=black] (1) circle[radius=0.1]; \path[draw=black, fill=black] (2) circle[radius=0.1]; \path[draw=black, fill=black] (3) circle[radius=0.1]; \end{feynman} \end{tikzpicture}
}}\quad 
\pp_2: \  
    \raisebox{-0.8cm}{\resizebox{2.5cm}{!}{%
\begin{tikzpicture} \begin{feynman} \vertex(1); \vertex[above right = 1.5cm and 2.5cm of 1](2); \vertex[below right = 1.5cm and 2.5cm of 1](3); \vertex[left = 1.5cm of 1](E1); \vertex[right = 1.5cm of 2](E2); \vertex[right = 1.5cm of 3](E3); \diagram*[large]{ (1) -- [->-,line width=0.6mm] (2) -- [-<-,line width=0.6mm] (3) -- [-<-,line width=0.6mm, col2] (1), (1) -- [-<-,line width=0.6mm, col2] (E1), (2) -- [->-,line width=0.6mm] (E1), (3) -- [->-,line width=0.6mm, col2] (E1), }; \path[draw=black, fill=white,thick] (E1) circle[radius=0.15]; \path[draw=black, fill=black] (1) circle[radius=0.1]; \path[draw=black, fill=black] (2) circle[radius=0.1]; \path[draw=black, fill=black] (3) circle[radius=0.1]; \end{feynman} \end{tikzpicture}
}}\quad 
\pp_3: \  
    \raisebox{-0.8cm}{\resizebox{2.5cm}{!}{%
\begin{tikzpicture} \begin{feynman} \vertex(1); \vertex[above right = 1.5cm and 2.5cm of 1](2); \vertex[below right = 1.5cm and 2.5cm of 1](3); \vertex[left = 1.5cm of 1](E1); \vertex[right = 1.5cm of 2](E2); \vertex[right = 1.5cm of 3](E3); \diagram*[large]{ (1) -- [->-,line width=0.6mm] (2) -- [-<-,line width=0.6mm, col3] (3) -- [-<-,line width=0.6mm, col3] (1), (1) -- [-<-,line width=0.6mm, col3] (E1), (2) -- [->-,line width=0.6mm, col3] (E1), (3) -- [->-,line width=0.6mm] (E1), }; \path[draw=black, fill=white,thick] (E1) circle[radius=0.15]; \path[draw=black, fill=black] (1) circle[radius=0.1]; \path[draw=black, fill=black] (2) circle[radius=0.1]; \path[draw=black, fill=black] (3) circle[radius=0.1]; \end{feynman} \end{tikzpicture}
}}.
\end{equation}
Each of these three cycles has exactly one edge that is not contained in any other cycle. For $\pp_1$, using the original labels, it is $e_4$. For $\pp_2$, it is $e_3$ and for $\pp_3$, it is $e_6$. 
This choice of edges gives exactly the basis of energies that we used to write eq.~\eqref{eq:basis_choice_triangle}, namely $E_3, \, E_4, \, E_6$. It turns out that for any strongly connected orientation such a choice can be made.
Furthermore, the three cycles above are \emph{canonical}. We can only find exactly these three cycles of the graph if we insist on the property of positive-energy flow. This is in contrast to the usual covariant momentum representation, where we have many choices to route the momenta through the diagram. In more mathematical terms, there is (up to permutation) a unique basis of the first homology\footnote{The first homology of a graph is the vector space spanned by all its loops.} of $G^\circ$ in which each basis vector is a simple, positively oriented cycle.
Opening up the $\circ$-vertex leads to an interpretation of $\pp_1,\pp_2,\pp_3$ as three paths through the diagram that connect different external vertices,
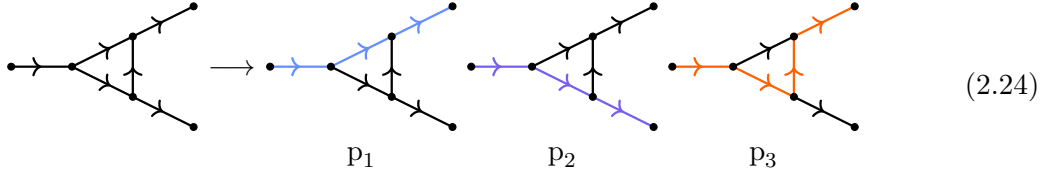
\begin{align} \label{eq:triangle_paths} \centering \begin{array}{cccccccccccc} &\def\scale{.8} \begin{tikzpicture}[baseline={([yshift=-0.7ex]0,0)}] \coordinate (i00) at (-2*\scale,0); \coordinate (i1) at (-\scale,0); \draw[->-,line width=0.3mm] (i00) -- (i1) node[midway,right] {}; \filldraw (i00) circle (1.3pt); \coordinate[] (i2) at (0,.5*\scale); \coordinate[] (i21) at (0,-.5*\scale); \coordinate[] (v1) at (\scale,\scale); \coordinate[] (v2) at (\scale,-\scale); \draw[-<-,line width=0.3mm] (i2) -- (i1) node[midway,above left] {}; \draw[-<-,line width=0.3mm] (i21) -- (i1) node[midway,below left] {}; \draw[->-,line width=0.3mm] (i21) -- (v2) node[midway,below left] {}; \draw[->-,line width=0.3mm] (i21) -- (i2) node[midway,right] {}; \draw[->-,line width=0.3mm] (i2) -- (v1) node[midway,above left] {}; \filldraw (v1) circle (1.3pt); \filldraw (i21) circle (1.3pt); \filldraw (v2) circle (1.3pt); \filldraw (i1) circle (1.3pt); \filldraw (i2) circle (1.3pt); \end{tikzpicture}&\longrightarrow&\def\scale{.8} \begin{tikzpicture}[baseline={([yshift=-0.7ex]0,0)}] \coordinate (i00) at (-2*\scale,0); \coordinate (i1) at (-\scale,0); \draw[col1,->-,line width=0.3mm] (i00) -- (i1) node[midway,right] {}; \filldraw (i00) circle (1.3pt); \coordinate[] (i2) at (0,.5*\scale); \coordinate[] (i21) at (0,-.5*\scale); \coordinate[] (v1) at (\scale,\scale); \coordinate[] (v2) at (\scale,-\scale); \draw[col1,-<-,line width=0.3mm] (i2) -- (i1) node[midway,above left] {}; \draw[-<-,line width=0.3mm] (i21) -- (i1) node[midway,below left] {}; \draw[->-,line width=0.3mm] (i21) -- (v2) node[midway,below left] {}; \draw[->-,line width=0.3mm] (i21) -- (i2) node[midway,right] {}; \draw[col1,->-,line width=0.3mm] (i2) -- (v1) node[midway,above left] {}; \filldraw (v1) circle (1.3pt); \filldraw (i21) circle (1.3pt); \filldraw (v2) circle (1.3pt); \filldraw (i1) circle (1.3pt); \filldraw (i2) circle (1.3pt); \end{tikzpicture}&\def\scale{.8} \begin{tikzpicture}[baseline={([yshift=-0.7ex]0,0)}] \coordinate (i00) at (-2*\scale,0); \coordinate (i1) at (-\scale,0); \draw[col2,->-,line width=0.3mm] (i00) -- (i1) node[midway,right] {}; \filldraw (i00) circle (1.3pt); \coordinate[] (i2) at (0,.5*\scale); \coordinate[] (i21) at (0,-.5*\scale); \coordinate[] (v1) at (\scale,\scale); \coordinate[] (v2) at (\scale,-\scale); \draw[-<-,line width=0.3mm] (i2) -- (i1) node[midway,above left] {}; \draw[col2,-<-,line width=0.3mm] (i21) -- (i1) node[midway,below left] {}; \draw[col2,->-,line width=0.3mm] (i21) -- (v2) node[midway,below left] {}; \draw[->-,line width=0.3mm] (i21) -- (i2) node[midway,right] {}; \draw[->-,line width=0.3mm] (i2) -- (v1) node[midway,above left] {}; \filldraw (v1) circle (1.3pt); \filldraw (i21) circle (1.3pt); \filldraw (v2) circle (1.3pt); \filldraw (i1) circle (1.3pt); \filldraw (i2) circle (1.3pt); \end{tikzpicture}&\def\scale{.8} \begin{tikzpicture}[baseline={([yshift=-0.7ex]0,0)}] \coordinate (i00) at (-2*\scale,0); \coordinate (i1) at (-\scale,0); \draw[col3,->-,line width=0.3mm] (i00) -- (i1) node[midway,right] {}; \filldraw (i00) circle (1.3pt); \coordinate[] (i2) at (0,.5*\scale); \coordinate[] (i21) at (0,-.5*\scale); \coordinate[] (v1) at (\scale,\scale); \coordinate[] (v2) at (\scale,-\scale); \draw[-<-,line width=0.3mm] (i2) -- (i1) node[midway,above left] {}; \draw[col3,-<-,line width=0.3mm] (i21) -- (i1) node[midway,below left] {}; \draw[->-,line width=0.3mm] (i21) -- (v2) node[midway,below left] {}; \draw[col3,->-,line width=0.3mm] (i21) -- (i2) node[midway,right] {}; \draw[col3,->-,line width=0.3mm] (i2) -- (v1) node[midway,above left] {}; \filldraw (v1) circle (1.3pt); \filldraw (i21) circle (1.3pt); \filldraw (v2) circle (1.3pt); \filldraw (i1) circle (1.3pt); \filldraw (i2) circle (1.3pt); \end{tikzpicture}\\
&&&\pp_1&\pp_2&\pp_3 \end{array} \end{align}
With our chosen basis, we are now ready to solve the delta functions in eq.~\eqref{eq:trangle_ori123}. Carrying out the 
$E_1,E_2,E_5$ integrals we obtain
\begin{gather} \begin{gathered} \label{eq:triangle_ori123_fin}  \AA_{G,\bb \sigma} = \frac{(-ig)^{3}}{i^6 (2\pi)^{9}} \left[ \int_{0}^\infty \frac{\mathrm{d}E_3\mathrm{d}E_4\mathrm{d}E_6}{\prod_{e\in\gE}2|\vec{z}_e|}\right] \times\\ \times e^{iE_4(|\vec{z}_1|+|\vec{z}_2|+|\vec{z}_4|+x_2^0-x_1^0+i\ep)} e^{iE_3(|\vec{z}_1|+|\vec{z}_3|+|\vec{z}_5|+x_3^0-x_1^0+i\ep)} e^{iE_6(|\vec{z}_1|+|\vec{z}_2|+|\vec{z}_5|+|\vec{z}_6|+x_2^0-x_1^0+i\ep)} . \end{gathered} \end{gather}
The remaining integrations are readily performed, which gives 
\begin{gather} \begin{gathered} \label{eq:triangle_fopt}  \AA_{G,\bb \sigma} = \frac{(-ig)^{3}}{i^6 (2\pi)^{9}} \frac{i^3}{\prod_{e\in \gE}2|\vec{z}_e|} \frac{1}{(\gamma_{\pp_1}+x_{12}^0+i\ep)(\gamma_{\pp_2}+x_{13}^0+i\ep)(\gamma_{\pp_3}+x_{12}^0+i\ep)} \end{gathered} \intertext{where $\gamma_{\pp_1},\gamma_{\pp_2},\gamma_{\pp_3}$ are the \emph{path lengths} associated to the cycles $\pp_1, \pp_2, \pp_3$,} \begin{aligned} \gamma_{\pp_1} &= |\vec{z}_1|+|\vec{z}_4|+|\vec{z}_2|,\\
\gamma_{\pp_2} &= |\vec{z}_1|+|\vec{z}_5|+|\vec{z}_3|,\\
\gamma_{\pp_3} &= |\vec{z}_1|+|\vec{z}_5|+|\vec{z}_6|+|\vec{z}_2|. \end{aligned} \end{gather}
Eq.~\eqref{eq:triangle_fopt} is the full FOPT expression associated 
to the digraph~\eqref{eq:triangle_right_orientation}.

\subsection{Derivation of the general FOPT Feynman rules}
\label{sec:derivation}
Having discussed the triangle diagram in detail, we can proceed to the general derivation of the 
FOPT Feynman rules. 
\subsubsection{Cauchy integrations}
We will perform the integrals over the time components $y^0_v$ of eq.~\eqref{eq:FI} analytically via the residue theorem. In fact, we are only interested in the partially integrated version $\AA_G$ of $A_G$ which we already discussed in the triangle example:
\begin{gather} \begin{gathered} \label{eq:mathcal_A} \AA_G(x_1,\ldots,x_{|\gVE|},\vec{y}_1, \ldots, \vec{y}_{|\gVI|})= \frac{(-ig)^{|\gVI|}}{(2\pi)^{2|\gE|}} \left[ \prod_{v \in \gVI} \int \dd y_v^0 \right] \prod_{e \in \gE}\frac{1}{-z_e^2+i\ep}\;, \end{gathered} \end{gather}
with the relation 
$A_G(x_1,\ldots,x_{|\gVE|}) = \left[ \prod_{v \in \gVI} \int \dd^3 \vec y_v \right] \AA_G(x_1,\ldots,x_{|\gVE|},\vec{y}_1, \ldots, \vec{y}_{|\gVI|})$.

To perform the integration in eq.~\eqref{eq:mathcal_A} in full generality, it is convenient to introduce some additional notation.
The edge displacement four-vectors $z_e^\mu$ can be written  as
\begin{equation}
z_e^\mu= \sum_{v \in \gVI} \EI_{e,v} \, y_v^\mu + \sum_{a \in \gVE} \EE_{e,a} \, x_a^\mu\;,
\end{equation}
where $\EI_{e,v}$ and $\EE_{e,a}$ are \emph{incidence matrices} of the graph $G$. To calculate these matrices, we have to pick some arbitrary orientation of the edges of the underlying graph and set $\EI_{e,v} = +1$ $(-1)$ if the edge $e$ is directed away from (towards) the internal vertex $v$. A matrix entry $\EI_{e,v}$ is $0$ if the edge $e$ is not incident to the vertex $v$. The matrix $\EE_{e,a}$ is defined analogously, but only for external vertices labeled by the $a$ index. The initial choice of an orientation of the edges, which is necessary to define these matrices, is arbitrary and the value of the integral does not depend on this choice.

Slightly abusing the previous notation, we can introduce one auxiliary integration variable $z_e^0$ for each edge, fixed to be the time difference between its incident vertices,
\begin{equation}
    \AA_G=
\frac{(-ig)^{|\gVI|}}{(2\pi)^{2|\gE|}}
\left( \prod_{v \in \gVI} \int \dd y_v^0 \right)
\left( \prod_{e \in \gE} 
\int_{-\infty}^{\infty}
\dd z^0_e 
\frac{
\delta
\left( z^0_e - 
 \EI_{e,v} \, y_v^0 -  \EE_{e,a} \, x_a^0 \right)
}{-{z_e^0}^2 + {\vec z_e}^{\,2} +i\ep}
\right)\,,
\end{equation}
where we implicitly sum over the indices $v$ and $a$ in the argument of the delta function.
We then again use the integral representation of the $\delta$ function, $\delta(z) = \int_{-\infty}^\infty \frac{\dd E}{2 \pi} e^{i E z}$ (where $E$ is an auxiliary variable with dimensions of energy),  to arrive at
\begin{equation}
    \AA_G=
\frac{(-ig)^{|\gVI|}}{(2\pi)^{2|\gE|}}
\left( \prod_{v \in \gVI} \int \dd y_v^0 \right)
\left( \prod_{e \in \gE} 
\int_{-\infty}^{\infty}
\dd z^0_e 
\int_{-\infty}^{\infty}
\frac{\dd E_e}{2\pi}
\frac{
e^{
 iE_e
\left( z^0_e - 
\EI_{e,v} \, y_v^0 - \EE_{e,a} \, x_a^0 \right)
}
}{-{z_e^0}^2 + {\vec z_e}^{\,2} +i\ep}
\right)\;.
\end{equation}
Carrying out the $z_e^0$ integrations using Cauchy's theorem is now straightforward (see eq.~\eqref{eq:contour}). It gives rise to the sum of two terms that we will interpret as a positive (closing the contour in the upper half-plane) and a negative energy contribution (closing the contour in the lower half-plane):
\begin{gather} \AA_G= \frac{(-ig)^{|\gVI|} (-i)^{|\gE|}}{(2\pi)^{|\gE|}} \left( \prod_{v \in \gVI} \int \dd y_v^0 \right) \nonumber \\
\times \left[ \prod_{e \in \gE} \int_{-\infty}^{\infty} \frac{\dd E_e}{2 |\vec z_e|} e^{ -iE_e \left( \EI_{e,v} \, y_v^0 + \EE_{e,a} \, x_a^0 \right) } \left( \Theta(E_e) e^{ iE_e (|\vec z_e|+ i\ep) } + \Theta(-E_e) e^{ -iE_e (|\vec z_e| + i\ep) } \right) \right],\label{eq:flowseparation} \end{gather}
where the $i \ep$ can be dropped in the 
denominators as it only matters if $\vec z_e = 0$, which is an end-point singularity. 
\subsubsection{Energy flows and digraphs}
We now interpret the terms of the integral in eq.~(\ref{eq:flowseparation}) as energy flows and split the integral into $2^{|E|}$ terms:
\begin{equation}
\label{eq:flowsum}
    \AA_G(x_1,\ldots,x_{|\gVE|},\vec{y}_1, \ldots, \vec{y}_{|\gVI|})
=\sum_{\bb{\sigma}\in\{\pm 1\}^{|E|}}\AA_{G,{\bb\sigma}}(x_1,\ldots,x_{|\gVE|},\vec{y}_1, \ldots, \vec{y}_{|\gVI|})\,,
\end{equation}
where (with suppressed dependence on the arguments of $\AA_{G,\bb\sigma}$),
\begin{equation}
    \AA_{G,{\bb\sigma}}= 
\frac{(-ig)^{|\gVI|}}{(2\pi)^{|\gE|}}
\int\left[\prod_{v \in \gVI}  \dd y_v^0 \right]
\left[ \prod_{e \in \gE} 
\int_{-\infty}^{\infty}
\frac{\dd E_e }{2 i |\vec z_e|}
e^{ 
iE_e
\left(  
-\EI_{e,v} \, y_v^0 - \EE_{e,a} \, x_a^0 
+\sigma_e (|\vec z_e| + i \ep) \right)
}
\Theta(\sigma_e E_e)
\right].
\end{equation}
We can identify the $y^0_v$ integrations with Fourier representations of the $\delta$ function. These $\delta$ functions give rise to energy conservation constraints at each internal vertex and cast $\AA_{G,{\bb\sigma}}$ into the form
\begin{equation}
\AA_{G,{\bb\sigma}}=
\frac{(-2\pi ig)^{|\gVI|}}{(2\pi)^{|\gE|}}
\int
\prod_{e \in \gE}\left[\frac{\dd E_e}{2 i|\vec z_e|}
e^{ 
iE_e(
-\EE_{e,a} \, x_a^0 
+\sigma_e (|\vec z_e|  + i\ep)
)
}
\Theta(\sigma_e E_e)
\right]
\prod_{v \in \gVI}
\delta\left( \sum_{e \in \gE} E_e \, \EI_{e,v} \right).
\end{equation}
The change of variables $\sigma_e E_e \rightarrow E_e$ resolves the $\Theta$ function, and because $\sigma_e^2 = 1$ we get
\begin{equation}
\label{eq:AGnoori}
\AA_{G,{\bb\sigma}}=
\frac{(-2\pi ig)^{|\gVI|}}{(2\pi)^{|\gE|}}
\int_{\mathbb{R}_+^{|E|}}
\prod_{e \in \gE}\left[\frac{\dd E_e}{2i |\vec z_e|}
e^{ 
iE_e
(
-\sigma_e \EE_{e,a} \, x_a^0 
+ |\vec z_e|  + i \ep
)
}
\right]
\prod_{v \in \gVI}
\delta\left( \sum_{e \in \gE} \sigma_eE_e \, \EI_{e,v} \right),
\end{equation}
where we integrate over all positive energies $E_e$.
Note that the $\sigma_e$ only appears in front of incidence matrices $\EI_{e,v}$ or $\EE_{e,a}$. Flipping the sign of some $e$-indexed row in these incidence matrices just changes the previously chosen arbitrary orientation by flipping the direction of the $e$-th edge.
It is clear that the starting orientation does not matter, as we eventually sum over all orientations by flipping each edge in all possible ways. 
Therefore, we can forget about the $\sigma_e$-sums and sum over all different overall orientations of the graph instead by always changing the $\mathcal E$ and $\mathcal R$-matrices accordingly. Such an orientation of the graph shall be denoted by $\bb \sigma$, in the obvious way. For each orientation $\bb \sigma$, we have different $\mathcal E$ and $\mathcal R$-matrices. The data of the integrand is therefore combinatorially encoded in the graph $G$ with an assigned orientation $\bb \sigma$. As before we will denote the resulting digraph as $(G,\bb \sigma)$. 
We can thus rewrite the integral in eq.~\eqref{eq:AGnoori} as
\begin{equation}
\label{eq:orientation_with_deltas}
   \AA_{G,\bb\sigma}=
\frac{(-2\pi ig)^{|\gVI|}}{(2\pi)^{|\gE|}}
\int_{\mathbb{R}_+^{|E|}}\left[ \prod_{e \in \gE} 
\frac{\dd E_e}{2 i |\vec z_e|}
e^{ 
iE_e \left(
- \EE_{e,a}^{\bb \sigma} \, x_a^0 
+
|\vec z_e| + i \ep
\right)
}
\right]
\prod_{v \in \gVI}
\delta\left( \sum_{e \in \gE} E_e \, \EI_{e,v}^{\bb \sigma} \right),
\end{equation}
where we absorbed the entire dependence on $\bb \sigma$ into the incidence matrices. That means $\EI_{e,v}^{\bb \sigma}$ is $+1$ if under the orientation $\bb \sigma$ the edge $e$ is pointing away from  the vertex $v$,  $-1$ if it points towards $v$ and $0$ if $e$ is not incident to $v$, $\EE_{e,a}^{\bb \sigma}$ is defined analogously.

Standard symmetry factor arguments (i.e.~by the orbit-stabilizer theorem) also allow us to rewrite eq.~\eqref{eq:flowsum} as
\begin{equation}
\label{eq:AAsym}
\frac{\AA_G(x_1, \ldots, x_{|\gVE|}, \vec y_{1},\ldots, \vec y_{|\gVI|})}{\Sym G}
=
\sum_{ \langle \bb \sigma\rangle} \frac{\AA_{G,\bb \sigma}(x_1, \ldots, x_{|\gVE|},\vec y_{1},\ldots, \vec y_{|\gVI|})}{\Sym(G,\bb \sigma)} ,
\end{equation}
where we sum over all \emph{nonequivalent} orientations $\bb \sigma$ of the graph $G$.
$\Sym(G,\bb \sigma)$ is the symmetry factor of the digraph $(G,\bb \sigma)$. The calculation 
of a digraph symmetry factor is the same as for covariant diagrams if all edges were associated to charged particles.
\begin{figure}[t]
\begin{align} \begin{tikzpicture}[baseline={([yshift=10ex]current bounding box.south)}] \coordinate (v) at (0,0); \pgfmathsetmacro{\rad}{.6} \pgfmathsetmacro{\rud}{1.6} \draw[pattern=north west lines] (v) circle (\rad); \draw ([shift=(150:{\rad})]v) -- ([shift=(150:{\rud})]v) node[left] {$x_1$}; \draw ([shift=(180:{\rad})]v) -- ([shift=(180:{\rud})]v) node[left] {$x_2$}; \draw ([shift=(210:{\rad})]v) -- ([shift=(210:{\rud})]v) node[left] {$x_3$}; \draw ([shift=(30:{\rad})]v) -- ([shift=(30:{\rud})]v) node[right] {$x_{|\gVE|}$}; \draw ([shift=(0:{\rad})]v) -- ([shift=(0:{\rud})]v) node[right] {$x_{|\gVE|-1}$}; \draw ([shift=(-30:{\rad})]v) -- ([shift=(-30:{\rud})]v) node[right] {$x_{|\gVE|-2}$}; \node (w) at ([shift=(-90:{1.3})]v) {$\bullet \bullet \bullet$}; \filldraw ([shift=(30:{\rud})]v) circle (1.3pt); \filldraw ([shift=(0:{\rud})]v) circle (1.3pt); \filldraw ([shift=(-30:{\rud})]v) circle (1.3pt); \filldraw ([shift=(150:{\rud})]v) circle (1.3pt); \filldraw ([shift=(180:{\rud})]v) circle (1.3pt); \filldraw ([shift=(210:{\rud})]v) circle (1.3pt); \node (u) at ([shift=(225:{2})]v) {$G$}; \end{tikzpicture}& &\rightarrow&& \begin{tikzpicture}[baseline={([yshift=10ex]current bounding box.south)}] \coordinate (v) at (0,0); \coordinate (w) at (0,1.3); \pgfmathsetmacro{\rad}{.6} \pgfmathsetmacro{\rud}{1.6} \draw[pattern=north west lines] (v) circle (\rad); \draw ([shift=(150:{\rad})]v) to[looseness=1,out=150,in=210] (w); \draw ([shift=(180:{\rad})]v) to[looseness=2,out=180,in=180] (w); \draw ([shift=(210:{\rad})]v) to[looseness=3,out=210,in=150] (w); \draw ([shift=(30:{\rad})]v) to[looseness=1,out=30,in=-30] (w); \draw ([shift=(0:{\rad})]v) to[looseness=2,out=0,in=0] (w); \draw ([shift=(-30:{\rad})]v) to[looseness=3,out=-30,in=30] (w); \filldraw[fill=white] (w) circle (3pt); \node (w) at ([shift=(-90:{1.3})]v) {$\bullet \bullet \bullet$}; \node (u) at ([shift=(225:{2})]v) {$G^\circ$}; \end{tikzpicture} \end{align}
\caption{Illustration of the \emph{completed} graph $G^\circ$ that is obtained after adding the artificial vertex $\circ$ at infinity.}
\label{fig:completion}
\end{figure}
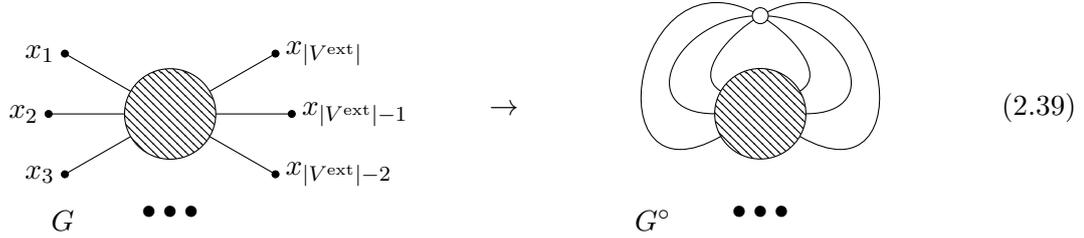
There is a positive-energy flow on each edge of the graph in the dictated direction, which is conserved at each vertex by the $\delta$-functions in eq.~\eqref{eq:orientation_with_deltas}. 
Due to this conservation law, not all possible orientations of a graph give a non-trivial contribution as we illustrated in detail in the last section and will expand upon next. 

\subsubsection{Canonical cycle basis and admissible paths}
Recall that for each graph $G$ we can form the associated completed graph $G^\circ$ by joining all the external vertices into a new vertex $\circ$. This process is illustrated in Figure~\ref{fig:completion} and works analogously for digraphs.
A digraph is \emph{strongly connected} if there is a positively oriented path between any ordered pair of vertices
\cite[Ch.~10]{bondy1976graph}. Such a strongly connected digraph comes with a \emph{canonical cycle basis}, which we can find as follows: start with some directed edge $e_1^{(1)}$ pointing from some vertex $v_a$ to another vertex $v_b$. If $v_a=v_b$, we have a tadpole cycle, which we take as one of our canonical base cycles. If we assume that $v_a \neq v_b$, then, by strong connectivity, there must be some oriented path in $(G,\boldsymbol{\sigma})^\circ$ that we can follow to go back from $v_b$ to $v_a$ in a full oriented cycle. Moreover, we can require that such a cycle consisting of a set of edges $\pp = \{e_1^{(\pp)}, e_2^{(\pp)},\ldots \}$ does not visit any vertex twice. We can pick such a path and declare it to be our first independent oriented cycle $\pp_1$. Next, we pick some edge that was not in this first cycle and construct another closed oriented cycle that contains this edge. For this second cycle $\pp_2$ we are allowed to revisit edges that have been in $\pp_1$, but clearly $\pp_1 \neq \pp_2$ as $\pp_2$ contains at least one edge that has not been in $\pp_1$. Continuing this, we start again with another edge that has neither been visited by the oriented cycle $\pp_1$, nor $\pp_2$, to find a third cycle and so on until no edges are left. In each step, we are guaranteed to find a closed oriented cycle by the strong connectivity requirement. The remarkable observation is that the resulting set of cycles $\pp_1,\ldots,\pp_L$ is unique up to renumbering of the path labels. 

The contribution of the directed graph $(G,\bb\sigma)$ will only be non-zero if the associated completed digraph $(G,\bb \sigma)^\circ$ is strongly connected. %
We will denote the canonical cycle basis of the completed digraph $(G,\bb \sigma)^\circ$ as $\Gamma = \{\pp_1,\ldots,\pp_{|\Gamma|}\}$ where we omit the explicit reference to the digraph $(G,\bb \sigma)$ if it is clear from the context.
Opening up the completed graph again, gives rise to a set of paths in the (uncompleted) digraph $(G,\bb \sigma)$. In this context we will also call the elements of $\Gamma$ the set of admissible paths of $(G,\bb \sigma)$. Note that every admissible path is either a completely internal and closed \emph{cycle} in $G$, which does not pass any external vertex, or it is a path that starts and ends at  external vertices without passing another external vertex in-between. We will call such open paths \emph{routes} through the graph.

Since the graph $(G,\bb \sigma)^\circ$ has $|\gVI|+1$ vertices including the special vertex $\circ$, and $|\gE|$ edges, one can conclude, using the graph's Euler characteristic, that the digraph has at most  $|\gE|-|\gVI|-1+1=|\gE|-|\gVI|$ independent cycles. That means there are $|\gE|-|\gVI|$ admissible paths of the digraph $(G,\bb \sigma)$. The energy integral in eq.~\eqref{eq:orientation_with_deltas} is effectively $|\gE|-|\gVI|$ dimensional, due to the $|\gVI|$ delta functions.  This calculation suggests therefore that we can associate each admissible path of $(G,\bb \sigma)$ with an independent energy integration.

It is instructive to compare this to the way usual momentum space Feynman integrals are written as unconstrained integrals over a set of independent loops. In the case of momentum space Feynman integrals it is known that there is a lot of ambiguity in the way the loop momenta can be routed through the graph. In this sense, the momentum Feynman space integral is thus a highly `non-canonical' object. This can cause a severe problem, as it quickly becomes hard to tell if one integral is equal to another one: there are non-trivial relations between these integrals. Fixing this problem is challenging and is usually done by solving complicated systems of integration by parts (IBP) relations between the integrals. Our observation indicates that this IBP problem could be much milder in our coordinate space framework.

Just as in the momentum space loop integral case, we can resolve the $\delta$ functions in eq.~\eqref{eq:orientation_with_deltas} by introducing one integration variable $E_{\pp}$ for each admissible path $\pp \in \Gamma$. 
The delta functions are resolved with the choice of coordinates
\begin{align} E_e = \sum_{\substack{\pp\in \Gamma \\ \text{s.t. } e \in \pp}} E_\pp\,, \end{align}
where we sum over all admissible paths of $(G,\bb \sigma)$, that contain the edge $e$. This is still analogous to the loop momentum integral case, except for the fact that we have a dictated orientation that we need to follow in our cycles.
Using these energy variables in eq.~\eqref{eq:orientation_with_deltas} resolves the delta functions and gives
\begin{gather} \begin{gathered} \AA_{G,\bb\sigma}= \frac{(-2\pi i g)^{|\gVI|}}{(2\pi i)^{|\gE|}} \int_{\mathbb{R}_+^{|\Gamma|}} \left[\prod_{\pp \in \Gamma} \dd E_\pp \right]\left[\prod_{e\in \gE} \frac{1}{2 |\vec z_e|}\right] \times \\
\exp\left( i \sum_{e\in \gE} \sum_{\substack{\pp\in \Gamma \\ \text{s.t. } e \in \pp}} E_\pp \left( -\EE_{e,a}^{\bb \sigma} \, x_a^0 + |\vec z_e| +i\ep \right) \right)\,. \end{gathered} \end{gather}
Recall that there is an implicit summation over $a$. Changing the order of summation in the exponential renders integration in the remaining energy variables straightforward, and gives a product of trivial oscillatory integrals
\begin{equation}
   \AA_{G,\bb\sigma}=
\frac{(-2\pi i g)^{|\gVI|}}{(2\pi i)^{|\gE|}}
    \left[\prod_{e\in \gE}
\frac{1}{2 |\vec z_e|}\right]
 \prod_{\pp \in \Gamma} 
\int_0^\infty
\dd E_\pp 
\exp\left(
i
E_\pp
\sum_{e\in \pp}
\left(
-\EE_{e,a}^{\bb \sigma} \, x_a^0 
+
|\vec z_e| 
+i\ep
\right)
\right)\,.
\end{equation}
We can now perform the $E_\pp$ integrations. The $\infty$ boundary term is going to vanish by the $\ep > 0$ assumption. The final expression for the FOPT representation of a digraph $(G,\bb \sigma)$ is remarkably simple and amounts to
\begin{equation}
   \AA_{G,\bb\sigma}=
\frac{(2\pi g)^{|\gVI|}}{(-2\pi)^{|\gE|}}
\left[
\prod_{e\in \gE}
\frac{1}{2 |\vec z_e|}
\right]
 \prod_{\pp \in \Gamma} 
\frac{1}
{
\sum_{e\in \pp}
\left(
- \EE_{e,a}^{\bb \sigma} \, x_a^0 
+
|\vec z_e|
+ i \ep
\right)
}.\label{eq:FOPTresult}
\end{equation}
The term $\EE_{e,a}^{\bb \sigma} \, x_a^0$ is only non-zero if the admissible path $\pp$ goes from external to external vertex. It can be identified with the \emph{time difference} of the two corresponding vertices. The other term in the denominator is the total Euclidean path length of the admissible path.
We will discuss this entirely combinatorial formula in the next section.

\subsection{FOPT Feynman rules}
\label{sec:flowfeynmanrules}
The procedure illustrated in the last section
generalizes to all Feynman diagrams. It provides an 
alternative perturbative decomposition of correlation functions:
\begin{align} \label{eq:FOPT_greens_functions} \Gamma(x_1,...,x_{|\gVE|})=\left\langle 0 | T ( \varphi(x_1) \cdots \varphi(x_{|\gVE|}) ) | 0 \right \rangle = \sum_{(G,\bb \sigma)} \frac{1}{\Sym (G,\bb \sigma)} A_{G,\bb \sigma}(x_1, \ldots, x_{|\gVE|}), \end{align}
where we sum over all topologically different digraphs $(G,\bb \sigma)$, i.e.~graphs $G$ from the given scalar QFT with a specified energy flow orientation on the propagators.
Note that in contrast to `old-fashioned', or time-ordered perturbation theory, where each covariant integral is replaced by $|\gVI|!$ time-ordered integrals, we get at most $2^{|\gE|}$ \emph{energy-flow-oriented}  integrals in our coordinate-space focused approach. 

By eq.~\eqref{eq:AAsym}, the FOPT Feynman rules provide a way of decomposing an individual covariant Feynman integral into its different flow-oriented components:
\begin{align} \frac{1}{\Sym G} A_{G}(x_1, \ldots, x_{|\gVE|}) = \sum_{\langle \bb \sigma \rangle} \frac{1}{\Sym (G,\bb \sigma)} A_{G,\bb \sigma}(x_1, \ldots, x_{|\gVE|}), \end{align}
where we sum over all nonequivalent ways to orient the graph $G$ via $\bb \sigma$.

An integral expression for $A_{G,\bb \sigma}(x_1, \ldots, x_{|\gVE|})$ can be found using the following, entirely combinatorial recipe:
\begin{enumerate}
        \item $A_{G,\bb \sigma} =0$ if the completed directed graph $(G,\bb \sigma)^\circ$ is not strongly connected.
        \item Multiply by a factor of $-i g$ for each interaction vertex. 
        \item For each edge $e$ of $G$ multiply by a factor $\frac{-i}{(8\pi^2)|\vec{z}_e|}$ where $\vec{z}_e = \vec{y}_{v} - \vec{y}_{u}$ and $\vec{y}_v,\vec{y}_u$ are the coordinates of the internal or external vertices to which the edge $e$ is incident. 
        \item For each admissible path $\pp$ of $(G,\bb \sigma)$ (i.e.~for each cycle in the 
        canonical cycle basis of $(G,\bb \sigma)^\circ$) multiply by a factor of $ i/\left( { \gamma_\pp + \tau_{\pp} + i\ep } \right), $ where
\begin{equation}
    \gamma_\mathrm{p} =\sum_{e\in \pp}
|\vec{z}_e|
\end{equation} is the sum over all edge lengths that are in the cycle and $\tau_{\pp}$ is either the time passed between the starting and ending external vertices of the path or vanishes if the cycle does not go through the $\circ$ vertex.
        \item For each internal vertex $v$ of the graph $G$ integrate over three-dimensional space $\int \dd^3 \vec{y}_v$ and multiply by $2 \pi$.
\end{enumerate}
Note that the $i\ep$ can be dropped for cycles that do not go through the special vertex $\circ$ as the denominator is only positive and the corresponding pole is an end-point singularity. %
We can summarize these Feynman rules as follows. For a given 
digraph $(G,\bb \sigma)$ with cycle basis $\Gamma$, where all interaction vertices in $G$ are internal vertices and vice-versa, we have
\begin{equation}
    A_{G,\bb \sigma}(x_1, \ldots, x_{|\gVE|})=
\frac{(2\pi g)^{|\gVI|}}{(-4\pi^2)^{|\gE|}}
\left( \prod_{v \in \gVI} \int \dd^3 \vec y_v 
\right)
\left(
\prod_{e\in \gE}
\frac{1}{2 |\vec z_e |}
\right)
 \prod_{\pp \in \Gamma} 
\frac{1}
{
\gamma_\pp
+\tau_{\pp} 
+ i \ep
}.
\label{eq:delta_free_rep}
\end{equation}

\subsection{A simple example: The bubble graph}
We now treat the simple example of the bubble graph to show the use of the FOPT Feynman rules. It will
also show how UV singularities are conveniently isolated in FOPT. The graph is
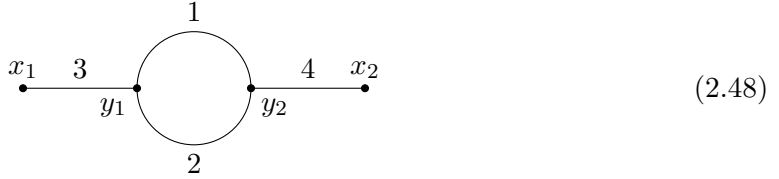
\begin{align} \label{eq:bubble_graph} \def\scale{1.5} \begin{tikzpicture}[baseline={([yshift=-0.7ex]0,0)}] \coordinate[label=above:$x_1$] (i1) at (-\scale,0); \coordinate[label=above:$x_2$] (i2) at (2*\scale,0); \coordinate[label=below left:$y_1$] (v1) at (0,0); \coordinate[label=below right:$y_2$] (v2) at (\scale,0); \draw (v1) arc (180:0:.5*\scale) node[midway,above] {$1$}; \draw (v2) arc (0:-180:.5*\scale) node[midway,below] {$2$}; \filldraw (v1) circle (1.3pt); \filldraw (v2) circle (1.3pt); \filldraw (i1) circle (1.3pt); \filldraw (i2) circle (1.3pt); \draw (v1) -- (i1) node[midway,above] {$3$}; \draw (v2) -- (i2) node[midway,above] {$4$}; \end{tikzpicture} \end{align}
The traditional coordinate-space Feynman integral associated to this diagram reads,
\begin{align} A_G(x_1,x_2) = \frac{(-i g)^2}{(4 \pi^2)^4} \int \dd^4 y_1 \dd^4 y_2 \frac{1}{-z_1^2+i \ep} \frac{1}{-z_2^2+i \ep} \frac{1}{-z_3^2+i \ep} \frac{1}{-z_4^2+i \ep}\,. \end{align}
The symmetry factor of the graph is $2$, because we can permute both edges of the bubble. 
Adding the $\circ$ vertex that joins the external vertices results in 
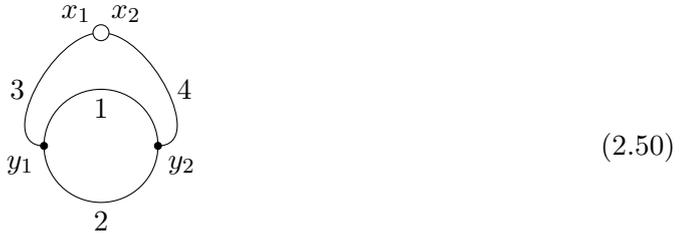
\begin{align} \def\scale{1.5} \begin{tikzpicture}[baseline={([yshift=-0.7ex]0,0)}] \coordinate (w) at (.5*\scale,\scale); \coordinate[label=below left:$y_1$] (v1) at (0,0); \coordinate[label=below right:$y_2$] (v2) at (\scale,0); \draw (v1) arc (180:0:.5*\scale) node[midway,below] {$1$}; \draw (v2) arc (0:-180:.5*\scale) node[midway,below] {$2$}; \filldraw (v1) circle (1.3pt); \filldraw (v2) circle (1.3pt); \draw (v1) to[out=180,in=180] node[midway,left] {$3$} (w); \draw (v2) to[out=0,in=0] node[midway,right] {$4$} (w); \filldraw[fill=white] (w) circle (3pt); \coordinate[label=above left:$x_1$] (w) at (.5*\scale,\scale); \coordinate[label=above right:$x_2$] (w) at (.5*\scale,\scale); \end{tikzpicture} \end{align}
There are 12 topologically distinguished ways to give an orientation to the edges of this graph:
{
\def\scale{.8}
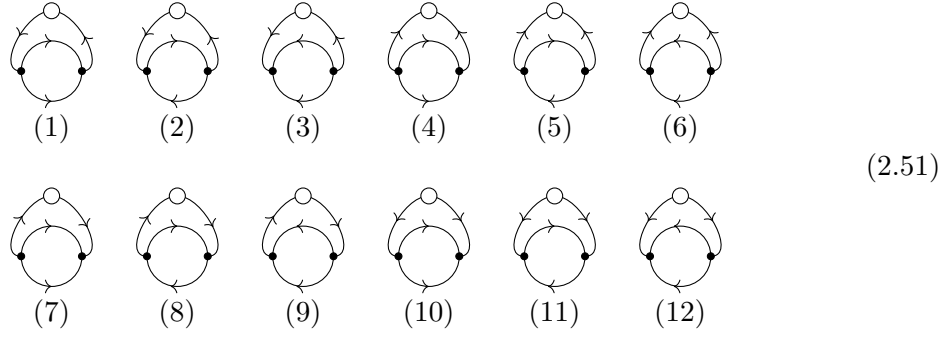
\begin{align} \begin{array}{cccccc} \begin{tikzpicture}[baseline={([yshift=-0.7ex]0,0)}] \coordinate (w) at (.5*\scale,\scale); \coordinate (v1) at (0,0); \coordinate (v2) at (\scale,0); \draw [->-] (v1) arc (180:0:.5*\scale); \draw [->-] (v1) arc (-180:0:.5*\scale); \filldraw (v1) circle (1.3pt); \filldraw (v2) circle (1.3pt); \draw [->-] (w) to[out=180,in=180] (v1); \draw [->-] (v2) to[out=0,in=0] (w); \filldraw[fill=white] (w) circle (3pt); \coordinate (w) at (.5*\scale,\scale); \coordinate (w) at (.5*\scale,\scale); \end{tikzpicture} & \begin{tikzpicture}[baseline={([yshift=-0.7ex]0,0)}] \coordinate (w) at (.5*\scale,\scale); \coordinate (v1) at (0,0); \coordinate (v2) at (\scale,0); \draw [->-] (v1) arc (180:0:.5*\scale); \draw [-<-] (v1) arc (-180:0:.5*\scale); \filldraw (v1) circle (1.3pt); \filldraw (v2) circle (1.3pt); \draw [->-] (w) to[out=180,in=180] (v1); \draw [->-] (v2) to[out=0,in=0] (w); \filldraw[fill=white] (w) circle (3pt); \coordinate (w) at (.5*\scale,\scale); \coordinate (w) at (.5*\scale,\scale); \end{tikzpicture} & \begin{tikzpicture}[baseline={([yshift=-0.7ex]0,0)}] \coordinate (w) at (.5*\scale,\scale); \coordinate (v1) at (0,0); \coordinate (v2) at (\scale,0); \draw [-<-] (v1) arc (180:0:.5*\scale); \draw [-<-] (v1) arc (-180:0:.5*\scale); \filldraw (v1) circle (1.3pt); \filldraw (v2) circle (1.3pt); \draw [->-] (w) to[out=180,in=180] (v1); \draw [->-] (v2) to[out=0,in=0] (w); \filldraw[fill=white] (w) circle (3pt); \coordinate (w) at (.5*\scale,\scale); \coordinate (w) at (.5*\scale,\scale); \end{tikzpicture} & \begin{tikzpicture}[baseline={([yshift=-0.7ex]0,0)}] \coordinate (w) at (.5*\scale,\scale); \coordinate (v1) at (0,0); \coordinate (v2) at (\scale,0); \draw [->-] (v1) arc (180:0:.5*\scale); \draw [->-] (v1) arc (-180:0:.5*\scale); \filldraw (v1) circle (1.3pt); \filldraw (v2) circle (1.3pt); \draw [-<-] (w) to[out=180,in=180] (v1); \draw [-<-] (w) to[out=0,in=0] (v2); \filldraw[fill=white] (w) circle (3pt); \coordinate (w) at (.5*\scale,\scale); \coordinate (w) at (.5*\scale,\scale); \end{tikzpicture} & \begin{tikzpicture}[baseline={([yshift=-0.7ex]0,0)}] \coordinate (w) at (.5*\scale,\scale); \coordinate (v1) at (0,0); \coordinate (v2) at (\scale,0); \draw [->-] (v1) arc (180:0:.5*\scale); \draw [-<-] (v1) arc (-180:0:.5*\scale); \filldraw (v1) circle (1.3pt); \filldraw (v2) circle (1.3pt); \draw [-<-] (w) to[out=180,in=180] (v1); \draw [-<-] (w) to[out=0,in=0] (v2); \filldraw[fill=white] (w) circle (3pt); \coordinate (w) at (.5*\scale,\scale); \coordinate (w) at (.5*\scale,\scale); \end{tikzpicture} & \begin{tikzpicture}[baseline={([yshift=-0.7ex]0,0)}] \coordinate (w) at (.5*\scale,\scale); \coordinate (v1) at (0,0); \coordinate (v2) at (\scale,0); \draw [-<-] (v1) arc (180:0:.5*\scale); \draw [-<-] (v1) arc (-180:0:.5*\scale); \filldraw (v1) circle (1.3pt); \filldraw (v2) circle (1.3pt); \draw [-<-] (w) to[out=180,in=180] (v1); \draw [-<-] (w) to[out=0,in=0] (v2); \filldraw[fill=white] (w) circle (3pt); \coordinate (w) at (.5*\scale,\scale); \coordinate (w) at (.5*\scale,\scale); \end{tikzpicture} \\
(1)&(2)&(3)&(4)&(5)&(6) \\
&&&&& \\
\begin{tikzpicture}[baseline={([yshift=-0.7ex]0,0)}] \coordinate (w) at (.5*\scale,\scale); \coordinate (v1) at (0,0); \coordinate (v2) at (\scale,0); \draw [->-] (v1) arc (180:0:.5*\scale); \draw [->-] (v1) arc (-180:0:.5*\scale); \filldraw (v1) circle (1.3pt); \filldraw (v2) circle (1.3pt); \draw [-<-] (w) to[out=180,in=180] (v1); \draw [-<-] (v2) to[out=0,in=0] (w); \filldraw[fill=white] (w) circle (3pt); \coordinate (w) at (.5*\scale,\scale); \coordinate (w) at (.5*\scale,\scale); \end{tikzpicture} & \begin{tikzpicture}[baseline={([yshift=-0.7ex]0,0)}] \coordinate (w) at (.5*\scale,\scale); \coordinate (v1) at (0,0); \coordinate (v2) at (\scale,0); \draw [->-] (v1) arc (180:0:.5*\scale); \draw [-<-] (v1) arc (-180:0:.5*\scale); \filldraw (v1) circle (1.3pt); \filldraw (v2) circle (1.3pt); \draw [-<-] (w) to[out=180,in=180] (v1); \draw [-<-] (v2) to[out=0,in=0] (w); \filldraw[fill=white] (w) circle (3pt); \coordinate (w) at (.5*\scale,\scale); \coordinate (w) at (.5*\scale,\scale); \end{tikzpicture} & \begin{tikzpicture}[baseline={([yshift=-0.7ex]0,0)}] \coordinate (w) at (.5*\scale,\scale); \coordinate (v1) at (0,0); \coordinate (v2) at (\scale,0); \draw [-<-] (v1) arc (180:0:.5*\scale); \draw [-<-] (v1) arc (-180:0:.5*\scale); \filldraw (v1) circle (1.3pt); \filldraw (v2) circle (1.3pt); \draw [-<-] (w) to[out=180,in=180] (v1); \draw [-<-] (v2) to[out=0,in=0] (w); \filldraw[fill=white] (w) circle (3pt); \coordinate (w) at (.5*\scale,\scale); \coordinate (w) at (.5*\scale,\scale); \end{tikzpicture} & \begin{tikzpicture}[baseline={([yshift=-0.7ex]0,0)}] \coordinate (w) at (.5*\scale,\scale); \coordinate (v1) at (0,0); \coordinate (v2) at (\scale,0); \draw [->-] (v1) arc (180:0:.5*\scale); \draw [->-] (v1) arc (-180:0:.5*\scale); \filldraw (v1) circle (1.3pt); \filldraw (v2) circle (1.3pt); \draw [->-] (w) to[out=180,in=180] (v1); \draw [->-] (w) to[out=0,in=0] (v2); \filldraw[fill=white] (w) circle (3pt); \coordinate (w) at (.5*\scale,\scale); \coordinate (w) at (.5*\scale,\scale); \end{tikzpicture} & \begin{tikzpicture}[baseline={([yshift=-0.7ex]0,0)}] \coordinate (w) at (.5*\scale,\scale); \coordinate (v1) at (0,0); \coordinate (v2) at (\scale,0); \draw [->-] (v1) arc (180:0:.5*\scale); \draw [-<-] (v1) arc (-180:0:.5*\scale); \filldraw (v1) circle (1.3pt); \filldraw (v2) circle (1.3pt); \draw [->-] (w) to[out=180,in=180] (v1); \draw [->-] (w) to[out=0,in=0] (v2); \filldraw[fill=white] (w) circle (3pt); \coordinate (w) at (.5*\scale,\scale); \coordinate (w) at (.5*\scale,\scale); \end{tikzpicture} & \begin{tikzpicture}[baseline={([yshift=-0.7ex]0,0)}] \coordinate (w) at (.5*\scale,\scale); \coordinate (v1) at (0,0); \coordinate (v2) at (\scale,0); \draw [-<-] (v1) arc (180:0:.5*\scale); \draw [-<-] (v1) arc (-180:0:.5*\scale); \filldraw (v1) circle (1.3pt); \filldraw (v2) circle (1.3pt); \draw [->-] (w) to[out=180,in=180] (v1); \draw [->-] (w) to[out=0,in=0] (v2); \filldraw[fill=white] (w) circle (3pt); \coordinate (w) at (.5*\scale,\scale); \coordinate (w) at (.5*\scale,\scale); \end{tikzpicture} \\
(7)&(8)&(9)&(10)&(11)&(12) \end{array} \end{align}%
}%
Note that we are not allowed to permute the edges that are incident to the $\circ$ vertex as this would correspond to a permutation of the external vertices of the original graph. We can think of the $\circ$ vertex and the edges that are incident to it as fixed while computing the symmetry factor. The edge-oriented graphs $(1),(3),(4),(6),(7),(9),(10),(12)$ have therefore a symmetry factor of $2$ and the other graphs have a trivial symmetry factor of $1$. 
Of these directed graphs only $(1),(2),(8)$ and $(9)$ are strongly connected as one can easily check. In the remaining directed graphs, we can always find a vertex (including $\circ$) that has only in-going or out-going incident edges. These graphs are forbidden as a consequence of energy conservation and positivity.
At this point it is good to remark that there are also non-strongly connected graphs that do not have a vertex with only in- or out-going incident edges. Examples are graphs that consist of two cycles, each with positive cyclic energy flow, which are connected  by edges only pointing from cycle 1 to cycle 2. 

The completed directed graph $(1)$ has two independent cycles, $314$ and $324$, with the edge numbering as indicated in~\eqref{eq:bubble_graph}.

{
\def\scale{.8}
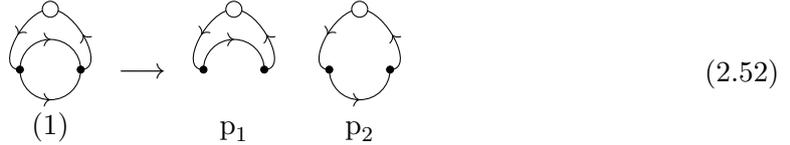
\begin{align} \begin{array}{cccc} \begin{tikzpicture}[baseline={([yshift=-0.7ex]0,0)}] \coordinate (w) at (.5*\scale,\scale); \coordinate (v1) at (0,0); \coordinate (v2) at (\scale,0); \draw [->-] (v1) arc (180:0:.5*\scale); \draw [->-] (v1) arc (-180:0:.5*\scale); \filldraw (v1) circle (1.3pt); \filldraw (v2) circle (1.3pt); \draw [->-] (w) to[out=180,in=180] (v1); \draw [->-] (v2) to[out=0,in=0] (w); \filldraw[fill=white] (w) circle (3pt); \coordinate (w) at (.5*\scale,\scale); \coordinate (w) at (.5*\scale,\scale); \end{tikzpicture} & \longrightarrow & \begin{tikzpicture}[baseline={([yshift=-0.7ex]0,0)}] \coordinate (w) at (.5*\scale,\scale); \coordinate (v1) at (0,0); \coordinate (v2) at (\scale,0); \draw [->-] (v1) arc (180:0:.5*\scale); \filldraw (v1) circle (1.3pt); \filldraw (v2) circle (1.3pt); \draw [->-] (w) to[out=180,in=180] (v1); \draw [->-] (v2) to[out=0,in=0] (w); \filldraw[fill=white] (w) circle (3pt); \coordinate (w) at (.5*\scale,\scale); \coordinate (w) at (.5*\scale,\scale); \end{tikzpicture} & \begin{tikzpicture}[baseline={([yshift=-0.7ex]0,0)}] \coordinate (w) at (.5*\scale,\scale); \coordinate (v1) at (0,0); \coordinate (v2) at (\scale,0); \draw [->-] (v1) arc (-180:0:.5*\scale); \filldraw (v1) circle (1.3pt); \filldraw (v2) circle (1.3pt); \draw [->-] (w) to[out=180,in=180] (v1); \draw [->-] (v2) to[out=0,in=0] (w); \filldraw[fill=white] (w) circle (3pt); \coordinate (w) at (.5*\scale,\scale); \coordinate (w) at (.5*\scale,\scale); \end{tikzpicture} \\
(1) & &\pp_1 & \pp_2 \end{array} \end{align}
}
 Both cycles pass the $\circ$ vertex.
We find by the FOPT Feynman rules from sect.~\ref{sec:flowfeynmanrules} that
\begin{gather} \begin{gathered} A_{G,\bb \sigma_{(1)}}(x_1,x_2) = \\
\frac{(2\pi g)^{2}}{(8\pi^2)^{4}} \int \frac{\dd^3 \vec y_1 \dd^3 \vec y_2}{|\vec z_1| | \vec z_2 | | \vec z_3 | | \vec z_4 |} \frac{1}{| \vec z_3| + | \vec z_1 | + | \vec z_4 | + \tau + i \ep} \frac{1}{| \vec z_3| + | \vec z_2 | + | \vec z_4 | + \tau + i \ep}, \end{gathered} \end{gather}
where we defined $\tau = x_{2}^0 - x_{1}^0$.
The graph $(2)$ has the independent cycles $314$ and $12$. The latter cycle does not pass the $\circ$ vertex. Therefore,
\begin{gather} \begin{gathered} A_{G,\bb \sigma_{(2)}} (x_1,x_2)= \frac{(2\pi g)^{2}}{(8\pi^2)^{4}} \int \frac{\dd^3 \vec y_1 \dd^3 \vec y_2}{|\vec z_1| | \vec z_2 | | \vec z_3 | | \vec z_4 |} \frac{1}{| \vec z_3| + | \vec z_1 | + | \vec z_4 | + \tau + i \ep} \frac{1}{| \vec z_1| + | \vec z_2 |} \end{gathered}\,. \end{gather}
For the graph $G$ we have $\vec z_1 = \vec z_2$. 
By changing integration variables from $\vec y_1,\vec y_2$ to $\vec y_1, \vec z_1 = \vec y_2 - \vec y_1$, we find that
\begin{gather} \begin{gathered} A_{G,\bb \sigma_{(2)}} (x_1,x_2)= \frac{(2\pi g)^{2}}{(8\pi^2)^{4}} \int \frac{\dd^3 \vec y_1 \dd^3 \vec z_1}{ | \vec z_3 | | \vec z_4 |} \frac{1}{| \vec z_3| + | \vec z_1 | + | \vec z_4 | + \tau + i \ep} \frac{1}{2| \vec z_1|^3}\,. \end{gathered} \end{gather}
Basic power counting reveals that the integrand features 
a UV singularity for $\vec z_1 \rightarrow 0$.
The cycle that loops between the vertices $\vec y_1$ and $\vec y_2$ 
is associated to such a singularity. The intuitive explanation of this in the FOPT formalism is that the energy that flows through this cycle is unbounded and can lead to a UV divergence\footnote{We do not address the issue of regularization of divergent FOPT graphs in this paper. We comment on possible approaches to UV renormalisation in the conclusions.}. 

The directed graph $(8)$ has exactly the flipped orientation of $(2)$. Therefore
\begin{gather} \begin{gathered} A_{G,\bb \sigma_{(8)}}(x_1,x_2) = \frac{(2\pi g)^{2}}{(8\pi^2)^{4}} \int \frac{\dd^3 \vec y_1 \dd^3 \vec y_2}{|\vec z_1| | \vec z_2 | | \vec z_3 | | \vec z_4 |} \frac{1}{| \vec z_3| + | \vec z_1 | + | \vec z_4 | - \tau + i \ep} \frac{1}{| \vec z_1| + | \vec z_2 |}\,. \end{gathered} \end{gather}
Analogously the graph $(9)$ has the flipped orientation of $(1)$
\begin{gather} \begin{gathered} A_{G,\bb \sigma_{(9)}}(x_1,x_2) = \\
= \frac{(2\pi g)^{2}}{(8\pi^2)^{4}} \int \frac{\dd^3 \vec y_1 \dd^3 \vec y_2}{|\vec z_1| | \vec z_2 | | \vec z_3 | | \vec z_4 |} \frac{1}{| \vec z_3| + | \vec z_1 | + | \vec z_4 | - \tau + i \ep} \frac{1}{| \vec z_3| + | \vec z_2 | + | \vec z_4 | - \tau + i \ep}. \end{gathered} \end{gather}
Observe that flipping the complete orientation only results in a sign change of all external time differences.
By collecting the overall and individual symmetry factors, we have
\begin{equation}
\frac{1}{2}A(x_1,x_2)=\frac{1}{2} A_{G,\bb \sigma_{(1)}}+ A_{G,\bb \sigma_{(2)}}+A_{G,\bb \sigma_{(8)}}+\frac{1}{2} A_{G, \bb \sigma_{(9)}}\;.
\end{equation}
The digraphs $(G,\bb \sigma_{(2)})$ and $(G,\bb \sigma_{(8)})$ feature UV singularities, as expected as $G$ is a UV singular graph in $D=4$; the other contributions are finite.

\subsection{Routes, cycles and UV singularities}
\label{sec:cycles_singularities}
As seen in the last section, UV singularities in the FOPT representation follow the intuition expected from momentum space: they can be associated with a set of internal vertices collapsing to a single point. In the FOPT representation, these limits correspond to the vanishing of denominators corresponding to closed cycles in $(G,\boldsymbol{\sigma})$. These in turn lead to a divergence of the integrand. A basic power counting argument helps decide whether the resulting singularity is integrable or not.

In this section we will briefly explain the general power counting procedure in the FOPT representation. To do so, we first recall that the admissible paths $\Gamma$ of a digraph $({G,\bb \sigma})$ (having a strongly connected completed digraph $({G,\bb \sigma})^\circ$) fall into two sets: \emph{routes} $\gGE$, which connect two external vertices and pass internal vertices in between, and \emph{cycles} $\gGI$, which are closed simple cycles that consist of internal vertices only. We have $\Gamma = \gGE \cup \gGI$.

Even though the FOPT representation can also deal with the more general case, we assume from now on that external vertices of our graph are attached to one edge only. 
Note that in this case we can also separate the external vertices of our digraphs $(G,\bb \sigma)$ into two subsets: The set of in-going external vertices $\gVEI$ and the set of out-going external vertices $\gVEO$ such that $\gVEI \cup \gVEO = \gVE$. They are defined such that in/out-going external vertices have energy flowing in/out of them.

\begin{figure}
\centering
\begin{subfigure}[b]{2in}
\begin{center}
\resizebox{5cm}{!}{%
\begin{tikzpicture} \begin{feynman} \vertex(1); \vertex[below = 6cm of 1](2); \vertex[right = 8cm of 2](3); \vertex[above = 6cm of 3](4); \vertex[above = 3cm of 3](M34); \vertex[right = 1.414cm of M34](E34); \vertex[above left = 1cm and 1cm of 1](E1); \vertex[above right = 1cm and 1cm of 4](E4); \vertex[below left = 1cm and 1cm of 2](E2); \vertex[below right = 1cm and 1cm of 3](E3); \vertex[above left = 1.2cm and 1.2cm of 1](L1) {\scalebox{2.5}{$x_1$}}; \vertex[above right = 1.2cm and 1.2cm of 4](L4) {\scalebox{2.5}{$x_5$}}; \vertex[below left = 1.2cm and 1.2cm of 2](L2) {\scalebox{2.5}{$x_2$}}; \vertex[below right = 1.2cm and 1.2cm of 3](L3) {\scalebox{2.5}{$x_3$}}; \vertex[right = 1.7cm of M34](L34) {\scalebox{2.5}{$x_4$}}; \vertex[below = 4cm of 1](M12); \vertex[right = 3cm of 1](T141); \vertex[below right = 2cm and 1.5 cm of 1](T142); \vertex[left = 2.3cm of 4](R4); \vertex[left = 2.3cm of 3](DR4); \vertex[below = 3.7cm of R4](MR4); \diagram*[large]{ (1) -- [->-,line width=0.8mm, blue!70!white] (M12) -- [->-,line width=0.6mm] (2) -- [->-,line width=0.6mm] (DR4) -- [->-,line width=0.6mm] (3) -- [->-,line width=0.6mm] (M34) -- [->-,line width=0.6mm] (4) -- [-<-,line width=0.8mm, blue!70!white] (R4) -- [->-,line width=0.6mm] (T141) -- [->-,line width=0.6mm] (1), (M34) -- [->-,line width=0.6mm] (E34), (1) -- [-<-,line width=0.8mm,, blue!70!white] (E1), (2) -- [-<-,line width=0.6mm] (E2), (3) -- [->-,line width=0.6mm] (E3), (4) -- [->-,line width=0.8mm,, blue!70!white] (E4), (R4) -- [-<-,line width=0.8mm, blue!70!white] (MR4) -- [-<-,line width=0.6mm] (DR4), (M12) -- [->-,line width=0.8mm, blue!70!white] (T142) -- [->-,line width=0.6mm] (T141), (MR4) -- [-<-,line width=0.8mm, blue!70!white] (T142), }; \path[draw=black, fill=black] (E1) circle[radius=0.15]; \path[draw=black, fill=black] (E2) circle[radius=0.15]; \path[draw=black, fill=black] (E3) circle[radius=0.15]; \path[draw=black, fill=black] (E4) circle[radius=0.15]; \path[draw=black, fill=black] (E34) circle[radius=0.15]; \end{feynman} \end{tikzpicture}
}%
\caption{Route, $\rr\in\gGE$}
\end{center}%
\end{subfigure}  \hspace{1cm}
\begin{subfigure}[b]{2in}
\label{fig:3to1}
\begin{center}
\resizebox{5cm}{!}{%
\begin{tikzpicture} \begin{feynman} \vertex(1); \vertex[below = 6cm of 1](2); \vertex[right = 8cm of 2](3); \vertex[above = 6cm of 3](4); \vertex[above = 3cm of 3](M34); \vertex[right = 1.414cm of M34](E34); \vertex[above left = 1cm and 1cm of 1](E1); \vertex[above right = 1cm and 1cm of 4](E4); \vertex[below left = 1cm and 1cm of 2](E2); \vertex[below right = 1cm and 1cm of 3](E3); \vertex[above left = 1.2cm and 1.2cm of 1](L1) {\scalebox{2.5}{$x_1$}}; \vertex[above right = 1.2cm and 1.2cm of 4](L4) {\scalebox{2.5}{$x_5$}}; \vertex[below left = 1.2cm and 1.2cm of 2](L2) {\scalebox{2.5}{$x_2$}}; \vertex[below right = 1.2cm and 1.2cm of 3](L3) {\scalebox{2.5}{$x_3$}}; \vertex[right = 1.7cm of M34](L34) {\scalebox{2.5}{$x_4$}}; \vertex[below = 4cm of 1](M12); \vertex[right = 3cm of 1](T141); \vertex[below right = 2cm and 1.5 cm of 1](T142); \vertex[left = 2.3cm of 4](R4); \vertex[left = 2.3cm of 3](DR4); \vertex[below = 3.7cm of R4](MR4); \diagram*[large]{ (1) -- [->-,line width=0.8mm, red!60!yellow] (M12) -- [->-,line width=0.6mm] (2) -- [->-,line width=0.6mm] (DR4) -- [->-,line width=0.6mm] (3) -- [->-,line width=0.6mm] (M34) -- [->-,line width=0.6mm] (4) -- [-<-,line width=0.6mm] (R4) -- [->-,line width=0.8mm, red!60!yellow] (T141) -- [->-,line width=0.8mm, red!60!yellow] (1), (1) -- [-<-,line width=0.6mm] (E1), (2) -- [-<-,line width=0.6mm] (E2), (3) -- [->-,line width=0.6mm] (E3), (4) -- [->-,line width=0.6mm] (E4), (M34) -- [->-,line width=0.6mm] (E34), (R4) -- [-<-,line width=0.8mm, red!60!yellow] (MR4) -- [-<-,line width=0.6mm] (DR4), (M12) -- [->-,line width=0.8mm, red!60!yellow] (T142) -- [->-,line width=0.6mm] (T141), (MR4) -- [-<-,line width=0.8mm, red!60!yellow] (T142), }; \path[draw=black, fill=black] (E1) circle[radius=0.15]; \path[draw=black, fill=black] (E2) circle[radius=0.15]; \path[draw=black, fill=black] (E3) circle[radius=0.15]; \path[draw=black, fill=black] (E4) circle[radius=0.15]; \path[draw=black, fill=black] (E34) circle[radius=0.15]; \end{feynman} \end{tikzpicture}
}%
\caption{Cycle, $\cc\in\gGI$}
\end{center}
\end{subfigure} 
\caption{Admissible paths of different types through a digraph $(G,\bb \sigma)$: A route and a cycle. The illustrated graph has in-going external vertices $\gVEI=\{1,2\}$ and out-going external vertices $\gVEO=\{3,4,5\}$. The route $\rr$ connects the in-going vertex $1$ to the out-going vertex $5$. Hence,
 $i(\rr) = 1$ and $f(\rr) = 5$.
\label{fig:cnnctdnss}
}
\end{figure}
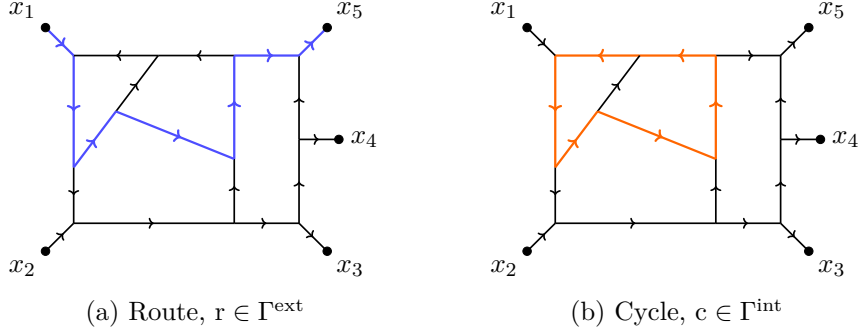

Because all admissible paths in the canonical cycle basis $\Gamma$ are  oriented, each route in $\gGE$ has to connect an in-going external vertex with an out-going external vertex. 
For a given route $\rr\in \gGE$, we will denote with $i(\rr)$ its initial in-going external vertex and with $f(\rr)$ its final out-going external vertex.
These notions are illustrated in Figure~\ref{fig:cnnctdnss}.

With this notation, we write the FOPT representation as
\begin{gather} \begin{gathered} \label{eq:FOPT_ext_explicit} A_{G,\boldsymbol{\sigma}}(x_1,\ldots,x_{|\gVE|})=\\
=\frac{(2\pi g)^{|\gVI|}}{(-4\pi^2)^{|\gE|}}\int \frac{ \prod_{v\in V_{\text{int}}} \mathrm{d}^3 \vec{y}_v }{\left(\prod_{e\in E} 2|\vec{z}_e|\right) \left( \prod_{\cc\in \gGI} \gamma_\cc \right) \left[ \prod_{\rr\in \gGE}(\gamma_\rr-x_{i(\rr)}^0 +x_{f(\rr)}^0 + i \ep ) \right] }, \end{gathered} \end{gather}
where we drop the $i\ep$ for the cycles, 
as $\gamma_\cc=\sum_{e\in \cc}|\vec{z}_e|$ and
$\gamma_\cc \rightarrow 0$ is an end-point singularity, which cannot be regulated via analytic continuation.
Only the contribution from the factor
\begin{equation}
  \frac{
\prod_{v\in V_{\text{int}}} \mathrm{d}^3 \vec{y}_v }{\left(\prod_{e\in E} 2|\vec{z}_e|\right)
\left( \prod_{\cc\in \gGI} \gamma_\cc \right)
}
\end{equation}
is relevant for the short distance behaviour of the integrand, since the denominators corresponding to routes in eq.~(\ref{eq:FOPT_ext_explicit}) will not vanish if the time difference between in- and out-vertices is finite. 
This fact makes the divergent terms in FOPT either having a UV or IR (i.e.~long distance) divergent nature, but not both at the same time. This feature allows us to analyze UV and IR divergences separately.

In this section, we are interested in the UV case, i.e.\ limits in which the coordinates of a collection of internal vertices $V' \in \gVI$, which we can assume to be connected, coincide. We can approach this limit by picking a reference vertex $w \in V'$ and changing to the relative coordinates $\dd^3 \vec{y}_v \rightarrow \dd^3 \vec{z}_{wv}$ for all $v\in V'$ with $v \neq w$, where $\vec{z}_{wv} = \vec{y}_v - \vec{y}_w$. Rescaling $\vec{z}_{wv} \rightarrow \lambda \vec{z}_{wv}$ and considering $\lambda \rightarrow 0$ corresponds to the desired UV limit.

The power counting procedure for the integrand above is now straightforward.
The part of the measure with nontrivial scaling behaviour becomes $\prod_{v \in V',v \neq w} \dd^3 \vec{z}_{wv}$. 
Each edge $e$ that connects any pair of vertices in $V'$ gives a nontrivial scaling in the denominator, as $|\vec{z}_e|$ scales like $\lambda$.
Furthermore, $\gamma_\cc$ scales as $\lambda$ if the cycle $\cc$ traverses only vertices in $V'$.

Let $E'$ be the set of edges between the vertices $V'$, and $\gGI'$ the set of cycles that only contain vertices in $V'$. The relevant part of the integrand scales as $\lambda^{3(|V'|-1) - |E'| - |\gGI'|}$ and we find that the degree of divergence of the integral reads,
\begin{align} \label{eq:dod} \text{(degree of divergence)} = 3(|V'|-1) - |E'| - |\gGI'|. \end{align}
We will encounter a UV singularity only if the degree of divergence is $\leq 0$. We find the FOPT version of \emph{Weinberg's theorem} \cite{Weinberg:1959nj}:
a FOPT Feynman integral is UV-finite, if for each subset of internal vertices $V' \subset \gVI$, we have 
$3(|V'|-1) - |E'| - |\gGI'| > 0$.
Likewise, the degree of divergence of a coordinate-space covariant Feynman integral as in eq.~\eqref{eq:FI} is given by $4(|V'|-1) - 2|E'|$ for an induced subgraph given by $V'$ and $E'$ (see e.g.~\cite[Proposition~11]{gfe} for a more detailed statement). 

We can now make a connection between the FOPT and covariant UV divergences. The claim is that if the subgraph of a digraph $(G,\bb \sigma)$ induced by $V'$ (i.e.~the subgraph given by the vertices in $V'$ 
and all edges connecting them) is 
strongly connected, then  the degree of divergence coincides with the usual Feynman integral degree of divergence. To see this, observe that for a strongly connected graph we have the maximum number of cycles allowed in it and every edge of the graph is part of some cycle. 
Strong connectivity implies connectivity and hence we can apply Euler's formula to determine the number of its cycles as $|\gGI'| = |E'|-|V'|+ 1$. From this we find that
$3(|V'|-1) - |E'| - |\gGI'| = 4(|V'|-1) - 2|E'|$, coinciding with the covariant power counting of eq.~\eqref{eq:FI}. Therefore, the full UV scaling of the original Feynman integral is only realized if the subgraph is strongly connected.  A non-strongly connected subgraph has less singular UV scaling than a strongly connected orientation of the same subgraph, since it has fewer cycles. Briefly summarized, we find that the fewer cycles a FOPT diagram has, the more moderate its UV behaviour is.

\section{Finite and long distance singularity structure of FOPT diagrams}\label{sec:singularitiesFOPT}

In this section we proceed to discuss the finite and long distance singularities within the FOPT representation. We do so in order to compare with previous studies of the coordinate space singularities of Green's functions, as well as to provide a motivation for a study of the singular structure at the S-matrix level in the next section. Our findings are as follows:
\begin{itemize}
    \item The vanishing of denominators corresponding to routes, $\gamma_\rr$ with $\rr\in\Gamma^{\text{ext}}$, leads to singularities that are located at finite values of the distance between vertices in the graph. They are compatible with the studies performed in refs.~\cite{Sterman3,ErdoganCS,StermanErdogan}. However, their relation with momentum space infrared singularities is not direct. We present two examples in which the behaviour of the integrand at these finite distance singularities does not conform to the common expectation based on a momentum space analysis.

    \item The FOPT expression of the triangle diagram is mostly well-behaved at large values of the distances between vertices. The only non-integrable singularity we find does not have a direct interpretation in terms of the usual collinear and soft singularities of the triangle.
\end{itemize}

\subsection{Finite distance singularities}

The finite distance singularity structure of coordinate space diagrams has been extensively studied in refs.~\cite{Sterman3,ErdoganCS,StermanErdogan}. In particular, the authors identify pinched singularities that are finite distance in nature (that is, neither short distance nor long distance). The investigation of ref.~\cite{ErdoganCS}, in particular, is based on a coordinate space analogue of the Landau pinching conditions (a complete proof of the latter conditions can be found in \cite{Collins:2020euz}). Consider the integral
\begin{equation}
    I=\int \left[\prod_{i=1}^n \mathrm{d}w_i \right] \frac{N(\{w_j\}_{j=1}^n)}{ \prod_{i=1}^m D_i(\{w_j\}_{j=1}^n)}\;,
\end{equation}
where $N$ is an entire function. %
For the purposes of this discussion, we work with a sufficient condition for pinching: an $n$-vector $\vec{w}^{\,\star}$ is said to be located on a pinch singularity if, for some index $i$,
\begin{align} D_i(\{w_j\}_{j=1}^n)\big|_{\vec{w}=\vec{w}^{\,\star}}&=0 \notag\\
 \nabla_{\vec{w}}D_i(\{w_j\}_{j=1}^n)\big|_{\vec{w}=\vec{w}^{\,\star}}&=0 \,. \label{eq:landau_equations_2} \end{align}
Eq.~\eqref{eq:landau_equations_2} implies the impossibility of regulating the singularity through a contour deformation\footnote{Note that pinching conditions take different forms in different
formalisms
(i.e.\ covariant perturbation theory including
Feynman parameters, three-dimensional representations, TOPT, etc).}. Since the FOPT representation of eq.~\eqref{eq:delta_free_rep} has denominators that are in direct correspondence with oriented paths of the graph, reproducing the result of ref.~\cite{ErdoganCS} is rather straightforward. We consider the denominator $\prod_{\rr\in\Gamma^{\text{ext}}}(\gamma_\rr-x_{i(\rr)}^0 +x_{f(\rr)}^0 + i \ep)$ so that, using eq.~\eqref{eq:landau_equations_2}, we obtain for any route  $\rr\in\Gamma^{\text{ext}}$
\begin{align} \label{eq:landau_routes} &\gamma_\rr(\{\vec{y}_v\}_{v\in V^{\text{int}}})=-x_{f(\rr)}^0+x_{i(\rr)}^0 \notag \\
 &\nabla_{\vec{y}_{v'}}\gamma_\rr(\{\vec{y}_v\}_{v\in V^{\text{int}}})=0, \quad \quad \forall v'\in V^{\textrm{int}} \,. \end{align}
Labelling the vertices according to their order of appearance in the path, we have
\begin{equation}
    \gamma_\rr=|\vec{x}_{i(\rr)}-\vec{y}_1|+|\vec{x}_{f(\rr)}-\vec{y}_n|+\sum_{j=1}^{n-1}|\vec{y}_j-\vec{y}_{j+1}|,
\end{equation}
the solution eq.~\eqref{eq:landau_routes} is
\begin{gather} \vec{y}_j^{\,\star}=\vec{x}_i+\alpha_j(\vec{x}_{f(\rr)}-\vec{x}_{i(\rr)}), \quad \quad \alpha_j>\alpha_{j-1}, \quad \alpha_j\in[0,1], \quad \quad j=1,...,n\,. \notag\\
 -x_{f(\rr)}^0+x_{i(\rr)}^0=|\vec{x}_{f(\rr)}-\vec{x}_{i(\rr)}| . \label{eq:collinear_red} \end{gather}
The first condition establishes that all the vertices in the route $\rr$ must lie on a line, and that their order of appearance on that line must be the same as the order of appearance in the route. The second condition requires the four-distance between an initial vertex $x_{i(\rr)}$ and a final vertex $x_{f(\rr)}$ to be light-like. Assuming eq.~\eqref{eq:collinear_red} holds, we then find that the behaviour of $\gamma_\rr$ around its pinch points is quadratic 
\begin{equation}
\label{eq:quadratic_scaling}
    \gamma_\rr(\{\vec{y}_j^{\,\star}+\delta \vec{y}_j\}_{j=1}^n)-|\vec{x}_{i(\rr)}-\vec{x}_{f(\rr)}|={\cal O}(|\delta \vec{y}_j|^2).
\end{equation}
While this seems a promising candidate for collinear singularity, a few wrinkles in the previous argument give pause:
\begin{enumerate}
    \item We expect IR singularities of momentum space Feynman integrals to be associated to long distance singularities in coordinate space. The singular location of eq.~\eqref{eq:collinear_red} describes a finite distance singularity, as no integration variable is set to be large. 
    \item There is no reason why $-x_{f(\rr)}^0+x_{i(\rr)}^0=|\vec{x}_{f(\rr)}-\vec{x}_{i(\rr)}|$ should hold; since we know that, in momentum space, external momenta are localised on their mass-shell, we would expect external vertices not to be localised, since the two are related by a Fourier transform (or equivalently, by the uncertainty principle).
    \end{enumerate} 
Let us look at two examples which highlight the ambiguous nature of such finite distance singularities:
\begin{enumerate}
    \item \textbf{The hen's claw}.  We consider the FOPT diagram
    \begin{equation}
    \raisebox{-0.7cm}{\resizebox{2cm}{!}{%
\begin{tikzpicture} \begin{feynman} \vertex(1); \vertex[above = 1.39cm of 1](2); \vertex[below right = 1cm and 1cm of 1](3); \vertex[below left = 1cm and 1cm of 1](4); \vertex[left = 0.2cm of 2](L1) {\scalebox{1.5}{$x_1$}}; \vertex[right = 0.2cm of 3](L2) {\scalebox{1.5}{$x_2$}}; \vertex[left = 0.2cm of 4](L3) {\scalebox{1.5}{$x_3$}}; \vertex[below = 0.2cm of 1](L1_I) {\scalebox{1.5}{$y$}}; \diagram*[large]{ (1) -- [->-,line width=0.6mm] (2), (1) -- [-<-,line width=0.6mm] (3), (1) -- [-<-,line width=0.6mm] (4) }; \path[draw=black, fill=black] (2) circle[radius=0.1]; \path[draw=black, fill=black] (3) circle[radius=0.1]; \path[draw=black, fill=black] (4) circle[radius=0.1]; \end{feynman} \end{tikzpicture}
}}=\int \mathrm{d}^3 \vec{y} \frac{1}{\left[\prod_{j=1}^3 2|\vec{y}-\vec{x}_j|\right]\prod_{\ell=2}^3(|\vec{x}_\ell-\vec{y}|+|\vec{y}-\vec{x}_1|-|\vec{x}_\ell-\vec{x}_1|)}.
    \end{equation}
    According to eq.~\eqref{eq:collinear_red}, there is a pinched singularity when $\vec{y}^{\,\star}=\vec{x}_3+\alpha(\vec{x}_1-\vec{x}_3)$, $\alpha\in[0,1]$ (and analogously when substituting $2$ with $3$). Using the parameterization $\vec{y}^{\,\star}=\vec{x}_3+\alpha(\vec{x}_1-\vec{x}_3)+\vec{y}^\perp$, we see that the measure contributes with two powers of the transverse variable $\vec{y}^\perp$, while eq.~\eqref{eq:quadratic_scaling} makes it clear that the denominator also contributes with two powers. Overall, this gives a logarithmically divergent behaviour in this limit.
    This finding is odd: in momentum space, such a diagram would be completely finite and, for a cubic scalar theory, it would (after truncation of external propagators) amount to a constant. 
    \item \textbf{The triangle}: We look specifically at the orientation
    \begin{equation}
    \resizebox{4.cm}{!}{\resizebox{5cm}{!}{%
\begin{tikzpicture} \begin{feynman} \vertex(1); \vertex[above right = 1.5cm and 2.5cm of 1](2); \vertex[below right = 1.5cm and 2.5cm of 1](3); \vertex[left = 1.5cm of 1](E1); \vertex[right = 1.5cm of 2](E2); \vertex[right = 1.5cm of 3](E3); \vertex[left = 0.2cm of E1](L1) {\scalebox{1.5}{$x_1$}}; \vertex[right = 0.2cm of E2](L2) {\scalebox{1.5}{$x_2$}}; \vertex[right = 0.2cm of E3](L3) {\scalebox{1.5}{$x_3$}}; \vertex[below = 0.2cm of 1](L1_I) {\scalebox{1.5}{$y_1$}}; \vertex[above = 0.2cm of 2](L2_I) {\scalebox{1.5}{$y_2$}}; \vertex[below = 0.2cm of 3](L3_i) {\scalebox{1.5}{$y_3$}}; \diagram*[large]{ (1) -- [-<-,line width=0.6mm] (2) -- [-<-,line width=0.6mm] (3) -- [->-,line width=0.6mm] (1), (1) -- [->-,line width=0.6mm] (E1), (2) -- [-<-,line width=0.6mm] (E2), (3) -- [-<-,line width=0.6mm] (E3), }; \path[draw=black, fill=black] (E1) circle[radius=0.15]; \path[draw=black, fill=black] (E2) circle[radius=0.15]; \path[draw=black, fill=black] (E3) circle[radius=0.15]; \path[draw=black, fill=black] (1) circle[radius=0.1]; \path[draw=black, fill=black] (2) circle[radius=0.1]; \path[draw=black, fill=black] (3) circle[radius=0.1]; \end{feynman} \end{tikzpicture}
}}
    \label{eq:triangle_FOPT}
\end{equation}
    and at the pinched singular surface identified by the implicit equation
    \begin{equation*}
        \gamma_\rr-|\vec{x}_1-\vec{x}_2|=|\vec{x}_2-\vec{y}_2|+|\vec{y}_2-\vec{y}_1|+|\vec{y}_1-\vec{x}_1|-|\vec{x}_1-\vec{x}_2|=0,
    \end{equation*}
    which has the solution
    \begin{equation*}
        \vec{y}_1^{\,\star}(\alpha_1)=\vec{x}_1+\alpha_1(\vec{x}_2-\vec{x}_1), \quad \vec{y}_2^{\,\star}(\alpha_2)=\vec{x}_1+\alpha_2(\vec{x}_2-\vec{x}_1), \quad \alpha_1,\alpha_2\in[0,1], \,\,  \alpha_2>\alpha_1.
    \end{equation*}
    In order to approach the limit, we consider the parameterization $\vec{y}_1=\vec{y}_1^{\,\star}(\alpha_1)+\vec{y}^\perp_1$, $\vec{y}_2=\vec{y}_2^{\,\star}(\alpha_2)+\vec{y}^\perp_2$. A simple power-counting procedure leads us to conclude that the integration measure scales as $\mathrm{d}^2\vec{y}_2^\perp\mathrm{d}^2\vec{y}_1^\perp$ while the singular surface only gives a contribution of two powers in the denominator. Overall, this gives an integrable degree of divergence. Carrying out the same argument for all other pinched surfaces of all triangle orientations as well as the intersections of pinched surfaces, one finds that the triangle has no finite distance singularity\footnote{In contrast, in ref.~\cite{ErdoganCS}, a logarithmic finite distance singularity is found for the triangle. However, in their derivation, one of the three internal vertices is set to be at the origin, and the corresponding external leg is truncated.}.
\end{enumerate}
The fact that these two diagrams do not reproduce the expectations concerning their IR structure invites us to change our perspective, and shift our attention to the S-matrix. First, we note that the S-matrix is constructed from truncated Green's functions, while here we considered FOPT expressions for non-truncated Green's functions. Indeed this is the reason that the hen's claw has singular denominators in the first place. On top of this, we  must set $x_{i(\rr)}^0-x_{f(\rr)}^0=|\vec{x}_{f(\rr)}-\vec{x}_{i(\rr)}|$, which is at odds with localization of momenta on the mass-shell. These two observations lead us to examine the S-matrix to correctly identify IR singularities in the FOPT formalism. The S-matrix is a truncated object and involves integration over the external coordinates $x_{a}^0$, for all $a\in V^{\text{ext}}$, via the Fourier transform to momentum space.

\subsection{Long distance singularities}

We now briefly discuss long distance singularities of FOPT orientations, which we would expect to map one-to-one to their momentum space analogues. In order to do so, let us look again at the orientation of eq.~\eqref{eq:triangle_FOPT} in the triangle diagram, and comment on the absence of such long distance singularities. In particular, consider the following limits
\begin{enumerate}
    \item Single limits: the limit $\vec{y}_1\rightarrow\infty$ has the measure scaling as $|\vec{y}_1|^2\mathrm{d}\vec{y}_1$ and the denominator scaling as $|\vec{y}_1|^6$, giving integrable behaviour. The limit $\vec{y}_2\rightarrow\infty$ has the measure scaling as $|\vec{y}_2|^2\mathrm{d}\vec{y}_2$ and the denominator scaling as $|\vec{y}_1|^5$, giving integrable behaviour. Finally, the limit $\vec{y}_3\rightarrow\infty$ has the measure scaling like $|\vec{y}_3|^2\mathrm{d}\vec{y}_3$ and the denominator scaling as $|\vec{y}_1|^5$, giving integrable behaviour.

    \item Double limits: consider taking the limit $\vec{y}_1, \vec{y}_2\rightarrow\infty$. The scaling is independent from $\vec{y}_1$ and $\vec{y}_2$ being collinear, as $|\vec{y}_1-\vec{y}_2|$ is finite in the collinear case. But the angle between $\vec{y}_1$ and $\vec{y}_2$ has to be fixed too, giving an extra suppressing power from the integration measure. Thus, let us study the non-collinear case: we have the measure scaling as $|\vec{y}_1|^2|\vec{y}_2|^2\mathrm{d}\vec{y}_1\mathrm{d}\vec{y}_2$, while the denominator scales as $|\vec{y}_1-\vec{y}_2||\vec{y}_1|^3|\vec{y}_2|^2\text{max}(|\vec{y}_1|^2,|\vec{y}_2|^2)$, giving integrable behaviour. Analogous integrable behaviour can be found in the remaining two double limits.
\end{enumerate}
The only limit that yields non-integrable behaviour is the one in which all three vertices of the triangle approach infinity in different directions, which has no clear analogue in momentum space.

The observations pointed out above motivate changing focus to the S-matrix in order to correctly analyze IR singularities in FOPT diagrams. We will do this by using the FOPT representation from the last section to construct a \emph{hybrid} representation for the S-matrix, where external parameters are given in momentum space, while internal integrations are performed in coordinate space.

\section{The S-matrix and its \texorpdfstring{$p$-$x$}{p-x} representation}
\label{sec:s_matrix}

The FOPT representation, derived in sect.~\ref{sec:derivation}, gives a 
canonical representation of QFT correlation functions in terms of three-dimensional integrals. However, its external data is given in coordinate space, which makes it ill-suited for the direct computation of realistic collider observables. This is reflected in the absence of a clear counterpart to IR singularities in the FOPT representation, which in turn has repercussions on how manifestly physical principles such as parton degeneracy and collinear mass factorization are realised within it.

In this section we will address this problem by constructing a new perturbative representation of the \emph{S-matrix} in momentum space. It is expressed as an integral over spatial coordinates, even though the external data is given in momentum space. We therefore call it the \emph{$p$-$x$} representation of the S-matrix, as a reference to its hybrid nature.

Recall that the  retarded and advanced propagators in momentum space can be written as 
\begin{align} \label{eq:ret_adv} \widetilde{\Delta}_{\genfrac{}{}{0pt}{3}{R}{A}}(p) =\frac{i}{(p_0\pm i\ep)^2 - \vec{p}^2} = \frac{1}{2\pi i} \int_{\R^4} \dd^4 z e^{iz \cdot p} \delta^{(\pm)}(z^2) = \frac{1}{2\pi i} \int_{\R^3} \frac{\dd^3 \vec{z}}{2|\vec{z}\,|} e^{\pm i|\vec{z}| p^0-i\vec{z}\cdot \vec{p} } \;. \end{align}
To give an expression for the S-matrix, we have to fix in- and out-going momenta. In anticipation of our FOPT treatment of the S-matrix we index the in-going momenta by $\gVEI$, $\{p_i\}_{i\in \gVEI}$ and the out-going momenta by $\gVEO$, $\{p_f\}_{f\in \gVEO}$. 
The S-matrix element in terms of the in- and out-going momenta is given by %
\begin{equation}
S(\{p_i\}_{i\in \gVEI}, \{p_f\}_{f\in \gVEO})= Z^{|V_\text{ext}|/2} \widetilde{\Gamma}_T(\{p_a\}_{a\in \gVE})\,,
\end{equation}
where $\widetilde{\Gamma}_T$ is the truncated Green's function, with the typical convention that energy directed towards the graph is positive, such that 
$p_i^0 > 0$ and $p_f^0 < 0$, and $Z^{|V_\text{ext}|/2}$ is the factor accounting for the truncation of external self-energies, as established by the LSZ formula.
The external momenta should be set on-shell eventually, i.e.~$p_i^0\rightarrow |\vec{p}_i|$, for $i\in \gVEI$ and $p_f^0\rightarrow -|\vec{p}_f|$, for $f\in \gVEO$. 
The truncated Green's function can be obtained from the ordinary Green's function with no external self-energy insertions by the rule
\begin{equation}
\label{eq:truncation_s_matrix}
    \widetilde{\Gamma}_T(\{p_a\}_{a\in \gVE})=\Bigg[\prod_{i\in \gVEI}\widetilde{\Delta}_R(p_i)\Bigg]^{-1}\Bigg[\prod_{f\in \gVEO}\widetilde{\Delta}_A(p_f)\Bigg]^{-1}\widetilde{\Gamma}(\{p_a\}_{a\in \gVE}).
\end{equation}
Notice that we use advanced and retarded propagators for in- and out-going particles respectively, instead of the Feynman propagator $\widetilde{\Delta}_F$. This is justified by the simple equation $\widetilde{\Delta}_F(p)=\Theta(p^0)\widetilde{\Delta}_R(p)+\Theta(-p^0)\widetilde{\Delta}_A(p)$ and the fact that in-coming (out-going) particles are taken to have positive (negative) energy $p^0$.
We stress that $\widetilde{\Gamma}$ has no external self-energy insertions, but it does have external propagators. $\widetilde{\Gamma}$ can then be obtained by Fourier transforming its coordinate space counterpart, which we can express using FOPT. In particular, we have
\begin{align} \label{eq:tildeGamma} \widetilde{\Gamma}(\{p_a\}_{a\in \gVE})=\int \left[\prod_{a\in V^{\text{ext}}} \mathrm{d}^4 x_a \, e^{ix_a\cdot p_a}\right]\Gamma(\{x_a\}_{a\in\gVE}), \end{align}
where $\Gamma(\{x_a\}_{a\in\gVE})$ is the coordinate space correlation function in eq.~(\ref{scalarcorrelator}) and, in agreement to the combinatorial notation of sect.~\ref{sec:cycles_singularities}, $\gVE = \gVEI \cup \gVEO$.

\subsection{Derivation of the \texorpdfstring{$p$-$x$}{p-x} representation}

In this section we derive the $p$-$x$ representation of the S-matrix. 
It is a representation of eq.~\eqref{eq:truncation_s_matrix} in terms of coordinate space integrals 
that are modulated by the Fourier transformation of a certain polytope. 
We will show that this polytope is a variant of the \emph{flow polytope} (see e.g.~\cite{baldoni2008kostant}).

We use the FOPT representation of the correlation function in eq.~\eqref{eq:FOPT_greens_functions} to express the Fourier transformed correlation function in eq.~\eqref{eq:tildeGamma}:
\begin{align} \label{eq:tildeGamma2} \widetilde{\Gamma}(\{p_a\}_{a\in \gVE})= \sum_{(G,\bb \sigma)} \frac{1}{\Sym (G,\bb \sigma)} \widetilde A_{G,\bb \sigma}(\{p_a\}_{a\in \gVE}), \end{align}
where the Fourier transform of a FOPT orientation is given by
\begin{align} \widetilde A_{G,\bb \sigma}(\{p_a\}_{a\in \gVE}) = \int \left[\prod_{a\in V^{\text{ext}}} \mathrm{d}^4 x_a \, e^{ix_a\cdot p_a}\right] A_{G,\bb \sigma}(\{x_a\}_{a\in \gVE}). \end{align}
We can use the FOPT representation with explicit separation of external and internal paths, i.e.\ cycles and routes, in eq.~\eqref{eq:FOPT_ext_explicit} to write $\widetilde A_{G,\bb \sigma}(\{p_a\}_{a\in \gVE})$ as 
\begin{gather} \begin{gathered} \label{eq:AG_f} \widetilde A_{G,\bb \sigma}(\{p_a\}_{a\in \gVE}) = \\
\frac{(2\pi g)^{|\gVI|}}{(-4\pi^2)^{|\gE|}} \int \frac{ \left( \prod_{a\in \gVE} \mathrm{d}^3 \vec{x}_a e^{-i\vec{x}_a \cdot \vec{p}_a} \right) \left( \prod_{v\in V^{\text{int}}} \mathrm{d}^3 \vec{y}_v \right) }{\left(\prod_{e\in E} 2|\vec{z}_e|\right) \left( \prod_{\cc\in \gGI} \gamma_\cc \right) } f_{G,\bb \sigma}(\{p_a^0\}_{a\in \gVE}, \{\gamma_\rr\}_{\rr \in \gGE}) , \end{gathered} \end{gather}
where we split off the nontrivial time integration, which we plan on performing analytically, through the definition %
\begin{align} f_{G,\bb \sigma}(\{p_a^0\}_{a\in \gVE}, \{\gamma_\rr\}_{\rr \in \gGE}) = \int \frac{ \prod_{a\in \gVE} \mathrm{d} x_a^0 e^{i x_a^0 p_a^0} }{ \prod_{\rr\in \gGE}(\gamma_\rr-x_{i(\rr)}^0 +x_{f(\rr)}^0 + i \ep ) }. \end{align}
Let us perform the energy integrations over the external time variables $x_a^0$ analytically. To this end we follow steps analogous to those performed in the derivation of sect.~\ref{sec:derivation}. 
We introduce one auxiliary integration variable $z_\rr^0$ for each route $\rr$, parameterizing the time difference between the external vertices $i(\rr)$ and $f(\rr)$,
\begin{equation}
    f_{G,\bb{\sigma}}=
\int \left[ \prod_{a\in V^{\text{ext}}} 
\mathrm{d} x_a^0 \, e^{i x_a^0p_a^0} \right] 
 \prod_{\rr\in \gGE}   \mathrm{d}z_{\rr}^0 \frac{  \delta(z_{\rr}^0-x_{i(\rr)}^0+x_{f(\rr)}^0)}{\gamma_\rr-z_{\rr}^0 + i\ep  }.
\end{equation}
We then rewrite the delta functions using their Fourier representation, thus introducing an extra integration in the energy $E_\rr$,
\begin{equation}
    f_{G,\boldsymbol{\sigma}}=\int \left[ \prod_{a\in V^{\text{ext}}} \mathrm{d} x_a^0 \, e^{i x_a^0p_a^0} \right]  \prod_{\rr\in \gGE}   \frac{\mathrm{d}z_{\rr}^0 \mathrm{d}E_\rr}{2\pi} \frac{  e^{iE_\rr\left(z_{\rr}^0-x_{i(\rr)}^0+x_{f(\rr)}^0\right)}}{\gamma_\rr-z_{\rr}^0 +i\ep }.
\end{equation}
Integration over $z_{\rr}^0$ is performed trivially using the residue theorem. For $E_\rr>0$, we close the contour in the upper-half of complex plane, where the pole $z_\rr^0=\gamma_\rr+i\varepsilon$ is located. For $E_\rr<0$, we close the contour in the lower-half of complex plane, where no pole is located. In summary, this gives 
\begin{equation}
    f_{G,\boldsymbol{\sigma}}=
(-i)^{|\gGE|}
\int \left[ \prod_{a\in V^{\text{ext}}} \mathrm{d} x_a^0 \, e^{i x_a^0p_a^0} \right]  \prod_{\rr\in \gGE}    \mathrm{d}E_\rr  e^{iE_\rr\left(\gamma_{\rr}-x_{i(\rr)}^0+x_{f(\rr)}^0 + i\ep\right)}\Theta(E_\rr).
\end{equation}
To perform the integration over $x_a^0$, we may write
\begin{gather} f_{G,\boldsymbol{\sigma}}= (-i)^{|\gGE|} \int \prod_{\rr\in \gGE} \left[ \mathrm{d}E_\rr e^{iE_\rr(\gamma_{\rr}+i\ep)}\Theta(E_\rr)\right] \nonumber \\ \times \left[ \prod_{i\in \gVEI} \int\mathrm{d} x_i^0 \, \exp\left({i x_i^0p_i^0 - i\sum_{\rr \ni i}E_\rr x_i^0 }\right) \right] \left[\prod_{f\in \gVEO} \int\mathrm{d} x_f^0 \, \exp\left({i x_f^0p_f^0 + i\sum_{\rr\ni f }E_\rr x_f^0 }\right) \right], \end{gather}
where we sum over all routes that include the indicated in- or out-going vertex in the last two exponentials.
Using the Fourier representation of the $\delta$ function, we can resolve the integrals over the $x_a^0$ variables, yielding
\begin{gather} \begin{gathered} \label{eq:fexpression_pre_poly} f_{G,\boldsymbol{\sigma}}= \frac{(-i)^{|\gGE|}}{ (2\pi)^{|\gVE|}} \int \left[ \prod_{\rr\in \gGE} \mathrm{d}E_\rr e^{iE_\rr(\gamma_{\rr}+i\ep)}\Theta(E_\rr)\right] \\ \times \left[ \prod_{i\in \gVEI} \delta\left( p_i^0 - \sum_{\rr \ni i}E_\rr \right) \right] \left[ \prod_{f\in \gVEO} \delta\left( p_f^0 + \sum_{\rr \ni f }E_\rr \right) \right]. \end{gathered} \end{gather}
The $\Theta$ and $\delta$ functions in the above integral describe a polytope in the $|\gGE|$ dimensional space of the energy integrations. This polytope is a variant of the \emph{flow polytope} which has various applications in combinatorics. 
We conclude that $f_{G,\bb \sigma}$ is an evaluation of the \emph{Fourier transform} of this polytope, as we integrate $\exp\left( i \sum_{\rr \in \gGE} E_\rr (\gamma_\rr+i\ep) \right)$.
The full form of the $p$-$x$ S-matrix representation is most conveniently given using this polytope. 
For this reason, we will describe this polytope in detail in the next section.

\subsection{The flow polytope}

To every digraph $(G,\bb \sigma)$ with a given set of in- and out-going energies $\{p^0_{i}\}_{i\in \gVEI}$ and $\{p^0_{f}\}_{f\in \gVEO}$, 
we associate a polytope,
$\mathcal{F}_{G,\bb \sigma}^{\{p^0_a\}}\subset\mathbb{R}^{|\gGE|}$.
The polytope is defined by linear equalities and inequalities that are given in the energy coordinates $E_\rr$ associated to the routes $\rr \in \gGE$. For each path, the respective energy represents the amount of energy that flows through the specified path. Accordingly, the delta functions in eq.~\eqref{eq:fexpression_pre_poly} enforce energy conservation at each external vertex. In summary, $\mathcal{F}_{G,\bb \sigma}^{\{p^0_a\}}$
is swept out by all tuples $(E_{\rr})_{\rr \in \gGE}$ which fulfill%
\begin{align} \begin{aligned} \label{eq:flow_poly_def} E_{\rr} &\geq 0 \text{ for all } \rr \in \gGE\,, \\
\sum_{\rr \ni i} E_\rr &= \phantom{-}p^0_{i} \text{ for all } i \in \gVEI, \\
\sum_{\rr \ni f} E_\rr &= -p^0_{f} \text{ for all } f \in \gVEO. \end{aligned} \end{align}
The inequalities ensure a positive energy flow through each route and the two sets of equalities ensure energy conservation at in- and out-going external vertices. Note that the equations above imply overall energy conservation for the external data: summing all the equalities with $i \in \gVEI$ and subtracting those with $f \in \gVEO$, one obtains that $\sum_{a \in \gVE} p^0_a =0$, since each path has exactly one in- and one out-going vertex.

The eqs.~\eqref{eq:flow_poly_def} define a convex polytope, obtained from the intersection of the positive orthant $\{(E_\rr)_{\rr \in \gGE} \in \R^{|\gGE|} : E_\rr \geq 0 \}$ with the $|\gVE|$ hyperplanes given by energy conservation constraints at the external vertices. It is moreover bounded, since the energies are contained in the hypercube, $0 \leq E_\rr \leq \min\left(p_{i(\rr)}^0, -p_{f(\rr)}^0\right)$.
We can also find, depending on the value of the external data, configurations of external energies where the eqs.~\eqref{eq:flow_poly_def} cannot be fulfilled. In those cases, the flow polytope has no support and the contribution of the digraph $(G,\bb \sigma)$ to the S-matrix is zero.

To simplify the notation, in the following we denote vectors in the route space $\R^{|\gGE|}$ with bold letters. The tuples $(E_\rr)_{\rr \in \gGE}\in\R^{|\gGE|}$ are denoted as $\bb E \in \R^{|\gGE|}$ and tuples of path lengths $(\gamma_\rr)_{\rr \in \gGE} \in \R^{|\gGE|}$ as $\bb \gamma \in \R^{|\gGE|}$. We also can define the standard scalar product $\bb E \cdot \bb \gamma = \sum_{\rr \in \gGE} E_\rr \gamma_\rr$.
We may use the definition of the polytope and the boldface notation to rewrite eq.~\eqref{eq:fexpression_pre_poly} in a more compact form,
\begin{gather} \label{eq:fexpression_post_poly} f_{G,\bb \sigma}(\{p_a^0\}_{a\in \gVE}, \bb \gamma) = \frac{(-i)^{|\gGE|}}{ (2\pi)^{|\gVE|}} \delta\left( \sum_{a \in \gVE} p^0_a \right) \widehat{\mathcal{F}}_{G,\bb \sigma}^{\{p^0_a\}}(\bb \gamma + i \ep \bb{1}) \end{gather}
where $\bb 1=(1,\ldots,1)$ is the all-ones-vector in $\R^{|\gGE|}$.
The \emph{Fourier transform}, $ \widehat{\mathcal{F}}_{G,\bb \sigma}^{\{p^0_a\}}(\bb \gamma) $
 of the polytope $\mathcal{F}_{G,\bb \sigma}^{\{p^0_a\}}$ is given by
\begin{gather} \label{eq:FF_def} \widehat{\mathcal{F}}_{G,\bb \sigma}^{\{p^0_a\}}(\bb \gamma + i \ep \bb{1}) = \int_{\mathcal{F}_{G,\bb \sigma}^{\{p^0_a\}}} \dd \bb E\, e^{i \bb E \cdot (\bb \gamma + i \ep \bb{1} ) }. \end{gather}
The integration over the bounded domain $\mathcal{F}_{G,\bb \sigma}^{\{p^0_a\}}$, 
allows us to drop the $i\ep$ prescription in the generic case. 
We will however keep it in place now to study singular configurations later on.
The Fourier transform of a polytope enjoys many remarkable properties. 
For instance it is bounded by the volume of the polytope. In our case we have
\begin{equation}
\left|
\widehat{\mathcal{F}}_{G,\bb \sigma}^{\{p^0_a\}}(\bb \gamma + i \ep \bb{1})
\right|
\leq \mathrm{Vol}\left( 
\mathcal{F}_{G,\bb \sigma}^{\{p^0_a\}}
\right)
\text{ for all } \bb \gamma \in \R^{|\gGE|}.
\end{equation}
Equality is only  attained for $\bb \gamma= 0$.
Fourier transforms of polytopes are always linear combinations of rational functions with phase factors as coefficients. An analytic expression for such Fourier transforms can always be obtained algorithmically by computing a \emph{facet presentation} of the associated polytope (see e.g.~\cite{polytope_fourier}). 

\subsection{Truncated routes}

In this section we will relate FOPT integrals to \emph{truncated FOPT integrals}. 
This relation is a special case of a general \emph{factorization} property of 
the Fourier transform of the flow polytope.

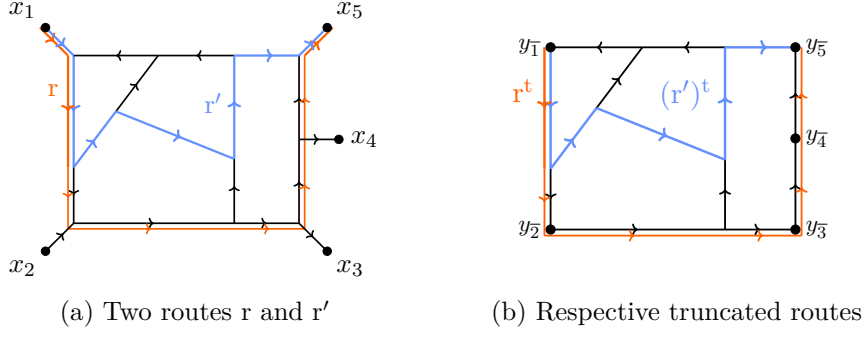
\begin{figure}[t!]
\centering
\begin{subfigure}[b]{2in}
\begin{center}
\resizebox{5cm}{!}{%
\begin{tikzpicture} \begin{feynman} \vertex(1); \vertex[below = 6cm of 1](2); \vertex[right = 8cm of 2](3); \vertex[above = 6cm of 3](4); \vertex[above = 3cm of 3](M34); \vertex[right = 1.414cm of M34](E34); \vertex[above left = 1cm and 1cm of 1](E1); \vertex[above right = 1cm and 1cm of 4](E4); \vertex[below left = 1cm and 1cm of 2](E2); \vertex[below right = 1cm and 1cm of 3](E3); \vertex[above left = 1.2cm and 1.2cm of 1](L1) {\scalebox{2.5}{$x_1$}}; \vertex[above right = 1.2cm and 1.2cm of 4](L4) {\scalebox{2.5}{$x_5$}}; \vertex[below left = 1.2cm and 1.2cm of 2](L2) {\scalebox{2.5}{$x_2$}}; \vertex[below right = 1.2cm and 1.2cm of 3](L3) {\scalebox{2.5}{$x_3$}}; \vertex[right = 1.7cm of M34](L34) {\scalebox{2.5}{$x_4$}}; \vertex[below left = 0.9cm and 0.4cm of 1](path1) {\scalebox{2.5}{$\color{col3} \rr$}}; \vertex[below = 4cm of 1](M12); \vertex[right = 3cm of 1](T141); \vertex[below right = 2cm and 1.5 cm of 1](T142); \vertex[left = 2.3cm of 4](R4); \vertex[left = 2.3cm of 3](DR4); \vertex[below = 3.7cm of R4](MR4); \vertex[above left = 1.5cm and 0.3cm of MR4](path2) {\scalebox{2.5}{$\color{col1} \rr'$}}; \vertex[left = 0.2cm of E1](shiftE1); \vertex[left = 0.2cm of 1](shift1); \vertex[left = 0.2cm of M12](shiftM12); \vertex[below left = 0.2cm and 0.2cm of 2](shift2); \vertex[below = 0.2cm of DR4](shiftDR4); \vertex[below right = 0.2cm and 0.2cm of 3](shift3); \vertex[right = 0.2cm of R4](shiftR4); \vertex[right = 0.2cm of M34](shiftM34); \vertex[right = 0.2cm of 4](shift4); \vertex[right = 0.2cm of E4](shiftE4); \diagram*[large]{ (1) -- [->-,line width=0.8mm, col1] (M12) -- [->-,line width=0.6mm, ] (2) -- [->-,line width=0.6mm, ] (DR4) -- [->-,line width=0.6mm, ] (3) -- [->-,line width=0.6mm, ] (M34) -- [->-,line width=0.6mm, ] (4) -- [-<-,line width=0.8mm, col1] (R4) -- [->-,line width=0.6mm] (T141) -- [->-,line width=0.6mm] (1), (shiftE1) -- [->-,line width=0.8mm, col3] (shift1) -- [->-,line width=0.8mm, col3] (shiftM12) -- [->-,line width=0.6mm, col3] (shift2) -- [->-,line width=0.6mm, col3] (shiftDR4) -- [->-,line width=0.6mm, col3] (shift3) -- [->-,line width=0.6mm, col3] (shiftM34) -- [->-,line width=0.6mm, col3] (shift4) -- [->-,line width=0.8mm, col3] (shiftE4), (M34) -- [->-,line width=0.6mm] (E34), (1) -- [-<-,line width=0.8mm, col1] (E1), (2) -- [-<-,line width=0.6mm] (E2), (3) -- [->-,line width=0.6mm] (E3), (4) -- [->-,line width=0.8mm, col1] (E4), (R4) -- [-<-,line width=0.8mm, col1] (MR4) -- [-<-,line width=0.6mm] (DR4), (M12) -- [->-,line width=0.8mm, col1] (T142) -- [->-,line width=0.6mm] (T141), (MR4) -- [-<-,line width=0.8mm, col1] (T142), }; \path[draw=black, fill=black] (E1) circle[radius=0.15]; \path[draw=black, fill=black] (E2) circle[radius=0.15]; \path[draw=black, fill=black] (E3) circle[radius=0.15]; \path[draw=black, fill=black] (E4) circle[radius=0.15]; \path[draw=black, fill=black] (E34) circle[radius=0.15]; \end{feynman} \end{tikzpicture}
}%
\caption{Two routes $\rr$ and $\rr'$}
\end{center}%
\end{subfigure}  \hspace{1cm}
\begin{subfigure}[b]{2in}
\begin{center}
\resizebox{4.3cm}{!}{%
\begin{tikzpicture} \begin{feynman} \vertex(1); \vertex[below = 6cm of 1](2); \vertex[right = 8cm of 2](3); \vertex[above = 6cm of 3](4); \vertex[above = 3cm of 3](M34); \vertex[below = 4cm of 1](M12); \vertex[right = 3cm of 1](T141); \vertex[below right = 2cm and 1.5 cm of 1](T142); \vertex[left = 2.3cm of 4](R4); \vertex[left = 2.3cm of 3](DR4); \vertex[below = 3.7cm of R4](MR4); \vertex[left = 0.2cm of 1](L1) {\scalebox{2}{$y_{\overline{1}}$}}; \vertex[left = 0.2cm of 2](L2) {\scalebox{2}{$y_{\overline{2}}$}}; \vertex[right = 0.2cm of 3](L3) {\scalebox{2}{$y_{\overline{3}}$}}; \vertex[right = 0.2cm of 4](L4) {\scalebox{2}{$y_{\overline{5}}$}}; \vertex[right = 0.2cm of M34](LM34) {\scalebox{2}{$y_{\overline{4}}$}}; \vertex[left = 0.2cm of 1](shift1); \vertex[left = 0.2cm of M12](shiftM12); \vertex[below left = 0.2cm and 0.2cm of 2](shift2); \vertex[below = 0.2cm of DR4](shiftDR4); \vertex[below right = 0.2cm and 0.2cm of 3](shift3); \vertex[right = 0.2cm of R4](shiftR4); \vertex[right = 0.2cm of M34](shiftM34); \vertex[right = 0.2cm of 4](shift4); \vertex[below left = 0.9cm and 0.4cm of 1](path1) {\scalebox{2.5}{$\color{col3} \rr^\mathrm{t}$}}; \vertex[above left = 1.5cm and 0.3cm of MR4](path2) {\scalebox{2.5}{$\color{col1} (\rr')^\mathrm{t}$}}; \diagram*[large]{ (1) -- [->-,line width=0.8mm, col1] (M12) -- [->-,line width=0.6mm] (2) -- [->-,line width=0.6mm] (DR4) -- [->-,line width=0.6mm] (3) -- [->-,line width=0.6mm] (M34) -- [->-,line width=0.6mm] (4) -- [-<-,line width=0.8mm, col1] (R4) -- [->-,line width=0.6mm] (T141) -- [->-,line width=0.6mm] (1), (shift1) -- [->-,line width=0.8mm, col3] (shiftM12) -- [->-,line width=0.6mm, col3] (shift2) -- [->-,line width=0.6mm, col3] (shiftDR4) -- [->-,line width=0.6mm, col3] (shift3) -- [->-,line width=0.6mm, col3] (shiftM34) -- [->-,line width=0.6mm, col3] (shift4), (R4) -- [-<-,line width=0.8mm, col1] (MR4) -- [-<-,line width=0.6mm] (DR4), (M12) -- [->-,line width=0.8mm, col1] (T142) -- [->-,line width=0.6mm] (T141), (MR4) -- [-<-,line width=0.8mm, col1] (T142), }; \path[draw=black, fill=black] (1) circle[radius=0.15]; \path[draw=black, fill=black] (2) circle[radius=0.15]; \path[draw=black, fill=black] (3) circle[radius=0.15]; \path[draw=black, fill=black] (4) circle[radius=0.15]; \path[draw=black, fill=black] (M34) circle[radius=0.15]; \end{feynman} \end{tikzpicture}
}
\vspace{0.5cm}%
\caption{Respective truncated routes}
\end{center}
\end{subfigure} 
\caption{Illustration of the truncation of routes. 
}
\label{fig:truncated}
\end{figure}

Let $\rr$ be a route joining the external vertices $i(\rr)\in \gVEI$ and $f(\rr)\in \gVEO$. Furthermore, let $\overline{i(\rr)}$ and $\overline{f(\rr)}$ denote the only vertices adjacent to $i(\rr)$ and $f(\rr)$, respectively. The truncated version of the route $\rr$ is given by all edges of $\rr$ except for the initial and final one. We will denote the length of the truncated route of a route $\rr$ as $\gamma_\rr^\tt$ (see Figure~\ref{fig:truncated}). The relation between the length $\gamma_\rr$ of the original route and the truncated route is  
\begin{equation}
    \gamma_{\rr}=\gamma_{\rr}^\mathrm{t}+|\vec{x}_{i(\rr)}-\vec{y}_{\,\overline{i(\rr)}}|+|\vec{x}_{f(\rr)}-\vec{y}_{\,\overline{f(\rr)}}|.
\end{equation}
It also follows that the length of the truncated route, $\gamma_\rr^\tt$, does not depend on the position of the external vertices $\vec{x}_a$, $a\in \gVE$. 
Using this new notation we get, for scalar products in route space,
\begin{equation}
\bb E \cdot \bb \gamma = 
    \sum_{\rr\in\gGE}E_\rr\gamma_\rr=\sum_{\rr\in\gGE}E_\rr\left(|\vec{x}_{i(\rr)}-\vec{y}_{\,\overline{i(\rr)}}|+|\vec{x}_{f(\rr)}-\vec{y}_{\,\overline{f(\rr)}}|\right)+\sum_{\rr\in\gGE}E_\rr\gamma_\rr^\mathrm{t}.
\end{equation}
We can require that $\bb E \in \mathcal{F}_{G,\bb \sigma}^{\{p^0_a\}}$ 
and use the constraints in eqs.~\eqref{eq:fexpression_pre_poly} to get
\begin{equation}
    \sum_{\rr\in\gGE}E_\rr|\vec{x}_{i(\rr)}-\vec{y}_{\,\overline{i(\rr)}}|=\sum_{i\in \gVEI} |\vec{x}_{i}-\vec{y}_{\,\overline{i}}|\sum_{\rr \ni i} E_{\rr} = \sum_{i\in \gVEI} p_i^0 \,|\vec{x}_{i}-\vec{y}_{\,\overline{i}}|,
\end{equation}
and similarly,
$  \sum_{\rr\in\gGE}E_\rr|\vec{x}_{f(\rr)}-\vec{y}_{\,\overline{f(\rr)}}|= -\sum_{f\in \gVEI} p_f^0 \,|\vec{x}_{f}-\vec{y}_{\,\overline{f}}|. $ Hence,
\begin{gather} \bb E \cdot \bb \gamma = \sum_{i\in \gVEI} p_i^0 \,|\vec{x}_{i}-\vec{y}_{\,\overline{i}}| -\sum_{f\in \gVEI} p_f^0 \,|\vec{x}_{f}-\vec{y}_{\,\overline{f}}| + \bb E \cdot \bb \gamma^\tt\,. \end{gather}
Applying these observations to eq.~\eqref{eq:fexpression_post_poly} 
and using the definition of the Fourier transform of the flow polytope in eq.~\eqref{eq:FF_def} gives
\begin{gather} \begin{gathered} \label{eq:f_truncated} f_{G,\bb \sigma} =  \frac{(-i)^{|\gGE|}}{ (2\pi)^{|\gVE|}} \delta\left( \sum_{a \in \gVE} p^0_a \right) \left[ \prod_{i \in \gVEI} e^{ip_i^0 |\vec{x}_{i}-\vec{y}_{\,\overline{i}}| } \right] \left[ \prod_{f \in \gVEO} e^{ -i p_f^0 |\vec{x}_{f}-\vec{y}_{\,\overline{f}}| } \right] \widehat{\mathcal{F}}_{G,\bb \sigma}^{\{p^0_a\}}(\bb \gamma^\tt). \end{gathered} \end{gather}
We can now use the definition of the retarded and advanced propagators in eq.~\eqref{eq:ret_adv} to relate the oscillating exponentials that we manifestly extracted from the Fourier transform of the polytope to advanced and retarded propagators. In particular, we have
\begin{align} \label{eq:advanced_retarded_poly} \frac{1}{2\pi i} \int \frac{\mathrm{d}^3 \vec{x}_a}{2|\vec{x}_{a}-\vec{y}_{\,\overline{a}}|} e^{\pm i{p_a^0} |\vec{x}_{a}-\vec{y}_{\,\overline{a}}| -i\vec{x}_a \cdot \vec{p}_a } = e^{ -i\vec{y}_{\,\overline{a}} \cdot \vec{p}_a } \widetilde{\Delta}_{\genfrac{}{}{0pt}{3}{R}{A}}(p). \end{align}
Substituting eq.~\eqref{eq:f_truncated} into eq.~\eqref{eq:AG_f}, and using eq.~\eqref{eq:advanced_retarded_poly}, results in
\begin{gather}  \label{eq:truncated_AG} \widetilde A_{G,\bb \sigma}(\{p_a\}) = \frac{ (2\pi g)^{|\gVI|} i^{|\gVE|} }{(-4\pi^2)^{|\gE|}i^{|\gGE|}} \Bigg[\prod_{i\in \gVEI}\widetilde{\Delta}_R(p_i)\Bigg]\Bigg[\prod_{f\in \gVEO}\widetilde{\Delta}_A(p_f)\Bigg] \delta\left( \sum_{a \in \gVE} p^0_a \right) \nonumber \\ \times \int \frac{ \left[ \prod_{v\in V^{\text{int}}} \mathrm{d}^3 \vec{y}_v \right] \left[ \prod_{a\in \gVE} e^{-i\vec{y}_{\,\overline{a}} \cdot \vec{p}_a} \right] }{\left[\prod_{e\in \gEI} 2|\vec{z}_e|\right] \left[ \prod_{\cc\in \gGI} \gamma_\cc \right] } \widehat{\mathcal{F}}_{G,\bb \sigma}^{\{p^0_a\}}(\bb \gamma^\tt + i \ep \bb 1) ,  \end{gather}
where $\gEI$ is the set of internal edges of $(G,\bb \sigma)$. 

Similar factorization of the Fourier transform of the flow polytope arises when the underlying graph 
has a separating edge, i.e.~an edge whose removal disconnects the graph. It can be checked that in these cases, the Fourier transform of the whole graph factors into the Fourier transforms of the flow polytopes associated to the 1PI components of the graph and a couple of trivial phase factors. For instance, for a tree-level graph, the conditions in eq.~\eqref{eq:flow_poly_def} have at most one solution. This solution is uniquely determined by enforcing energy conservation at all vertices of the tree. Hence, the associated Fourier transform then consists exclusively of phase factors.

\subsection{The \texorpdfstring{$p$-$x$}{p-x} S-matrix representation}
\label{sec:px_derivation}

The $p$-$x$ representation of the S-matrix follows by 
comparing eq.~\eqref{eq:truncated_AG} with the FOPT expansion of the correlation function in eq.~\eqref{eq:tildeGamma2} and the definition of the S-matrix in eq.~\eqref{eq:truncation_s_matrix}:
\begin{equation}
    S(\{p_i\}_{i\in \gVEI}, \{p_f\}_{f\in \gVEO})=
\sum_{(G,\bb \sigma)} \frac{1}{\Sym (G,\bb \sigma)} 
    S_{G,\bb \sigma}(\{p_i\}_{i\in \gVEI}, \{p_f\}_{f\in \gVEO}),
\end{equation}
where we sum over all FOPT graphs. Following the partition of in- and out-going external vertices as indicated by the in- and out-going energies $\{p_a^0\}$, the S-matrix element $S_{G,\bb \sigma}(\{p_i\}_{i\in \gVEI}, \{p_f\}_{f\in \gVEO})$ is given by
\begin{align} S_{G,\bb \sigma} = &\frac{Z^{|V^{\text{ext}}|/2} (2\pi g)^{|\gVI|} i^{|\gVE|} }{(-4\pi^2)^{|\gE|}i^{|\gGE|}} \delta\left( \sum_{a \in \gVE} p^0_a \right) \nonumber \\
&\times \int \frac{ \left[ \prod_{v\in V^{\text{int}}} \mathrm{d}^3 \vec{y}_v \right] \left[ \prod_{a\in \gVE} e^{-i\vec{y}_{\,\overline{a}} \cdot \vec{p}_a} \right] }{\left[\prod_{e\in \gEI} 2|\vec{z}_e|\right] \left[ \prod_{\cc\in \gGI} \gamma_\cc \right] } \widehat{\mathcal{F}}_{G,\bb \sigma}^{\{p^0_a\}}(\bb \gamma^\tt + i \ep \bb 1). \end{align}
Except for the oscillating exponential term, the integral above is invariant under joint translations of all internal coordinates. We may pick some root vertex $w \in \gVI$ of the graph and shift all other $\vec y$-vectors as $\vec{y}_v \rightarrow \vec{y}_w + \vec{y}_v$. This way, we remove all nontrivial dependence of the $\vec{y}_w$ vector from the integrand. The integration over $\vec{y}_w$ gives a momentum-conservation-ensuring $\delta$ function. In equations,
\begin{align} \label{eq:x-p_representation} S_{G,\bb \sigma} = \frac{Z^{|V^{\text{ext}}|/2} (2\pi)^3 (2\pi g)^{|\gVI|} i^{|\gVE|} }{(-4\pi^2)^{|\gE|}i^{|\gGE|}} \delta^{(4)}\left( \sum_{a \in \gVE} p_a \right) s_{G,\bb \sigma}(\{p_i\}_{i\in \gVEI}, \{p_f\}_{f\in \gVEO}) , \end{align}
where $ s_{G,\bb \sigma}=(\{p_i\}_{i\in \gVEI}, \{p_f\}_{f\in \gVEO}) $ is the \emph{reduced} S-matrix element without trivial prefactors
\begin{gather} \begin{gathered} \label{eq:sG_red} s_{G,\bb \sigma} = \int \left[ \frac{ \prod_{v\in V^{\text{int}}\setminus\{w\}} \mathrm{d}^3 \vec{y}_v }{ \prod_{e\in \gEI} 2|\vec{z}_e| } \right] \left[ \frac{ \prod_{a\in \gVE} e^{-i\vec{y}_{\,\overline{a}} \cdot \vec{p}_a} }{ \left[ \prod_{\cc\in \gGI} \gamma_\cc \right] } \widehat{\mathcal{F}}_{G,\bb \sigma}^{\{p^0_a\}}(\bb \gamma^\tt + i \ep \bb 1) \right] \Bigg|_{\vec y_w = 0}. \end{gathered} \end{gather}
Note that this formula is, up to the contribution of the cycles, a 3-dimensional coordinate space Feynman integral 
that is modulated by a Fourier transformation of the flow polytope.
We stress that this is a \emph{hybrid} representation for the S-matrix contributions: the external kinematics are given in momentum space, but internal integrations are performed in coordinate space. We will call eq.~\eqref{eq:x-p_representation} the \emph{$p$-$x$ representation of the S-matrix} in order to emphasise this fact. This hybrid aspect is crucial for our investigation of collinear and soft singularities. 

\subsection{Example: The \texorpdfstring{$p$-$x$}{p-x} representation of a triangle diagram}

To illustrate the $p$-$x$ representation of the S-matrix we discuss here the contribution to the S-matrix from our recurring example, the oriented triangle diagram
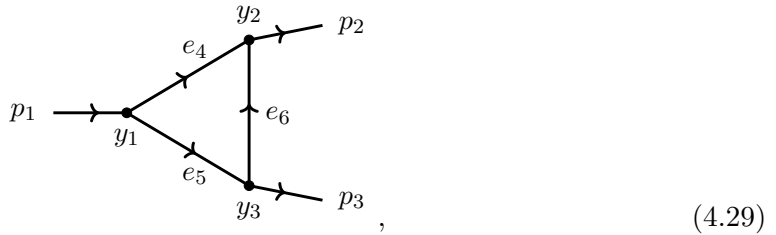
\begin{align} \label{eq:Striangle} \resizebox{5cm}{!}{ \begin{tikzpicture} \begin{feynman} \vertex(1); \vertex[above right = 1.5cm and 2.5cm of 1](2); \vertex[below right = 1.5cm and 2.5cm of 1](3); \vertex[left = 1.5cm of 1](E1); \vertex[above right = 0.3cm and 1.5cm of 2](E2); \vertex[below right = 0.3cm and 1.5cm of 3](E3); \vertex[left = 0.2cm of E1](L1) {\scalebox{1.5}{$p_1$}}; \vertex[right = 0.2cm of E2](L2) {\scalebox{1.5}{$p_2$}}; \vertex[right = 0.2cm of E3](L3) {\scalebox{1.5}{$p_3$}}; \vertex[below = 0.2cm of 1](L1_I) {\scalebox{1.5}{$y_1$}}; \vertex[above = 0.2cm of 2](L2_I) {\scalebox{1.5}{$y_2$}}; \vertex[below = 0.2cm of 3](L3_i) {\scalebox{1.5}{$y_3$}}; \vertex[above right = 1cm and 1cm of 1](edge4) {\scalebox{1.5}{$e_4$}}; \vertex[below right = 1cm and 1cm of 1](edge5) {\scalebox{1.5}{$e_5$}}; \vertex[below right = 1.25cm and 0.2cm of 2](edge6) {\scalebox{1.5}{$e_6$}}; \diagram*[large]{ (1) -- [->-,line width=0.6mm] (2) -- [-<-,line width=0.6mm] (3) -- [-<-,line width=0.6mm] (1), (1) -- [-<-,line width=0.6mm] (E1), (2) -- [->-,line width=0.6mm] (E2), (3) -- [->-,line width=0.6mm] (E3), }; \path[draw=black, fill=black] (1) circle[radius=0.1]; \path[draw=black, fill=black] (2) circle[radius=0.1]; \path[draw=black, fill=black] (3) circle[radius=0.1]; \end{feynman} \end{tikzpicture} }, \end{align}
where we now label the external vertex $x_i$ with its Fourier conjugate momentum $p_i$. In accordance with the all-in-going momentum convention we require that $p_1^0 > 0$ and $p_2^0, p_3^0<0$. The routes of this digraph have been illustrated in eq.~\eqref{eq:triangle_paths}. It has the three truncated routes $\rr_1^\tt = \{e_4\}$, $\rr_2^\tt=\{e_5\}$ and $\rr_3^\tt=\{e_5,e_6\}$. Hence, $\gamma^\tt_1 = |\vec z_4|, \gamma^\tt_2 = |\vec z_5|, \gamma^\tt_3 = |\vec z_5| + |\vec z_6|$. Let $E_1, E_2$ and $E_3$ be the energies that flow through the respective route.

The flow polytope $\mathcal{F}_{G,\bb \sigma}^{\{p^0_a\}}$ for this digraph is cut out by the conditions,
\begin{align} \label{eq:path_energy_cons} \begin{split} E_1, E_2, E_3 &\geq 0\,,\\
E_1 + E_2 + E_3 &= p_1^0\,,\\
E_1 + E_3 &= -p_2^0\,,\\
E_2 &= -p_3^0\,, \end{split} \end{align}
where one of the last three equations is redundant by overall momentum conservation, which is assumed. We can give an interpretation to the energy-conservation condition of eq.~\eqref{eq:path_energy_cons} as follows: for each external vertex $v\in V^{\text{ext}}$, enumerate the paths that start or end at that vertex, and correspondingly sum their energies. Then, set the sum of such energies to be $p_v^0$ if the vertex is the starting vertex for such paths or $-p_v^0$ if it is an ending vertex. For the triangle example, we can represent such constraints graphically as follows:
\begin{align} \resizebox{5cm}{!}{ \begin{tikzpicture} \begin{feynman} \vertex(1); \vertex[above right = 1.5cm and 2.5cm of 1](2); \vertex[below right = 1.5cm and 2.5cm of 1](3); \vertex[left = 1.5cm of 1](E1); \vertex[above right = 0.3cm and 1.5cm of 2](E2); \vertex[below right = 0.3cm and 1.5cm of 3](E3); \vertex[above=0.2cm of 1] (r11); \vertex[above= 0.2cm of 2](r12); \vertex[above = 0.2cm of E1](r1E1); \vertex[above = 0.2cm of E2](r1E2); \vertex[below=0.2cm of 1] (r21); \vertex[below= 0.2cm of 3](r22); \vertex[below = 0.2cm of E1](r2E1); \vertex[below = 0.2cm of E3](r2E2); \vertex[below = 0.75cm of E1](label1) {\LARGE $E_1+E_2+E_3=p_1^0$}; \vertex[above = 0.75cm of E2](label2) {\LARGE$E_1+E_3=-p_2^0$}; \vertex[below = 0.75cm of E3](label2) {\LARGE$E_2=-p_3^0$}; \vertex[above right = 1.2cm and 1cm of 1](edge4) {\scalebox{1.5}{$\color{col1}E_1$}}; \vertex[below right = 1.2cm and 1cm of 1](edge5) {\scalebox{1.5}{$\color{col2}E_2$}}; \vertex[below right = 1.25cm and 0.2cm of 2](edge6) {\scalebox{1.5}{$\color{col4}E_3$}}; \diagram*[large]{ (1) -- [->-,line width=0.6mm] (2) -- [-<-,line width=0.6mm, col4] (3) -- [-<-,line width=0.6mm, col4] (1), (1) -- [-<-,line width=0.6mm, col4] (E1), (2) -- [->-,line width=0.6mm, col4] (E2), (3) -- [->-,line width=0.6mm] (E3), (r1E1) -- [->-,line width=0.6mm, col1] (r11) -- [->-,line width=0.6mm, col1] (r12) -- [->-,line width=0.6mm, col1] (r1E2), (r2E1) -- [->-,line width=0.6mm, col2] (r21) -- [->-,line width=0.6mm, col2] (r22) -- [->-,line width=0.6mm, col2] (r2E2), }; \path[draw=black, fill=black] (1) circle[radius=0.1]; \path[draw=black, fill=black] (2) circle[radius=0.1]; \path[draw=black, fill=black] (3) circle[radius=0.1]; \path[draw=black] (E1) circle[radius=0.7]; \path[draw=black] (E2) circle[radius=0.7]; \path[draw=black] (E3) circle[radius=0.7]; \end{feynman} \end{tikzpicture} } \end{align}
We can parameterize the polytope by 
setting $(E_1,E_2,E_3) = (E,-p_3^0,-p_2^0-E)$ and let $E$ vary between $0$ and $-p_2^0$. The polytope $\mathcal{F}_{G,\bb \sigma}^{\{p^0_a\}}$ is therefore a 1-dimensional polytope, i.e.\ a line segment. Its volume is obviously $\textrm{Vol}(\mathcal{F}_{G,\bb \sigma}^{\{p^0_a\}}) = -p_2^0$ (recall that $p_2^0 <0$). 
Using this parameterization, we can explicitly evaluate the Fourier transformation of the flow polytope associated to the digraph above,
\begin{gather} \widehat{ \mathcal{F}}_{G,\bb \sigma}^{\{p^0_a\}} (\bb \gamma^\tt + i \ep \bb 1) = \int_{\mathcal{F}_{G,\bb \sigma}^{\{p^0_a\}}} \dd \bb E\, e^{i \bb E \cdot (\bb \gamma^\tt+ i \ep \bb 1 ) } = \int_{0}^{-p_2^0} \dd E \,e^{i E ( \gamma_1^\tt + i \ep) - ip_3^0 ( \gamma_2^\tt + i \ep) - i(p_2^0+E) (\gamma_3^\tt + i \ep) } \notag \\
= e^{- i(p_3^0 + p_2^0) (\gamma_2^\tt + i \ep)} \int_{0}^{-p_2^0} \dd E\,e^{i E (\gamma_1^\tt - \gamma_3^\tt)} = -i e^{- i(p_3^0 + p_2^0) (\gamma_2^\tt + i \ep)} \frac{ e^{-i p_2^0 (\gamma_1^\tt - \gamma_3^\tt) } - 1 }{\gamma_1^\tt - \gamma_3^\tt} \notag \\
= -p_2^0 e^{- ip_3^0 ( \gamma_2^\tt + i \ep) - ip_2^0 (\gamma_2^\tt + \frac12 \gamma_1^\tt - \frac12 \gamma_3^\tt + i \ep) } \sinc\left( \frac{p_2^0 (\gamma_1^\tt - \gamma_3^\tt)}{2} \right), \label{eq:triangleFourierFlow} \end{gather}
where $\sinc(x) = \frac{\sin(x)}{x}$. This expression is manifestly bounded as $\sinc(x) \leq 1$. The prefactor $-p_2^0$ is the volume of the flow polytope as expected.
The reduced S-matrix contribution of the digraph above is,
\begin{align} \label{eq:triangle_px_final} s_{G,\bb \sigma}(\{p_1\}, \{p_2,p_3\}) = \int \frac{ \left[ \prod_{v\in \{2,3\}} \mathrm{d}^3 \vec{y}_v \right] \left[ e^{-i\vec{y}_{2} \cdot \vec{p}_2 -i\vec{y}_{3} \cdot \vec{p}_3} \right] }{8|\vec{z}_4| |\vec{z}_5| |\vec{z}_6| } \widehat{\mathcal{F}}_{G,\bb \sigma}^{\{p^0_a\}}(\bb \gamma^\tt + i \ep \bb 1) \Big|_{\vec y_1 = 0}, \end{align}
where we used the freedom, guaranteed by translation invariance, to fix one vertex position at the origin, in this case $\vec{y}_1 = 0$.

\subsection{Example: The flow polytope of a pentagon digraph}

The previous example already illustrates that the only nontrivial part of the 
$p$-$x$ S-matrix representation is the Fourier transform of the flow polytope. The Fourier transform of a polytope is a \emph{perfect invariant}, i.e.~it characterizes the polytope completely. Geometric understanding of the shape of this polytope is equivalent to knowledge of the associated Fourier transform. 

It is a straightforward, algorithmic task to compute the shape of this polytope -- even for rather intricate FOPT diagrams. To illustrate this point we compute the flow polytope of an oriented pentagon digraph:
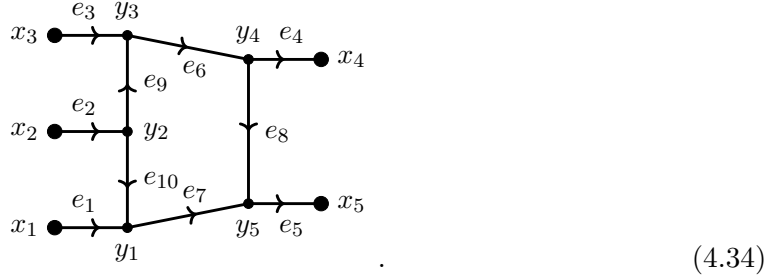
\begin{align} \resizebox{5cm}{!}{ \begin{tikzpicture} \begin{feynman} \vertex(1); \vertex[above = 2cm of 1](1up) ; \vertex[below = 2cm of 1](1down) ; \vertex[above right = 1.5cm and 2.5cm of 1](2); \vertex[below right = 1.5cm and 2.5cm of 1](3); \vertex[left = 1.5cm of 1](E1); \vertex[left = 1.5cm of 1up](E1up); \vertex[left = 1.5cm of 1down](E1down); \vertex[right = 1.5cm of 2](E2); \vertex[right = 1.5cm of 3](E3); \vertex[left = 0.2cm of E1](L1) {\scalebox{1.5}{$x_2$}}; \vertex[left = 0.2cm of E1up](L1up) {\scalebox{1.5}{$x_3$}}; \vertex[left = 0.2cm of E1down](L1down) {\scalebox{1.5}{$x_1$}}; \vertex[right = 0.2cm of E2](L2) {\scalebox{1.5}{$x_4$}}; \vertex[right = 0.2cm of E3](L3) {\scalebox{1.5}{$x_5$}}; \vertex[right = 0.2cm of 1](L1_I) {\scalebox{1.5}{$y_2$}}; \vertex[above= 0.2cm of 1up](L1_Iup) {\scalebox{1.5}{$y_3$}}; \vertex[below = 0.2cm of 1down](L1_Idown) {\scalebox{1.5}{$y_1$}}; \vertex[above = 0.2cm of 2](L2_I) {\scalebox{1.5}{$y_4$}}; \vertex[below = 0.2cm of 3](L3_i) {\scalebox{1.5}{$y_5$}}; \vertex[above left = 0.2cm and 0.5cm of 1](edge1) {\scalebox{1.5}{$e_2$}}; \vertex[above left = 0.2cm and 0.5cm of 1up](edge1up) {\scalebox{1.5}{$e_3$}}; \vertex[above left = 0.2cm and 0.5cm of 1down](edge1down) {\scalebox{1.5}{$e_1$}}; \vertex[above right = 0.2cm and 0.5cm of 2](edge2) {\scalebox{1.5}{$e_4$}}; \vertex[below right = 0.2cm and 0.5cm of 3](edge3) {\scalebox{1.5}{$e_5$}}; \vertex[above right = 1cm and 1cm of 1](edge4) {\scalebox{1.5}{$e_6$}}; \vertex[below right = 1cm and 1cm of 1](edge5) {\scalebox{1.5}{$e_7$}}; \vertex[below right = 1.25cm and 0.2cm of 2](edge6) {\scalebox{1.5}{$e_8$}}; \vertex[above right = 0.7cm and 0.2cm of 1](edge7) {\scalebox{1.5}{$e_9$}}; \vertex[below right = 0.7cm and 0.2cm of 1](edge8) {\scalebox{1.5}{$e_{10}$}}; \diagram*[large]{ (1) -- [->-,line width=0.6mm] (1up) -- [->-,line width=0.6mm] (2) -- [->-,line width=0.6mm] (3) -- [-<-,line width=0.6mm] (1down) -- [-<-,line width=0.6mm] (1), (1) -- [-<-,line width=0.6mm] (E1), (1up) -- [-<-,line width=0.6mm] (E1up), (1down) -- [-<-,line width=0.6mm] (E1down), (2) -- [->-,line width=0.6mm] (E2), (3) -- [->-,line width=0.6mm] (E3), }; \path[draw=black, fill=black] (E1) circle[radius=0.15]; \path[draw=black, fill=black] (E1up) circle[radius=0.15]; \path[draw=black, fill=black] (E1down) circle[radius=0.15]; \path[draw=black, fill=black] (E2) circle[radius=0.15]; \path[draw=black, fill=black] (E3) circle[radius=0.15]; \path[draw=black, fill=black] (1) circle[radius=0.1]; \path[draw=black, fill=black] (1up) circle[radius=0.1]; \path[draw=black, fill=black] (1down) circle[radius=0.1]; \path[draw=black, fill=black] (2) circle[radius=0.1]; \path[draw=black, fill=black] (3) circle[radius=0.1]; \end{feynman} \end{tikzpicture} }. \end{align}
The Fourier transform is parametric in the five external energies $p_1^0,...,p_5^0$. We choose the energy signs such that they coincide with the FOPT orientation shown above, that is
\begin{equation}
    p_1^0,\, p_2^0,\, p_3^0>0, \quad p_4^0,\, p_5^0<0.
\end{equation}
The six routes in $\gGE$ through the diagram are 
\begin{align} &\rr_1=\{e_4,e_6,e_3\}, \quad\nonumber &&\rr_2=\{e_4,e_6,e_9, e_2\}, \\
 &\rr_3=\{e_5,e_8,e_6,e_3\}, \quad &&\rr_4=\{e_5,e_8,e_6,e_9,e_2\}, \\ &\rr_5=\{e_5,e_7,e_1\}, \quad &&\rr_6=\{e_5,e_7,e_{10}, e_2\},\nonumber \end{align}
and the Fourier transform of the flow polytope reads
\begin{equation}
    \widehat{\mathcal{F}}_{G,\bb \sigma}^{\{p^0_a\}}(\bb \gamma)=\int_{\mathcal{F}_{G,\bb \sigma}^{\{p^0_a\}}} \mathrm{d}\boldsymbol{E} \, e^{i \boldsymbol{\gamma}\cdot\boldsymbol{E}}\,,
\end{equation}
with $\mathcal{F}_{G,\bb \sigma}^{\{p^0_a\}}$ being the polytope defined by the following constraints
\begin{align} & E_i>0, \quad i=1,...,6\nonumber \\
 & E_1+E_3=p_3^0, \quad E_5=p_1^0, \quad E_2+E_4+E_6=p_2^0\,, \\
 & E_1+E_2=-p_4^0, \quad E_3+E_4+E_5+E_6=-p_5^0\,.\nonumber \end{align}
Due to overall momentum conservation, 
one of the five equalities is linearly dependent on the others. The route space $\R^{|\gGE|}$ is obviously six dimensional. Hence, the flow polytope is a two dimensional object embedded in six-dimensional space. To be able to draw it, we will project it on the  plane spanned by the variables $E_1$ and $E_4$. In this $E_1$-$E_4$-plane, the linear inequalities defining the positive orthant read
\begin{equation}
    p_3^0-E_1>0, \quad -p_4^0-E_1>0, \quad p_2^0+p_4^0-E_4+E_1>0, \quad E_1>0, \quad E_4>0\,.
\end{equation}
The domain identified by these linear inequalities is a bounded polytope in the $E_1$-$E_4$-plane, as expected, and is drawn as follows

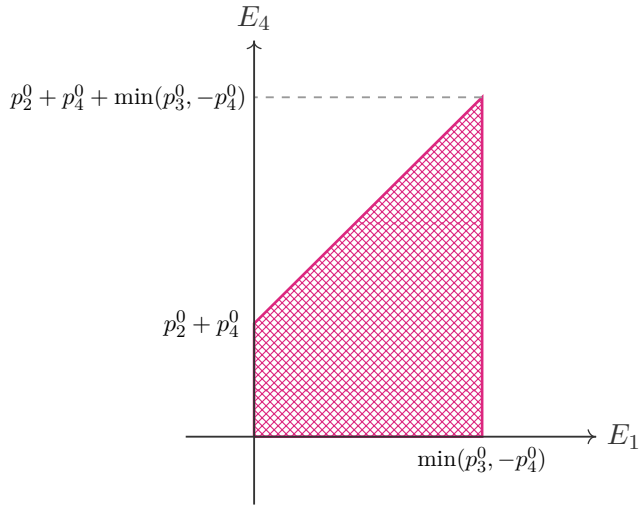
\begin{center}
\begin{tikzpicture}[thick,scale=3] \coordinate (A1) at (0,0); \coordinate (A2) at (1,0); \coordinate (A3) at (0,0.5); \coordinate (A4) at (1,1.5); \coordinate (A5) at (0,1.5); \coordinate (A6) at (1,1); \node[] at (1,-0.1) {\resizebox{1.7cm}{!}{$\text{min}(p_3^0,-p_4^0)$}}; \node[] at (-0.23,0.5) {\resizebox{1cm}{!}{$p_2^0+p_4^0$}}; \node[] at (-0.55,1.5) {\resizebox{3,1cm}{!}{$p_2^0+p_4^0+\text{min}(p_3^0,-p_4^0)$}}; \draw[draw=col4, fill=white, pattern=crosshatch, pattern color=col4][line width=1pt] (A1) -- (A2) -- (A4) -- (A3) -- (A1); \draw[dashed, black!40!white][line width=0.7pt] (A4) -- (A5); \draw[line width=0.7pt, black!80!white] (-0.3,0)--(1.5,0) node[right]{$E_1$}; \draw[line width=0.7pt, black!80!white] (0,-0.3)--(0,1.75) node[above]{$E_4$}; \draw[->,line width=0.5pt, black!80!white] (0,1.74)--(0,1.75); \draw[->,line width=0.5pt, black!80!white] (1.49,0)--(1.5,0); \end{tikzpicture}
\end{center}
The geometric properties of the polytope, such as the number of distinct edges and vertices, depend on the values of the external energies. If $p_2^0+p_4^0> 0$, the polytope is a quadrilateral, a triangle if $-\text{min}(p_3^0,-p_4^0)<p_2^0+p_4^0\le 0$, a point if $p_2^0+p_4^0+\text{min}(p_3^0,-p_4^0)=0$, and the empty set otherwise.

\subsection{Polytopes and spurious singularities}
\label{sec:spurious}

The $p$-$x$ S-matrix representation involves the Fourier transform of the flow polytope, an object that is both elegant and allows to use the existing mathematical literature on the subject to study its properties. One particular advantage is the absence of the spurious singularities that typically arise from contour integration of an integrand with multiple poles. A simple example of the appearance of spurious singularities is the following integral which depends on the parameters $a_1,\ldots,a_n \in \mathbb{C}$,
\begin{equation}
\label{eq:contour_integral}
    \oint \mathrm{d}z \frac{f(z)}{\prod_{i=1}^n (z-a_i)}=2\pi i\sum_{i=1}^n \frac{f(a_i)}{\prod_{j\neq i}(a_i-a_j)},
\end{equation}
where the contour of integration shall contain all poles. The integral is defined whenever $a_i\neq a_j$ for all $i\neq j$. One might expect the right-hand side of eq.~\eqref{eq:contour_integral} to be singular if two or more $a$-parameters coincide. However, if we make a subset of these parameters approach the same limit, i.e.\ $a_i\rightarrow a\in\mathbb{C}$, for all $i\in I\subseteq\{1,...,n\}$, then the right-hand side of eq.~\eqref{eq:contour_integral} remains finite as long as $f$ is $|I|-1$ times differentiable at $a$. For instance, in the case $n=2$, we have
\begin{equation}
    \frac{f(a_1)}{a_1-a_2}+\frac{f(a_2)}{a_2-a_1}\rightarrow f'(a),\quad \text{when }a_1,a_2\rightarrow a,
\end{equation}
showing that the singularity for $a_1,a_2\rightarrow a$ is spurious.

This simple model exemplifies the common occurrence of spurious cancellations in the study of IR singularities of amplitudes. Such cancellations are present in any method that is based on operating cuts or performing contour integration of Feynman integrals, including \textit{dual cancellations} in the LTD formalism,  and KLN cancellations~\cite{Kinoshita:1962ur,Lee:1964is} (see~\cite{Capatti_2020} for a study of KLN cancellations in terms of divided differences and~\cite{Hannesdottir:2022bmo,Frye:2018xjj} for an original approach on the KLN theorem) of IR singularities at the cross-section level. It is also not surprising that one of the motivating arguments in the amplituhedron approach~\cite{Arkani-Hamed:2012zlh,Arkani-Hamed:2013jha,Arkani-Hamed:2014dca,Hodges:2009hk,Hodges:2011wm} is its manifest realization of the cancellation of spurious singularities that are generated by the application of the BCFW recursion relation~\cite{Britto:2004ap,Britto:2005fq}. Spurious poles are also present in the $p$-$x$ representation of the S-matrix, as exemplified by the $\sinc$-function in eq.~\eqref{eq:triangleFourierFlow}, in the limit $\gamma_{\mathrm{r}_1} \rightarrow \gamma_{\mathrm{r}_3}$. In the general formulation, we have carefully avoided introducing spurious poles by directly casting our expressions in terms of the Fourier transform of the flow polytope. Had we been less careful in the derivation in sect.~\ref{sec:px_derivation}, where we introduced one helper variable $z_\rr$ for each route and, instead, introduced one variable $z_{ij}$ for each difference $x_i^0-x_f^0$, we would have obtained an expression with the following oscillating part for the reduced S-matrix element,
\begin{equation}
    s_{G,\boldsymbol{\sigma}}\propto\int_{\Delta'} \prod_{\substack{(i,f)\in \gVEI\times \gVEO \\ |\Gamma_{if}^{\text{ext}}|\neq 0}}\mathrm{d}E_{if} \sum_{\mathrm{r}\in \Gamma_{if}^{\text{ext}}} \frac{e^{i\gamma_\mathrm{r}E_{if}}}{\prod_{\substack{\mathrm{r}'\in \Gamma_{if}^{\text{ext}} \\ \mathrm{r}'\neq \mathrm{r}}} (\gamma_{\mathrm{r}}-\gamma_{\mathrm{r}'})},
\end{equation}
where $\Delta'$ is a polytope similar to the flow polytope and $\Gamma_{if}$ is the set of routes connecting the external vertices $i$ and $f$. This alternative expression for the oscillating part features many spurious singularities and a complicated pattern of cancellations for them. For the example of eq.~\eqref{eq:contour_integral}, it is possible to make the cancellation of spurious singularities manifest by using the Hermite-Genocchi representation of divided differences
\begin{equation}
    \oint \mathrm{d}z \frac{f(z)}{\prod_{i=1}^n (z-a_i)}=2\pi i\int_{\Sigma_n} \mathrm{d}\mathbf{r} f^{(n-1)}(\mathbf{r}\cdot \mathbf{a}) \,,
\end{equation}
which is expressed in terms of an integral over the standard simplex. The right hand side of this expression is obviously bounded. The simplex accommodates the simplest cancellation pattern of spurious singularities. More complicated patterns can be constructed by looking at generic polytopes. As an example of such cancellation patterns, let us consider the closed form expression for the Fourier transform of a simple $d$-dimensional polytope $\Delta$~\cite{polytope_fourier}:
\begin{equation}
\label{eq:fourier_transform_polytope}
    \int_\Delta \mathrm{d}\mathbf{r} \, e^{i \boldsymbol{\xi}\cdot \mathbf{r}}=\frac{1}{(2\pi i)^d} \sum_{\substack{\mathbf{v}\text{ vertex } \\ \text{of } \Delta}} \frac{e^{i \mathbf{v}\cdot\boldsymbol{\xi}} \, |\text{det}M_v|}{\prod_{j=1}^d \boldsymbol{w}_j(\mathbf{v})\cdot\boldsymbol{\xi}}\,,
\end{equation}
where $\mathbf{w}_k(\mathbf{v})$ is the $k$-th edge vector emanating from the vertex $\mathbf{v}$ (since the polytope is simple, there are exactly $d$ edge vectors emanating from each vertex), not necessarily normalised, and $M_v$ is the matrix containing all such vectors as columns. Much like the right-hand side of eq.~\eqref{eq:contour_integral}, the right-hand side of eq.~\eqref{eq:fourier_transform_polytope} exhibits spurious cancellation of singularities: if, for a summand corresponding to the vertex $\mathbf{v}$, the denominator $\bb w_j(\mathbf{v})\cdot\boldsymbol{\xi}$ vanishes, and if $\mathbf{v}'$ is the vertex that the edge vector $\bb w_j(\mathbf{v})$ points to, then the summand corresponding to the vertex $\mathbf{v}'$ also contains the same singular denominator and the sum of the two terms is finite. Such complicated considerations are, however, completely absent when one looks at the left-hand side of eq.~\eqref{eq:fourier_transform_polytope}, which is manifestly bounded. In this way,  mathematical identities that relate quantities with spurious poles to integrals over polytopes prove to be a powerful theoretical tool.

\section{IR singularities in the \texorpdfstring{$p$-$x$}{p-x} representation}
\label{sec:IRpxsing}

Having found a representation for the S-matrix as an integral over the spatial coordinates, we can now investigate where the infrared singularities of the S-matrix are located in the space of integration variables. This will yield a precise coordinate space equivalent of infrared singularities. We will start in sect.~\ref{sec:coll_sing_triangle} by deriving the collinear singularities for the triangle diagram, motivated by the intuition that collinear singularities in momentum space should correspond to long distance singularities in coordinate space that also embed a notion of collinearity. We then derive a diagram-level factorization formula for the leading behaviour in the singular limit. This is then generalised in sect.~\ref{sec:coll_sing_general} to arbitrary graphs. In sect.~\ref{sec:soft_singularities_triangle} we discuss the soft singularity of the triangle diagram. 

\subsection{Collinear singularity on the triangle digraph}
\label{sec:coll_sing_triangle}

We begin with 
a description of collinear singularities of the triangle diagram in the hybrid formalism. Following the intuition provided by momentum space, we expect the collinear singularity to appear when the length of the two edges adjacent to an external vertex becomes infinite. Furthermore, we expect that such distance vectors become collinear to $p_2$ in this limit. Indeed, we find a collinear singularity if the coordinate $\vec y_2$ goes to infinity \emph{in the direction} of $\vec{p}_2$:
\begin{align} \label{eq:infinitetriangle} \resizebox{4cm}{!}{ \begin{tikzpicture}[baseline={([yshift=-0.7ex]0,0)}] \begin{feynman} \vertex(1); \vertex[above right = 1.5cm and 2.5cm of 1](2); \vertex[below right = 1.5cm and 2.5cm of 1](3); \vertex[left = 1.5cm of 1](E1); \vertex[above right = 0.3cm and 1.5cm of 2](E2); \vertex[below right = 0.3cm and 1.5cm of 3](E3); \vertex[below = 0.2cm of 1](L1_I) {\scalebox{1.5}{$y_1$}}; \vertex[above = 0.2cm of 2](L2_I) {\scalebox{1.5}{$y_2$}}; \vertex[below = 0.2cm of 3](L3_i) {\scalebox{1.5}{$y_3$}}; \vertex[above left = 0.2cm and 0.5cm of 1](edge1) {\scalebox{1.5}{$p_1$}}; \vertex[above right = 0.2cm and 0.5cm of 2](edge2) {\scalebox{1.5}{$p_2$}}; \vertex[below right = 0.2cm and 0.5cm of 3](edge3) {\scalebox{1.5}{$p_3$}}; \diagram*[large]{ (1) -- [->-,line width=0.6mm] (2) -- [-<-,line width=0.6mm] (3) -- [-<-,line width=0.6mm] (1), (1) -- [-<-,line width=0.6mm, dashed] (E1), (2) -- [->-,line width=0.6mm,dashed] (E2), (3) -- [->-,line width=0.6mm,dashed] (E3), }; \path[draw=black, fill=black] (1) circle[radius=0.1]; \path[draw=black, fill=black] (2) circle[radius=0.1]; \path[draw=black, fill=black] (3) circle[radius=0.1]; \end{feynman} \end{tikzpicture} } \quad \rightarrow \quad \resizebox{7.5cm}{!}{ \begin{tikzpicture}[baseline={([yshift=-0.7ex]0,0)}] \begin{feynman} \vertex(1); \vertex[above right = 1.5cm and 8cm of 1](2); \vertex[below right = 1.5cm and 2.5cm of 1](3); \vertex[left = 1.5cm of 1](E1); \vertex[above right = 0.3cm and 1.5cm of 2](E2); \vertex[below right = 0.3cm and 1.5cm of 3](E3); \vertex[below = 0.2cm of 1](L1_I) {\scalebox{1.5}{$y_1$}}; \vertex[above = 0.2cm of 2](L2_I) {\scalebox{1.5}{$y_2$}}; \vertex[below = 0.2cm of 3](L3_i) {\scalebox{1.5}{$y_3$}}; \vertex[above left = 0.2cm and 0.5cm of 1](edge1) {\scalebox{1.5}{$p_1$}}; \vertex[above right = 0.2cm and 0.5cm of 2](edge2) {\scalebox{1.5}{$p_2$}}; \vertex[below right = 0.2cm and 0.5cm of 3](edge3) {\scalebox{1.5}{$p_3$}}; \diagram*[large]{ (1) -- [->-,line width=0.6mm] (2) -- [-<-,line width=0.6mm] (3) -- [-<-,line width=0.6mm] (1), (1) -- [-<-,line width=0.6mm, dashed] (E1), (2) -- [->-,line width=0.6mm,dashed] (E2), (3) -- [->-,line width=0.6mm,dashed] (E3), }; \path[draw=black, fill=black] (1) circle[radius=0.1]; \path[draw=black, fill=black] (2) circle[radius=0.1]; \path[draw=black, fill=black] (3) circle[radius=0.1]; \end{feynman} \end{tikzpicture} } \end{align}
Note that both edges of the graph that have infinite length in this limit have energy flowing into the vertex $y_2$. This is a necessary condition for the non-integrability of this singularity in the FOPT representation.
The total energy flowing through both edges must be equal to the energy flowing out of the $y_2$ vertex by 
energy conservation. This observation will play a role in the factorization of the integral in the soft and collinear limits. 

A useful parameterization for the collinear limit is 
\begin{align} \label{eq:param_triangle_coll} \vec{y}_2=\lambda \vec{p}_2/|\vec{p}_2|^2+\sqrt{|\lambda|}\vec{y}^\perp_2, \quad \vec{y}_3=\vec{y}_3\,, \end{align}
 where the variable $\lambda$ that measures the distance in the direction $\vec p_2$ and the perpendicular 
part of $\vec{y}_2$ is parameterized by $\vec{y}^\perp_2$.
We find from eq.~\eqref{eq:triangle_px_final}, 
\begin{align} \label{eq:trangle_split_lambda1} s_{G,\bb \sigma}(\{p_1\}, \{p_2,p_3\}) &= \int_{-\infty}^\infty \dd \lambda \, s_{G,\bb \sigma}(\lambda; \{p_1\}, \{p_2,p_3\})\,, \\
\label{eq:triangle_split_lambda} s_{G,\bb \sigma}(\lambda; \{p_1\}, \{p_2,p_3\}) &= \frac{| \lambda|}{|\vec p_2|} e^{-i\lambda} \int \frac{ \left[ \dd^2 \vec{y}^\perp_2 \mathrm{d}^3 \vec{y}_3 \right] \left[ e^{-i\vec{y}_{3} \cdot \vec{p}_3} \right] }{8 |\vec{z}_4| |\vec{z}_5| |\vec{z}_6| } \widehat{\mathcal{F}}_{G,\bb \sigma}^{\{p^0_a\}}(\bb \gamma^\tt + i \ep \bb 1) \Big|_{\vec y_1 = 0}\,, \end{align}
where we isolated the integration over $\lambda$. 
Let us study the behaviour of the expression under the $\lambda$ integration in the $\lambda \rightarrow \infty$ limit. We remark that the $\lambda \rightarrow -\infty$ limit could also be considered, but 
it does not give a singular contribution. The reason for this is that the oscillating exponent $-i\lambda$ in eq.~\eqref{eq:triangle_split_lambda} is not canceled and  retains its dampening role for the integration.
We have 
\begin{align} |\vec z_4| &= | \vec y_2 - \vec y_1 | = | \vec y_2 | = \sqrt{\lambda^2 /{|\vec{p}_2|^2}+\lambda(\vec{y}^\perp_2)^2 }  = \frac{\lambda}{|\vec{p}_2|} + \frac12 |\vec p_2| (\vec{y}^\perp_2)^2 + \mathcal O(\lambda^{-1})\,,\nonumber \\
 |\vec z_5| &= |\vec y_3|\,, \\
|\vec z_6| &=        \frac{\lambda}{|\vec{p}_2|} + \frac12 |\vec p_2| (\vec{y}^\perp_2)^2 - \frac{\vec{p}_2 \cdot \vec y_3}{|\vec{p}_2|} + \mathcal O(\lambda^{-\frac12})\,,\nonumber \end{align}
and hence
\begin{align} \nonumber \bb E \cdot \bb \gamma^\tt &= E_1 \gamma^\tt_1 + E_2 \gamma^\tt_2 + E_3 \gamma^\tt_3 = E_1 |\vec z_4| + E_2 |\vec z_5| + E_3 (|\vec z_5| + |\vec z_6|) \\
\nonumber &= (E_1+E_3) \frac{\lambda}{|\vec{p}_2|} + \frac12 (E_1+E_3) |\vec p_2| (\vec{y}^\perp_2)^2 + (E_2+E_3) |\vec y_3| - E_3 \frac{\vec{p}_2 \cdot \vec y_3}{|\vec{p}_2|} + \mathcal O(\lambda^{-\frac12}) \\
&= -p_2^0 \frac{\lambda}{|\vec{p}_2|} - \frac12 p_2^0 |\vec p_2| (\vec{y}^\perp_2)^2 + (-p_3^0 + E_3) |\vec y_3| - E_3 \frac{\vec{p}_2 \cdot \vec y_3}{|\vec{p}_2|} + \mathcal O(\lambda^{-\frac12})\,, \end{align}
where we applied the constraints from the flow polytope description in eq.~\eqref{eq:path_energy_cons} 
in the second line. Note that due to the total amount of energy flowing through the two infinite-length
lines being equal to $-p_2^0$, the leading term in $\lambda$ does not have any dependence on the coordinates 
in the flow polytope. This is a general feature which always occurs if a subset of vertices, which are only connected to the rest of the graph via two edges, are sent to 
spatial infinity in the way described above.

Substitution into the $\lambda$ dependent integrand 
$s_{G,\bb \sigma}(\lambda; \{p_1\}, \{p_2,p_3\})$
 gives
\begin{gather} \begin{gathered} \label{eq:triangle_lambda} s_{G,\bb \sigma}(\lambda; \{p_1\}, \{p_2,p_3\}) = \frac{|\vec p_2|}{4\lambda} e^{-i\lambda -ip_2^0 (\frac{\lambda}{|\vec{p}_2|} +i \ep) } \int \dd^2 \vec{y}^\perp_2 e^{ - \frac12 i p_2^0 ( |\vec p_2| (\vec{y}^\perp_2)^2 + i \ep ) } \\
\times \int \frac{ \mathrm{d}^3 \vec{y}_3 e^{-i p_3^0|\vec{y}_3|-i\vec{y}_{3} \cdot \vec{p}_3} }{2 |\vec{y}_3| } \int_{\mathcal{F}_{G,\bb \sigma}^{\{p^0_a\}}} \dd \bb E e^{ i E_3 \left( |\vec{y}_3| - \frac{\vec{p}_2 \cdot \vec y_3}{|\vec{p}_2|}\right) } +\mathcal O(\lambda^{-\frac12}). \end{gathered} \end{gather}
Let us examine this integrand in the limit $p_2^2 \rightarrow 0$. 
Recall that $p_2^0 < 0$. Hence, in this limit 
$-p_2^0 \rightarrow |\vec p_2|$.
The first term involving $\lambda$ leads to a 
logarithmically divergent integrand 
from the $\lambda \rightarrow \infty$ part of the integration.
The oscillatory part vanishes if we set $p_2^0 = - |\vec p_2|$ 
and we lose the dampening of the divergence due to the oscillation.
Explicitly, the divergent part in the on-shell limit is produced by the exponential integral,
\begin{align} \int_{1}^\infty \dd \lambda \frac{ e^{i \left(-\frac{p_2^0}{|\vec{p}_2|} -1\right) \lambda }} {\lambda}. \end{align}
Changing the lower boundary if this integral only results in a finite change of the integral in the singular limit 
$-p_2^0 \rightarrow |\vec p_2|$. By standard asymptotic expansions  \cite{abramowitz1964handbook} of the exponential integral we find
\begin{align}\label{Sudakovintegral1} \nonumber\int_{1}^\infty \dd \lambda \frac{ e^{i \left(-\frac{p_2^0}{|\vec{p}_2|} -1\right) \lambda }} {\lambda} & = -\log\left( -\frac{p_2^0}{|\vec{p}_2|} -1 \right) + (\text{terms finite when $p_2^2 \rightarrow 0$}) \\
&= -\log\left( \frac{p^2_2}{p_1^2} \right) + (\text{terms finite when $p_2^2 \rightarrow 0$})\,, \end{align}
where we normalized the argument of the logarithm using $p_1^2$.
The remaining part of the integral in eq.~\eqref{eq:trangle_split_lambda1} remains finite in the 
$p_2^2 \rightarrow 0$ limit.

In eq.~\eqref{eq:triangle_lambda} we find a Gaussian integral over the 
perpendicular contribution in the collinear limit. 
We may interpret this Gaussian as the 
interaction event at $\vec y_2$ being localized in a cylinder of 
radius $1/|p_2^0|$ around the collinear direction. 
The precise distribution is given by an imaginary Gaussian 
distribution with standard deviation given by this radius. The total contribution from the perpendicular part reads,
after using $p_2^0 = - |\vec p_2|$,
\begin{align}\label{perpendiculargaussian} \int \dd^2 \vec{y}^\perp_2 e^{ \frac12 i ((p_2^0)^2 (\vec{y}^\perp_2)^2 + i \ep ) } = \frac{ 2 \pi i} { (p_2^0)^2 }\,. \end{align}
 Using the evaluation of these two integrals in eqs.~(\ref{Sudakovintegral1}) and (\ref{perpendiculargaussian}), and dropping terms that vanish in the limit $p^2_2\rightarrow 0$, the reduced S-matrix in eq.~(\ref{eq:triangle_lambda}) takes the form
\begin{align} s_{G,\bb \sigma}(\{p_1\}, \{p_2,p_3\}) = -\frac{2\pi i}{4|\vec p_2|} \log\left(\frac{p_2^2}{p_1^2}\right) \int \frac{ \mathrm{d}^3 \vec{y}_3 e^{-i p_3^0|\vec{y}_3|-i\vec{y}_{3} \cdot \vec{p}_3} }{2 |\vec{y}_3| } \nonumber \\
\times \int_{\mathcal{F}_{G,\bb \sigma}^{\{p^0_a\}}} \dd \bb E e^{ i E_3 \left( |\vec{y}_3| - \frac{\vec{p}_2 \cdot \vec y_3}{|\vec{p}_2|}\right) } + (\text{terms finite when $p_2^2 \rightarrow 0$}). \end{align}
We can recast this expression in a more illuminating form. Since dependence on both $E_1$ and $E_2$ dropped out of the Fourier transform, we can integrate over them, changing variables to $x=E_3/p_2^0$ for the remaining integration. The variable $x$ parameterizes the proportion of energy that flows through the lower infinite line in \eqref{eq:infinitetriangle} in the collinear limit. As a result we obtain
\begin{align} \label{eq:S-matrix_with_x} s_{G,\bb \sigma}(\{p_1\}, \{p_2,p_3\}) = -\frac{2\pi i}{4} \log\left(\frac{p_2^2}{p_1^2}\right) \int_0^1 \mathrm{d}x \int \frac{ \mathrm{d}^3 \vec{y}_3 e^{-i p_3^0|\vec{y}_3|+ix|\vec{y}_{3}||\vec{p}_2|} }{2 |\vec{y}_3| } e^{-i\vec{y}_3\cdot\vec{p}_3-ix \vec{p}_2\cdot \vec{y}_3} \nonumber \\
+ (\text{terms finite when $p_2^2 \rightarrow 0$}). \end{align}
The integral over $\vec{y}_3$ can be identified with an integral over the 
reduced S-matrix element of the \emph{hard graph} $(G,\bb \sigma)_\text{hard}$. More precisely, consider the diagram 
\begin{align} \resizebox{5cm}{!}{ \begin{tikzpicture} \begin{feynman} \vertex(1); \vertex[below right = 1.5cm and 2.5cm of 1](3); \vertex[left = 1.5cm of 1](E1); \vertex[below right = 0.3cm and 1.5cm of 3](E3); \vertex[above right = 0.3cm and 1.5cm of 1](E21); \vertex[above right = 0.3cm and 1.5cm of 3](E22); \vertex[below = 0.2cm of 1](L1_I) {\scalebox{1.5}{$y_1$}}; \vertex[below = 0.2cm of 3](L3_I) {\scalebox{1.5}{$y_3$}}; \vertex[above left = 0.2cm and 0.5cm of 1](edge1) {\scalebox{1.5}{$p_1$}}; \vertex[below right = 0.2cm and 0.5cm of 3](edge3) {\scalebox{1.5}{$p_3$}}; \vertex[above right = 0.4cm and 0.5cm of 1](edge21) {\scalebox{1.5}{$(1-x) p_2$}}; \vertex[above right = 0.4cm and 0.5cm of 3](edge22) {\scalebox{1.5}{$x p_2$}}; \diagram*[large]{ (1) -- [->-,line width=0.6mm] (3), (1) -- [-<-,line width=0.6mm, dashed] (E1), (3) -- [->-,line width=0.6mm,dashed] (E3), (1) -- [->-,line width=0.6mm,dashed] (E21), (3) -- [->-,line width=0.6mm,dashed] (E22), }; \path[draw=black, fill=black] (1) circle[radius=0.1]; \path[draw=black, fill=black] (3) circle[radius=0.1]; \end{feynman} \end{tikzpicture} } \end{align}
which is the lower left part of the triangle diagram 
in eq.~\eqref{eq:infinitetriangle}, which stays \emph{hard}, i.e.~bounded, in the singular limit.
The S-matrix element corresponding to the hard graph is, by similar reasoning as above,
\begin{align} s_{(G,\bb \sigma)_\text{hard}}(\{p_1\}, \{(1-x) p_2, x p_2 ,p_3\}) &= \int \frac{ \mathrm{d}^3 \vec{y}_3 e^{ -i\vec{y}_{3} \cdot \vec{p}_3 -ix \vec{y}_{3} \cdot \vec{p}_2 } }{2|\vec{y}_3| } \widehat{\mathcal{F}}_{(G,\bb \sigma)_\text{hard}}^{\{p^0_a\}}(\bb \gamma^\tt + i \ep \bb 1) \Big|_{\vec y_1 = 0}\,. \end{align}
The diagram allows for three paths, starting at the incoming leg labelled by $p_1$ and ending up at the external legs labelled by $(1-x)p_2$, $xp_2$ and $p_3$. The paths have length $0$, $|\vec{y}_3|$ and $|\vec{y}_3|$, respectively (recall that for the reduced S-matrix, we have shifted $\vec{y}_1$ to be located at the origin). Evaluating the Fourier transform of the corresponding flow polytope gives
\begin{equation}
\widehat{\mathcal{F}}_{(G,\bb \sigma)_\text{hard}}^{\{p^0_a\}}(\bb \gamma^\tt + i \ep \bb 1)
\Big|_{\vec y_1 = 0}=\exp\left(-i(x p_2^0+p_3^0)|\vec{y}_3|\right)\,,
\end{equation}
where the reduced flow polytope is just a single point given by
$E_2 = p_1^0 + (1-x) p_2^0 = -p_3^0 - x p_2^0$. We labeled both the energy and the relevant coordinates in accordance with the labeling of the full digraph $(G,\bb \sigma)$. Hence,
\begin{align} \label{eq:collinear_sgx} s_{(G,\bb \sigma)_\text{hard}}(\{p_1\}, \{(1-x) p_2, x p_2 ,p_3\}) &= \int \frac{ \mathrm{d}^3 \vec{y}_3 e^{ -i\vec{y}_{3} \cdot \vec{p}_3 -ix \vec{y}_{3} \cdot \vec{p}_2 } }{2|\vec{y}_3| } e^{-i (p_3^0 + x p_2^0) (|\vec y_3| + i \ep) }\,. \end{align}
It follows by substitution of eq.~\eqref{eq:collinear_sgx} into \eqref{eq:S-matrix_with_x} that
\begin{gather} \begin{gathered} \label{eq:collinearsg} s_{G,\bb \sigma}(\{p_1\}, \{p_2,p_3\}) = - \frac{2 \pi i}{4} \log\frac{p_2^2}{p_1^2} \int_0^1 \dd x \, s_{(G,\bb \sigma)_\text{hard}}(\{p_1\}, \{(1-x) p_2, x p_2 ,p_3\}) +\mathcal O_{p_2^2 \rightarrow 0}(1)\,. \end{gathered} \end{gather}
We hence resolved the collinear singularity $p_2^2 \rightarrow 0$ of the oriented triangle 
explicitly. This simple example shows that the leading singular part of the S-matrix in the collinear limit under inspection can be expressed in a factorised manner. We can actually evaluate the hard S-matrix explicitly
\begin{equation}
    s_{(G,\bb \sigma)_\text{hard}}(\{p_1\}, \{(1-x) p_2, x p_2 ,p_3\})
=
\frac{-\pi}{(p_3^0+xp_2^0)^2-|\vec{p}_3+x \vec{p}_2|^2}\,.
\end{equation}
For massless on-shell, non-exceptional, values of $p_2$ and $p_3$ the S-matrix is only singular when $x\rightarrow 0$, with a logarithmic divergence. In sect.~\ref{sec:soft_singularities_triangle} we will see that $x\rightarrow 0$ corresponds to the soft singularity. 
We can now generalize the analysis of collinear singularities to more complicated integrals. 

We emphasise that the collinear singularity required the two collinear edges $e_4,e_6$ to have the same flow direction. If, for instance, we would have considered the limit where $\vec y_3$ is sent to infinity in the direction of $\vec p_3$ instead, then we would not have encountered a singularity. 
The reason is that the Fourier transform of the flow polytope does not factor in this case and the integral remains finite due to the rapid oscillation of the Fourier transform.

\subsection{General collinear singularities in the \texorpdfstring{$p$-$x$}{p-x} S-matrix representation}
\label{sec:coll_sing_general}

For a general FOPT graph, a collinear singularity is present whenever there exists a two-edge cut that divides the graph into two separate, connected components, such that each component has at least one in- or out-going momentum. Furthermore, the two edges being cut are required to \emph{have the same orientation}. More precisely, suppose we are interested in  
$s_{G,\bb \sigma}(\{p_1,\ldots,p_k\}, \{p_{k+1}, \ldots, p_n\})$,
the reduced S-matrix element
as defined in~\eqref{eq:sG_red}, associated to the digraph $({G,\bb \sigma})$. Then, consider the following bipartition into components $(G,\boldsymbol{\sigma})_\text{col}$ and $(G,\boldsymbol{\sigma})_\text{hard}$,
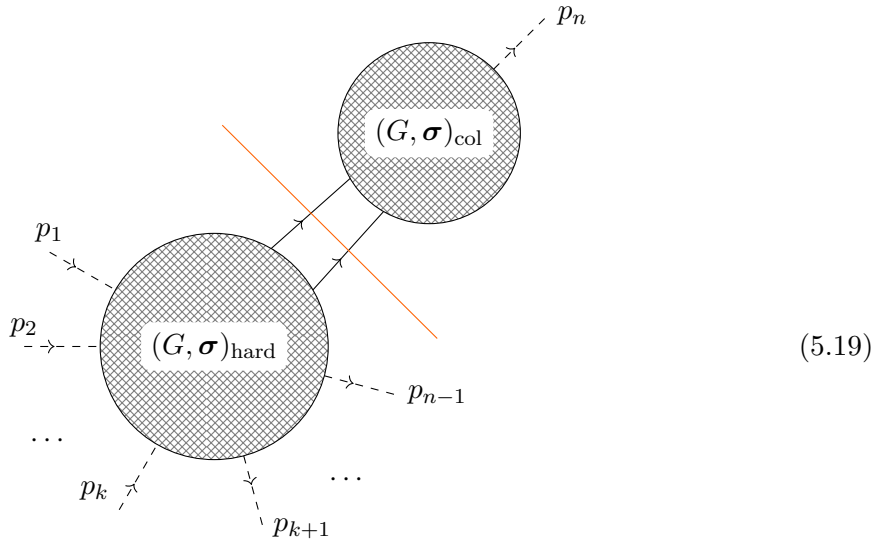
\begin{align} \begin{tikzpicture}[baseline={([yshift=-0.7ex]0,0)}] \coordinate (v); \coordinate (i1) at ([shift=(150:1.5)]v); \coordinate (j1) at ([shift=(150:2.5)]v); \coordinate (i2) at ([shift=(180:1.5)]v); \coordinate (j2) at ([shift=(180:2.5)]v); \coordinate (i3) at ([shift=(210:1.5)]v); \coordinate (j3) at ([shift=(210:2.5)]v); \coordinate (i4) at ([shift=(240:1.5)]v); \coordinate (j4) at ([shift=(240:2.5)]v); \coordinate (o1) at ([shift=(345:1.5)]v); \coordinate (u1) at ([shift=(345:2.5)]v); \coordinate (o2) at ([shift=(315:1.5)]v); \coordinate (u2) at ([shift=(315:2.5)]v); \coordinate (o3) at ([shift=(285:1.5)]v); \coordinate (u3) at ([shift=(285:2.5)]v); \coordinate (o4) at ([shift=(30:1.5)]v); \coordinate (o5) at ([shift=(60:1.5)]v); \coordinate (w) at ([shift=(45:4)]v); \coordinate (wo4) at ([shift=(240:1.2)]w); \coordinate (wo5) at ([shift=(210:1.2)]w); \coordinate (wo6) at ([shift=(45:1.2)]w); \coordinate (wu6) at ([shift=(45:2.2)]w); \coordinate (m) at ([shift=(45:2.15)]v); \coordinate (m1) at ([shift=(135:2)]m); \coordinate (m2) at ([shift=(-45:2)]m); \draw[pattern color=black!50,pattern=crosshatch] (w) circle (1.2); \draw[pattern color=black!50,pattern=crosshatch] (v) circle (1.5); \node[fill=white,rounded corners] at (v) {$(G,\bb \sigma)_{\text{hard}}$}; \node[fill=white,rounded corners] at (w) {$(G,\bb \sigma)_{\text{col}}$}; \node at (j3) {$\cdots$}; \node at (u2) {$\cdots$}; \draw[dashed,->] (j1) -- (i1); \draw[dashed,->] (j2) -- (i2); \draw[dashed,->] (j4) -- (i4); \draw[dashed,->] (o1) -- (u1); \draw[dashed,->] (o3) -- (u3); \draw[dashed,->] (wo6) -- (wu6); \draw[->] (o4) -- (wo4); \draw[->] (o5) -- (wo5); \draw[col3] (m1) -- (m2); \node[above] at (j1) {$p_1$}; \node[above] at (j2) {$p_2$}; \node[above left] at (j4) {$p_k$}; \node[right] at (u3) {$p_{k+1}$}; \node[right] at (u1) {$p_{n-1}$}; \node[right] at (wu6) {$p_n$}; \end{tikzpicture} \end{align}
which features an admissible two-cut, indicated by the orange line. For non-exceptional external momenta, the collinear singularity is attained when $(G,\boldsymbol{\sigma})_\text{col}$ contains only one external leg (in- or out-going), as in the picture above. 
As illustrated in the last section, the collinear singularity appears when the vertices of the  $(G,\boldsymbol{\sigma})_\text{col}$ part of the graph move to infinity in the direction of $\vec p_n$.
By energy conservation, the total energy flow through the cut is equal to the value of $p_n^0$.

Elementary power counting along the lines of the previous section shows that only two-cuts yield non-integrable 
collinear singularities. Three- and $n$-cuts are integrable in the \emph{single} (i.e.\ when no multiple overlapping collinear singularities are considered) collinear limit: to see this, observe that all the inverse distances associated with the cut edges lead to a cumulative scaling of $-n$, while the measure will scale the same as that for the parameterization of eq.~\eqref{eq:param_triangle_coll}, i.e.\ like two powers, independently of $n$, and consistently with the fact that we are considering a single limit. This gives an overall scaling of $2-n$, from which the power-counting argument follows.

In the limit where the split momenta goes on-shell, in our case $p_n^2 \rightarrow 0$, we have the diagrammatic factorization law:
\begin{gather} \begin{gathered} s_{G,\bb \sigma}(\{p_1,\ldots,p_k\}, \{p_{k+1}, \ldots, p_n\}) = - \frac{2 \pi i}{4} \log\frac{p_n^2}{Q^2} \int_0^1 \dd x \, s_{(G,\bb \sigma)_{\text{hard}}} s_{(G,\bb \sigma)_{\text{col}}} +\mathcal O_{p_n^2 \rightarrow 0}(1)\,, \end{gathered} \end{gather}
where $Q$ is an arbitrary external momentum scale (for example, $Q=p_1+...+p_k$) and $s_{(G,\bb \sigma)_{\text{hard}}}$ and $s_{(G,\bb \sigma)_{\text{col}}}$ are the reduced S-matrix elements associated to the subgraphs, i.e.
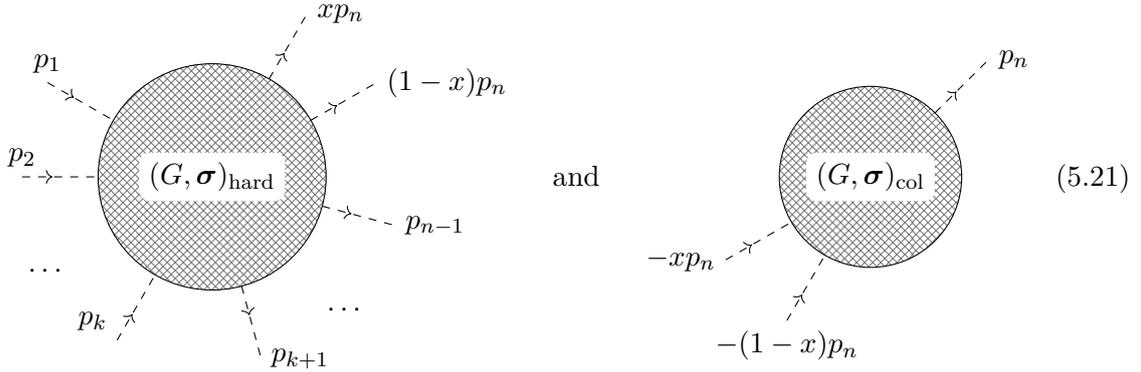
\begin{align} & \begin{tikzpicture}[baseline={([yshift=-0.7ex]0,0)}] \coordinate (v); \coordinate (i1) at ([shift=(150:1.5)]v); \coordinate (j1) at ([shift=(150:2.5)]v); \coordinate (i2) at ([shift=(180:1.5)]v); \coordinate (j2) at ([shift=(180:2.5)]v); \coordinate (i3) at ([shift=(210:1.5)]v); \coordinate (j3) at ([shift=(210:2.5)]v); \coordinate (i4) at ([shift=(240:1.5)]v); \coordinate (j4) at ([shift=(240:2.5)]v); \coordinate (o1) at ([shift=(345:1.5)]v); \coordinate (u1) at ([shift=(345:2.5)]v); \coordinate (o2) at ([shift=(315:1.5)]v); \coordinate (u2) at ([shift=(315:2.5)]v); \coordinate (o3) at ([shift=(285:1.5)]v); \coordinate (u3) at ([shift=(285:2.5)]v); \coordinate (o4) at ([shift=(30:1.5)]v); \coordinate (u4) at ([shift=(30:2.5)]v); \coordinate (o5) at ([shift=(60:1.5)]v); \coordinate (u5) at ([shift=(60:2.5)]v); \draw[pattern color=black!50,pattern=crosshatch] (v) circle (1.5); \node[fill=white,rounded corners] at (v) {$(G,\bb \sigma)_{\text{hard}}$}; \node at (j3) {$\cdots$}; \node at (u2) {$\cdots$}; \draw[dashed,->] (j1) -- (i1); \draw[dashed,->] (j2) -- (i2); \draw[dashed,->] (j4) -- (i4); \draw[dashed,->] (o1) -- (u1); \draw[dashed,->] (o3) -- (u3); \draw[dashed,->] (o4) -- (u4); \draw[dashed,->] (o5) -- (u5); \node[above] at (j1) {$p_1$}; \node[above] at (j2) {$p_2$}; \node[above left] at (j4) {$p_k$}; \node[right] at (u3) {$p_{k+1}$}; \node[right] at (u1) {$p_{n-1}$}; \node[right] at (u4) {$(1-x) p_n$}; \node[right] at (u5) {$x p_n$}; \end{tikzpicture} & \text{ and } & & \begin{tikzpicture}[baseline={([yshift=-0.7ex]0,0)}] \coordinate (w); \coordinate (wo4) at ([shift=(240:1.2)]w); \coordinate (wu4) at ([shift=(240:2.2)]w); \coordinate (wo5) at ([shift=(210:1.2)]w); \coordinate (wu5) at ([shift=(210:2.2)]w); \coordinate (wo6) at ([shift=(45:1.2)]w); \coordinate (wu6) at ([shift=(45:2.2)]w); \draw[pattern color=black!50,pattern=crosshatch] (w) circle (1.2); \node[fill=white,rounded corners] at (w) {$(G,\bb \sigma)_{\text{col}}$}; \draw[dashed,->] (wo6) -- (wu6); \draw[dashed,->] (wu4) -- (wo4); \draw[dashed,->] (wu5) -- (wo5); \node[left] at (wu5) {$- x p_n$}; \node[below] at (wu4) {$- (1-x) p_n$}; \node[right] at (wu6) {$p_n$}; \end{tikzpicture} \end{align}
with the dependence on the external momenta as indicated.
Note that in the previous example the contribution from the collinear part of the diagram was 
trivial as it just consisted of a single vertex.

\subsection{Soft-collinear singularity of the triangle diagram}
\label{sec:soft_singularities_triangle}

 We will illustrate the appearance of soft singularities in the triangle diagram of eq.~\eqref{eq:Striangle}. 
This soft singularity will be a \emph{nested singularity}, 
in the sense that it is contained in the previous collinear limit. %

By eqs.~\eqref{eq:collinear_sgx} and~\eqref{eq:collinearsg} we have the following expression for the S-matrix element in the collinear limit:
\begin{gather} \begin{gathered} \label{eq:collinearsg_full} s_{G,\bb \sigma}(\{p_1\}, \{p_2,p_3\}) = \\
- \frac{2 \pi i}{4} \log\frac{p_2^2}{p_1^2} \int_0^1 \dd x \, \int \frac{ \mathrm{d}^3 \vec{y}_3 e^{ -i\vec{y}_{3} \cdot \vec{p}_3 -ix \vec{y}_{3} \cdot \vec{p}_2 } }{2|\vec{y}_3| } e^{i (-p_3^0 - x p_2^0) (|\vec y_3| + i \ep) } +\mathcal O_{p_2^2 \rightarrow 0}(1)\,. \end{gathered} \end{gather}
The soft singularity is associated with the limit in which, in addition to the  vertex $\vec y_2$, the vertex $\vec y_3$ in the diagram below is sent to infinity along the direction $\vec p_3$ as indicated.
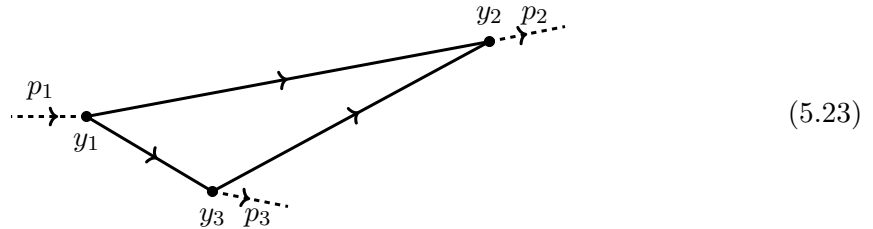
\begin{align} \label{eq:triangle_flow2} \resizebox{7.5cm}{!}{ \begin{tikzpicture}[baseline={([yshift=-0.7ex]0,0)}] \begin{feynman} \vertex(1); \vertex[above right = 1.5cm and 8cm of 1](2); \vertex[below right = 1.5cm and 2.5cm of 1](3); \vertex[left = 1.5cm of 1](E1); \vertex[above right = 0.3cm and 1.5cm of 2](E2); \vertex[below right = 0.3cm and 1.5cm of 3](E3); \vertex[below = 0.2cm of 1](L1_I) {\scalebox{1.5}{$y_1$}}; \vertex[above = 0.2cm of 2](L2_I) {\scalebox{1.5}{$y_2$}}; \vertex[below = 0.2cm of 3](L3_i) {\scalebox{1.5}{$y_3$}}; \vertex[above left = 0.2cm and 0.5cm of 1](edge1) {\scalebox{1.5}{$p_1$}}; \vertex[above right = 0.2cm and 0.5cm of 2](edge2) {\scalebox{1.5}{$p_2$}}; \vertex[below right = 0.2cm and 0.5cm of 3](edge3) {\scalebox{1.5}{$p_3$}}; \diagram*[large]{ (1) -- [->-,line width=0.6mm] (2) -- [-<-,line width=0.6mm] (3) -- [-<-,line width=0.6mm] (1), (1) -- [-<-,line width=0.6mm, dashed] (E1), (2) -- [->-,line width=0.6mm,dashed] (E2), (3) -- [->-,line width=0.6mm,dashed] (E3), }; \path[draw=black, fill=black] (1) circle[radius=0.1]; \path[draw=black, fill=black] (2) circle[radius=0.1]; \path[draw=black, fill=black] (3) circle[radius=0.1]; \end{feynman} \end{tikzpicture} } \end{align}
The soft singularity only appears if both $p_2,p_3$ are on-shell, i.e.~$p_2^2, p_3^2 \rightarrow 0$.
The large distance limit can be parameterized as follows in the integrand of eq.~\eqref{eq:collinearsg_full} 
\begin{align} \vec{y}_3=\lambda {\vec p_3}/{|\vec p_3|^2} +\sqrt{|\lambda|}\vec{y}^\perp_3\,,\;\;\;\; x = \frac{1}{|\lambda|} x'\,, \end{align}
where, considering $\lambda \rightarrow \infty$, we send $\vec y_3$ to infinity and let the energy flow through the bottom right edge in \eqref{eq:triangle_flow2}, which is parameterized by the energy fraction parameter $x$, go to zero. 
That means this edge is \emph{becoming soft}.
In the limit $\lambda \rightarrow \infty$ we find 
\begin{align} |\vec z_5| & = \frac{\lambda}{|\vec{p}_3|} + \frac12 |\vec p_3| (\vec{y}^\perp_3)^2 + \mathcal O(\lambda^{-1}). \end{align}
Applying this to the relevant part of eq.~\eqref{eq:collinearsg_full}, we obtain
\begin{gather} \begin{gathered} \label{eq:collinearsg_full2} s_{G,\bb \sigma}(\{p_1\}, \{p_2,p_3\}) = - \frac{2 \pi i}{8} \log\frac{p_2^2}{p_1^2} \\
\times \int_1^\infty \frac{\dd \lambda}{\lambda} \left( \int_0^\lambda \dd x' \, \int \mathrm{d}^2 \vec{y}^\perp_3 e^{ -i \lambda -ix' \frac{\vec{p}_{3} \cdot \vec{p}_2}{\vec p_3^2} - i \frac{p_3^0}{|\vec p_3|} \lambda - i \frac12 p_3^0 |\vec p_3| (\vec y^\perp_3)^2 - i x' (\frac{p_2^0}{|\vec p_3|} - i \ep) } +\mathcal O_{\lambda \rightarrow \infty}(\lambda^{-\frac12}), \right) \\
+\mathcal O_{p_2^2 \rightarrow 0}(1)\,. \end{gathered} \end{gather}
In the $p_3^2 \rightarrow 0$ limit, the integration over $x'$ results in,
\begin{align} \int_0^\lambda \dd x' \, e^{ -ix' \left(\frac{\vec{p}_{3} \cdot \vec{p}_2}{\vec p_3^2} + \frac{p_2^0}{|\vec p_3|} - i \ep \right) } = \frac{i (p^0_3)^2} { p^0_2 p^0_3 - \vec p_2 \cdot \vec p_3 } +\mathcal O_{\lambda \rightarrow \infty}(e^{-\ep \lambda}). \end{align}
The other parts of the integral have already been evaluated in the last section and we obtain
\begin{gather} \begin{gathered} s_{G,\bb \sigma}(\{p_1\}, \{p_2,p_3\}) = -i \frac{(2 \pi)^2}{8} \frac{\log\frac{p_2^2}{p_1^2} \log\frac{p_3^2}{p_1^2} }{ p_2 \cdot p_3 } +\mathcal O_{ \substack{ p_2^2 \rightarrow 0 \\
p_3^2 \rightarrow 0 } }(1). \end{gathered} \end{gather}
This expression gives the full contribution to the 
reduced S-matrix element of the oriented triangle diagram \eqref{eq:Striangle} in the limit $p_2^2,p_3^2 \rightarrow 0$. We find the expected double-$\log$ Sudakov structure. The expression stays well-defined as long as $p_2 \cdot p_3 = p^0_2 p^0_3 - \vec p_2 \cdot \vec p_3 \neq 0$. This is expected as $p_1^2= (p_2+p_3)^2 = 2 p_2 \cdot p_3$ is required to be non-vanishing by energy conservation. 
Similar soft-factorization phenomena can be observed for the $p$-$x$ S-matrix representations of other FOPT diagrams. We will leave the detailed analysis of general soft singularities to a future work.

\section{Conclusion}

In this paper we introduced a new version of diagram-based perturbation theory in quantum field theory: flow-oriented perturbation theory. Instead of the usual covariant momentum space perturbation theory in four-dimensional Minkowski space, it is effectively a three-dimensional coordinate space integral representation in which the time integrations are resolved via the residue theorem. This representation is dual to time-ordered perturbation theory. The main features of this new representation are that it is combinatorial and canonical, and that the $i\ep$ dependence simplified. Our derivation of the canonical FOPT Feynman rules in sect.~\ref{sec:fopt}, which associate a unique integral expression to each FOPT diagram, is illustrated by detailed examples.

We described UV singularities within the FOPT formalism and showed that it yields equivalent power-counting to the covariant picture. Interestingly, UV and IR singularities turn out to be more isolated within FOPT than in the covariant framework. It would be interesting to study this phenomenon in more detail by analyzing the nested structure of FOPT divergences e.g.~by using Hopf algebra or lattice techniques~\cite{Connes:1999yr,Figueroa:2004hb,Borinsky:2014xwa,Beekveldt:2020kzk}.

Our derivation relied on the underlying QFT being formulated in four-dimensional Minkowski spacetime. Especially, while treating  gauge theories it is desirable to be able to use dimensional regularization, i.e.~to work in $4-2 \epsilon$ dimensions. We leave the formulation of dimensionally regularized FOPT to a future work. An immediate alternative to dimensional regularization, which removes UV singularities and which can be used in conjunction with FOPT is a BHPZ subtraction scheme as it has been applied for three-dimensional momentum-space representations in \cite{Capatti:2022tit}.

The FOPT representation also features finite distance singularities whose structure we discussed in sect.~\ref{sec:singularitiesFOPT}, where we also compared them to analyses put forward in~\cite{Sterman3,StermanErdogan,ErdoganCS}. We concluded that these singularities are somewhat different from the usual collinear and soft IR divergences that are observed in momentum space. %

This observation led us to derive  a new representation for the perturbative expansion of the S-matrix in sect.~\ref{sec:s_matrix}. This $p$-$x$ S-matrix representation is of hybrid momentum and coordinate space nature, as external parameters are given in terms of momenta, but internal integrations are performed in coordinate space. This representation produces an effectively three-dimensional coordinate space integral, which is modulated by the Fourier transform of the \emph{flow polytope}.  
The flow polytope has many remarkable properties~\cite{gallo1978extreme,stanley2000acyclic,baldoni2004counting,meszaros2015flow,meszaros2017polytope,benedetti2019combinatorial}. 
We exposed the explicit resolution of spurious singularities within this Fourier transform. We provided some general remarks on such resolutions in sect.~\ref{sec:spurious}.

As expected from the covariant picture, the $p$-$x$ S-matrix representation features soft and collinear divergences. We illustrated their appearance in sect.~\ref{sec:IRpxsing} and showed the central role of the flow polytope and its Fourier transform. Moreover, we explained how manifest collinear factorization at the diagram level is attained in the $p$-$x$ S-matrix representation. Even though we also illustrate how soft divergences appear, we postpone the derivation of a general per-diagram factorization law for soft singularities to  future work. %

It would be interesting to study these singular structures with respect to their nesting e.g.~as in ref.~\cite{Borinsky:2015mga}, and it might be possible to design a new IR subtraction scheme along the lines of ref.~\cite{Herzog:2018ily,Ma:2019hjq} using the $p$-$x$ S-matrix representation. As we illustrated in sect.~\ref{sec:IRpxsing}, these singularities are naturally regularized with the value of the external kinematics and computations in singular limits are comparatively easy. 

As we show in the appendix, an interesting application of the FOPT representation in the context of explicit computations are phase space unitarity-cut integrals. The FOPT representation applied to these, turns out to have the same dimension for all cut diagrams in contrast to covariant momentum space, where the integration measure effectively has a different dimension for each cut diagram. Over all cut diagrams we expect cancellation of IR divergences by the KLN theorem and within the FOPT representation this cancellation should be local. In appendix~\ref{app:cutkosky}, we illustrate the application of the FOPT approach to unitarity-cut integrals and summarize the associated Feynman rules, but also here we leave the exploration of potential local cancellation properties with possible applications for numerical evaluations of cross-sections to a future work.
\acknowledgments{
MB was supported by Dr.\ Max Rössler, the Walter Haefner Foundation and the ETH Zürich Foundation.  ZC would like to thank Babis Anastasiou for his continuous support. ASB and EL thank the Institute for Theoretical Studies at the ETH Z\"urich for hospitality during the completion of this work. ASB received financial support from Spanish MICINN PID2019-108655GB-I00/AEI/10.13039/501100011033 grant and expresses their gratitude to Juan Ferrera for useful discussions at the early stages of this research.}

\appendix

\section{Unitarity, cut integrals and Cutkosky's Theorem}
\label{app:cutkosky}

The FOPT representation that we derived in sect.~\ref{sec:fopt} takes an interesting form for unitarity-cut based phase space integrals.  In this appendix, we  discuss this viewpoint along the lines of a classic exposition of phase space integrals focused on the largest time equation \cite{Veltman2}. The result is also related to Cutkosky's theorem \cite{Cutkosky:1960sp}.  The following application of the Cutkosky rules to the FOPT representation brings us close to results on the general analyticity properties of Feynman integrals, as described in \cite{Bloch:2015efx,Berghoff:2020bug,Berghoff:2022vah}.
This FOPT-cut representation has some remarkable properties in this regard. For instance the integrals associated to virtual and real corrections turn out to have the same dimension, so that both can be put on the same measure. Hence, the following considerations might be useful to pursue explicit computations of phase space integrals with manifest cancellation of real and virtual singularities. 

Here, we briefly present this FOPT-based representation of cut integrals.
Given a subset of vertices $V \subset V_G$ of a graph $G$, the \emph{subgraph induced} from $V$ is the subgraph of $G$ that contains all edges whose both end-vertices lie in $V$. A \emph{cut} $\mathfrak C$ of a Feynman graph $\Gamma$ is partition of the vertices $V_G$ into two parts $V_\sunny, V_\shady$ such that the respective induced subgraphs with edges $E_{\sunny},E_{\shady}$ from both parts are connected and each contains at least one external vertex. The edges that have one end in $V_\sunny$ and one in $V_\shady$ are \emph{cut edges}, $E_{\mathfrak{C}}$. Intuitively, we think of energy flowing from the $\sunny$-side to the $\shady$-side.

Recall that the \emph{cut propagator} is given by
\begin{gather*} \Delta^\pm(z) = \int \frac{\dd^{4} p}{(2\pi)^4} 2 \pi \delta(p^2) \theta(\pm p_0) e^{-i p \cdot z } = -\frac{1}{(2\pi)^{2}} \left( \frac{1}{(z^0\mp i\ep)^2 - |\vec z\,|^2 } \right). \end{gather*}
It follows, from an argument due to Veltman \cite{Veltman2}, that the real part\footnote{Due to an extra conventional factor of $i$ this real part of the integral contributes to the \emph{imaginary} part of the Feynman amplitude.} of a Feynman integral can be expressed in terms of a sum over \emph{cut integrals} (see \cite[Ch.~8]{Martinus} or \cite[Ch.~10.4]{le1991quantum} for a detailed derivation):
\begin{align} A_{G}(x_1, \ldots, x_{|\gVE|}) + A_{G}^*(x_1, \ldots, x_{|\gVE|}) = - \sum_{\text{cuts } \mathfrak C \text{ of } G} A_{G,\mathfrak C}(x_1, \ldots, x_{|\gVE|})\,, \end{align}
where the cut integrals $A_{G,\mathfrak C}(x_1, \ldots, x_{|\gVE|})$ are given by the expression
\begin{gather} \begin{gathered} \label{eq:cutintegral} A_{G,\mathfrak C}(x_1, \ldots, x_{|\gVE|})=\\= (-ig)^{|\gVI_\sunny|} (ig)^{|\gVI_\shady|} \left( \prod_{v\in \gVI} \int \dd^4 y_v \right) \left( \prod_{e \in E_{\sunny}} \Delta_F(z_e) \right) \left( \prod_{e \in E_{\mathfrak C}} \Delta^+(z_e) \right) \left( \prod_{e \in E_{\shady}} \Delta^*_F(z_e) \right). \end{gathered} \end{gather}
In contrast to the original Feynman integral, the propagators on the cut are replaced with the positive frequency $\Delta^+$ cut propagator and the Feynman propagators on the $\shady$-side of the cut are replaced with the complex conjugate Feynman propagator.

We have the following expressions for the Fourier transform of the coordinate space propagator
\begin{gather} \begin{gathered} \int_{-\infty}^\infty \dd z^0_e e^{i E_e z^0_e} \Delta^{+}(z_e) = - \frac{1}{(2 \pi)^2} \int_{-\infty}^\infty \dd z^0_e \frac{e^{i E_e z^0_e}}{(z^0_e - i \ep)^2 - \vec z_e^2} \\
= - \frac{1}{(2 \pi)^2} \int_{-\infty}^\infty \dd z^0_e \frac{e^{i E_e z^0_e}}{2|\vec z_e|} \left( \frac{1}{z^0_e - |\vec z_e| - i \ep} - \frac{1}{z^0_e + |\vec z_e| - i \ep} \right) \\
= - \frac{1}{(2 \pi)^2} \frac{2 \pi i}{2|\vec z_e|} \theta(E_e) \left( e^{i E_e (|\vec z_e| + i \ep)} - e^{i E_e (-|\vec z_e| + i \ep)} \right)\,, \end{gathered} \end{gather}
with the analogous Cauchy integrals
\begin{align} &\begin{aligned} \int_{-\infty}^\infty \dd z^0_e e^{i E_e z^0_e} \Delta_{F}(z_e)               &= - \frac{1}{(2 \pi)^2} \frac{2 \pi i}{2|\vec z_e|} \left( \theta(E_e) e^{i E_e (|\vec z_e| + i \ep)} + \theta(-E_e) e^{-i E_e (|\vec z_e| + i \ep)} \right)\,,\\
\int_{-\infty}^\infty \dd z^0_e e^{i E_e z^0_e} \Delta_{F}^*(z_e)               &= - \frac{1}{(2 \pi)^2} \frac{-2 \pi i}{2|\vec z_e|} \left( \theta(E_e) e^{i E_e (-|\vec z_e| + i \ep)} + \theta(-E_e) e^{-i E_e (-|\vec z_e| + i \ep)} \right)\,. \end{aligned} \end{align}
Repeating the derivation in sect.~\ref{sec:derivation}, now for the cut integral in eq.~\eqref{eq:cutintegral}, while using the Fourier transforms of the respective propagators, results in a representation of a cut integral as a sum over FOPT-cut integrals.
\subsection*{FOPT Feynman rules for cut integrals}
\label{sec:cutrules}
The result is the following set of FOPT-cut integral Feynman rules for a digraph $(G,\boldsymbol{\sigma})$ with a cut $\mathfrak C$.
\begin{enumerate}
\item The integral is $0$ if the closed directed graph $(G,\boldsymbol{\sigma})^\circ$ is not strongly connected or if the admissible paths on the cut do not go from the $\sunny$-side to the $\shady$-side of the graph.
\item Multiply a factor of $-ig$ ($ig$) for each $\sunny$-side ($\shady$-side) interaction vertex. 
\item For each internal vertex $v\in V^{\text{int}}$ of the digraph $(G,\boldsymbol{\sigma})$ integrate over $3$-dimensional space with the measure $2\pi \int \dd^3 \vec{y}_v$.
\item For each edge $e$ of the graph multiply a factor of $\frac{\mp i}{8\pi^2|\vec{z}_e|}$ with a $-$ sign for a $\sunny$-side or a cut edge, and a $+$ sign for a $\shady$-side edge. 
\item For each entirely uncut directed admissible path, $ \pp_\ell$, of $(G,\boldsymbol{\sigma})^\circ$ multiply a factor of 
\begin{align*} \frac{i} { \sum_{e\in \pp_\ell} |\vec z_e| + \tau_{\mathcal \pp_\ell} +i\ep } &&\text{ if $\pp_\ell$ consists entirely of $\sunny$-side edges} \\
\frac{i} { -\sum_{e\in \pp_\ell} |\vec z_e| + \tau_{\pp_\ell} +i\ep } &&\text{ if $\pp_\ell$ consists entirely of $\shady$-side edges} \end{align*}
where the sum in the denominator goes over all edges that are in the admissible path $\pp_\ell$ and $\tau_{\mathcal \pp_\ell}$ is the time difference that has passed while going through the $\circ$ vertex, or $0$ if the admissible path does not go through the $\circ$ vertex, i.e.~is a cycle.
\item For each directed admissible path $\pp_{\ell}$ of $(G,\boldsymbol{\sigma})^\circ$ that passes the cut $\mathfrak C$, multiply a factor of 
\begin{align*} \frac{-2 i |\vec z_{e_\mathfrak C}|} { \left( \sum_{e\in \pp_\ell^\sunny} |\vec z_e| -\sum_{e\in \pp_\ell^\shady} |\vec z_e| + \tau_{\pp_\ell} +i\ep \right)^2 - {\vec z_{e_\mathfrak C}}^{\,2} }\,, \end{align*}
where we sum over the uncut $\sunny$-side and $\shady$-side edges in $\pp_\ell$, $\pp_\ell^\sunny$ and $\pp_\ell^\shady$, and where $e_\mathfrak C$ denotes the unique edge of the admissible path that is on the cut. 
The edge is unique because, once the path passes over the cut edge, the energy cannot flow back through the cut.
\end{enumerate}

\subsection*{Example}

We consider the cut integrals associated to the following graph.
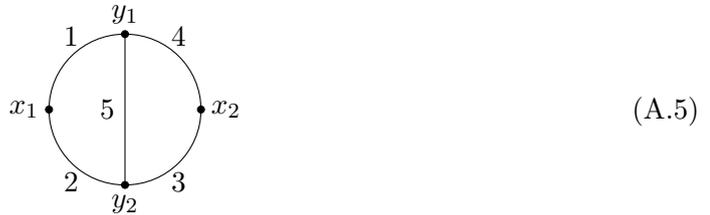
\begin{align} \label{eq:theta_graph} \def\scale{1} \begin{tikzpicture}[baseline={([yshift=-0.7ex]0,0)}] \coordinate[label=left:$x_1$] (i1) at (-\scale,0); \coordinate[label=right:$x_2$] (i2) at (+\scale,0); \coordinate[label=above:$y_1$] (v1) at (0,\scale); \coordinate[label=below:$y_2$] (v2) at (0,-\scale); \draw (i1) arc (180:90:\scale) node[midway,above] {$1$}; \draw (i1) arc (180:270:\scale) node[midway,below] {$2$}; \draw (i2) arc (0:-90:\scale) node[midway,below] {$3$}; \draw (i2) arc (0:90:\scale) node[midway,above] {$4$}; \filldraw (v1) circle (1.3pt); \filldraw (v2) circle (1.3pt); \filldraw (i1) circle (1.3pt); \filldraw (i2) circle (1.3pt); \draw (v1) -- (v2) node[midway,left] {$5$}; \end{tikzpicture} \end{align}
Note that $x_1$ and $x_2$ are both external and interaction vertices. 
The associated completed graph is
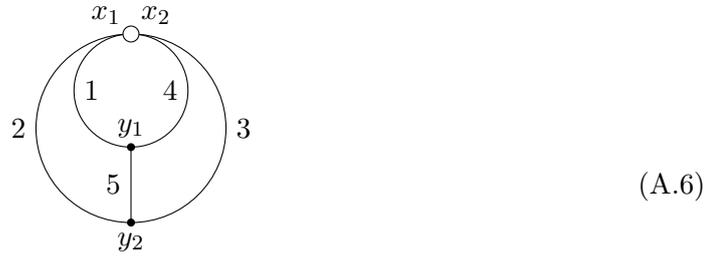
\begin{align} \label{eq:theta_graph_compl} \def\scale{1} \begin{tikzpicture}[baseline={([yshift=-0.7ex]0,0)}] \coordinate (m) at (0,2*\scale); \coordinate[label=above:$y_1$] (v1) at (0,\scale/2); \coordinate[label=below:$y_2$] (v2) at (0,-\scale/2); \draw (m) arc (90:270:\scale*3/4) node[midway,right] {$1$}; \draw (m) arc (90:270:\scale*5/4) node[midway,left] {$2$}; \draw (m) arc (90:-90:\scale*3/4) node[midway,left] {$4$}; \draw (m) arc (90:-90:\scale*5/4) node[midway,right] {$3$}; \filldraw[fill=white] (m) circle (3pt); \draw (v1) -- (v2) node[midway,left] {$5$}; \filldraw (v1) circle (1.3pt); \filldraw (v2) circle (1.3pt); \coordinate[label=above left:$x_1$] (m) at (0,2*\scale); \coordinate[label=above right:$x_2$] (m) at (0,2*\scale); \end{tikzpicture} \end{align}
and we have the following three different admissible cuts as permutations of the internal vertices result in topologically indistinguishable graphs,
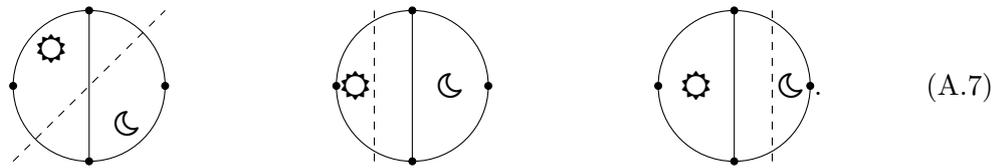
\begin{align} & \def\scale{1} \begin{tikzpicture}[baseline={([yshift=-0.7ex]0,0)}] \coordinate (i1) at (-\scale,0); \coordinate (i2) at (+\scale,0); \coordinate (v1) at (0,\scale); \coordinate (v2) at (0,-\scale); \draw (i1) arc (180:90:\scale); \draw (i1) arc (180:270:\scale); \draw (i2) arc (0:-90:\scale); \draw (i2) arc (0:90:\scale); \filldraw (v1) circle (1.3pt); \filldraw (v2) circle (1.3pt); \filldraw (i1) circle (1.3pt); \filldraw (i2) circle (1.3pt); \draw (v1) -- (v2); \draw[dashed] (-\scale,-\scale) -- (\scale,\scale); \node at (-\scale/2,\scale/2) {$\sunny$}; \node at (\scale/2,-\scale/2) {$\shady$}; \end{tikzpicture} & & \def\scale{1} \begin{tikzpicture}[baseline={([yshift=-0.7ex]0,0)}] \coordinate (i1) at (-\scale,0); \coordinate (i2) at (+\scale,0); \coordinate (v1) at (0,\scale); \coordinate (v2) at (0,-\scale); \draw (i1) arc (180:90:\scale); \draw (i1) arc (180:270:\scale); \draw (i2) arc (0:-90:\scale); \draw (i2) arc (0:90:\scale); \filldraw (v1) circle (1.3pt); \filldraw (v2) circle (1.3pt); \filldraw (i1) circle (1.3pt); \filldraw (i2) circle (1.3pt); \draw (v1) -- (v2); \draw[dashed] (-\scale/2,-\scale) -- (-\scale/2,\scale); \node at (-\scale/4*3,0) {$\sunny$}; \node at (\scale/2,0) {$\shady$}; \end{tikzpicture} & & \def\scale{1} \begin{tikzpicture}[baseline={([yshift=-0.7ex]0,0)}] \coordinate (i1) at (-\scale,0); \coordinate (i2) at (+\scale,0); \coordinate (v1) at (0,\scale); \coordinate (v2) at (0,-\scale); \draw (i1) arc (180:90:\scale); \draw (i1) arc (180:270:\scale); \draw (i2) arc (0:-90:\scale); \draw (i2) arc (0:90:\scale); \filldraw (v1) circle (1.3pt); \filldraw (v2) circle (1.3pt); \filldraw (i1) circle (1.3pt); \filldraw (i2) circle (1.3pt); \draw (v1) -- (v2); \draw[dashed] (\scale/2,-\scale) -- (\scale/2,\scale); \node at (-\scale/2,0) {$\sunny$}; \node at (\scale/4*3,0) {$\shady$}; \end{tikzpicture} . \intertext{Recall that in addition to the positivity requirements only energy flows from $\sunny$ to $\shady$ are allowed on cut edges. Therefore only the following energy flows are compatible with the cuts and the positive energy requirement:} \label{eq:theta_cuts1} & \def\scale{1} 
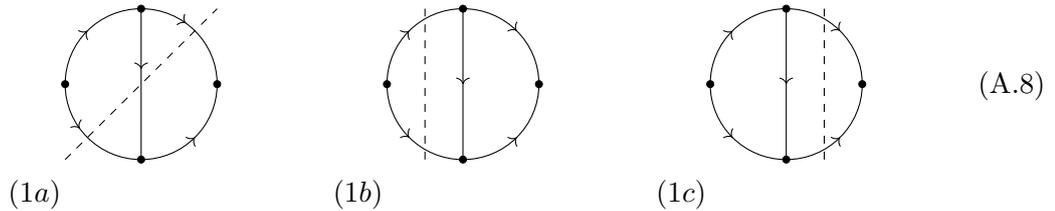
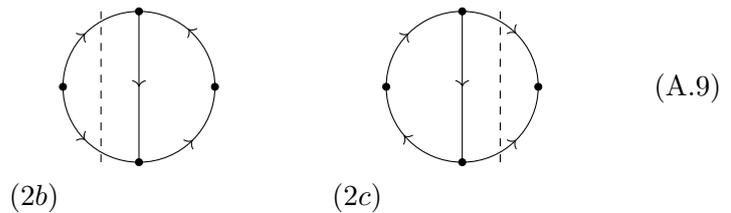
\begin{tikzpicture}[baseline={([yshift=-0.7ex]0,0)}] \coordinate (i1) at (-\scale,0); \coordinate (i2) at (+\scale,0); \coordinate (v1) at (0,\scale); \coordinate (v2) at (0,-\scale); \draw[->-] (i1) arc (180:90:\scale); \draw[->] (i1) arc (180:270:\scale); \draw[->] (v1) arc (90:0:\scale); \draw[->-] (v2) arc (-90:0:\scale); \draw[->] (v1) -- (v2); \filldraw (v1) circle (1.3pt); \filldraw (v2) circle (1.3pt); \filldraw (i1) circle (1.3pt); \filldraw (i2) circle (1.3pt); \draw[dashed] (-\scale,-\scale) -- (\scale,\scale); \end{tikzpicture} & & \def\scale{1} \begin{tikzpicture}[baseline={([yshift=-0.7ex]0,0)}] \coordinate (i1) at (-\scale,0); \coordinate (i2) at (+\scale,0); \coordinate (v1) at (0,\scale); \coordinate (v2) at (0,-\scale); \draw[->-] (i1) arc (180:90:\scale); \draw[->-] (i1) arc (180:270:\scale); \draw[->-] (v1) arc (90:0:\scale); \draw[->-] (v2) arc (-90:0:\scale); \draw[->-] (v1) -- (v2); \filldraw (v1) circle (1.3pt); \filldraw (v2) circle (1.3pt); \filldraw (i1) circle (1.3pt); \filldraw (i2) circle (1.3pt); \draw[dashed] (-\scale/2,-\scale) -- (-\scale/2,\scale); \end{tikzpicture} & & \def\scale{1} \begin{tikzpicture}[baseline={([yshift=-0.7ex]0,0)}] \coordinate (i1) at (-\scale,0); \coordinate (i2) at (+\scale,0); \coordinate (v1) at (0,\scale); \coordinate (v2) at (0,-\scale); \draw[->-] (i1) arc (180:90:\scale); \draw[->-] (i1) arc (180:270:\scale); \draw[->-] (v1) arc (90:0:\scale); \draw[->-] (v2) arc (-90:0:\scale); \draw[->-] (v1) -- (v2); \filldraw (v1) circle (1.3pt); \filldraw (v2) circle (1.3pt); \filldraw (i1) circle (1.3pt); \filldraw (i2) circle (1.3pt); \draw[dashed] (\scale/2,-\scale) -- (\scale/2,\scale); \end{tikzpicture} \\
(1a) && (1b) && (1c) \notag \\
\label{eq:theta_cuts2} && & \def\scale{1} \begin{tikzpicture}[baseline={([yshift=-0.7ex]0,0)}] \coordinate (i1) at (-\scale,0); \coordinate (i2) at (+\scale,0); \coordinate (v1) at (0,\scale); \coordinate (v2) at (0,-\scale); \draw[->-] (i1) arc (180:90:\scale); \draw[->-] (i1) arc (180:270:\scale); \draw[-<-] (v1) arc (90:0:\scale); \draw[->-] (v2) arc (-90:0:\scale); \draw[->-] (v1) -- (v2); \filldraw (v1) circle (1.3pt); \filldraw (v2) circle (1.3pt); \filldraw (i1) circle (1.3pt); \filldraw (i2) circle (1.3pt); \draw[dashed] (-\scale/2,-\scale) -- (-\scale/2,\scale); \end{tikzpicture} & & \def\scale{1} \begin{tikzpicture}[baseline={([yshift=-0.7ex]0,0)}] \coordinate (i1) at (-\scale,0); \coordinate (i2) at (+\scale,0); \coordinate (v1) at (0,\scale); \coordinate (v2) at (0,-\scale); \draw[->-] (i1) arc (180:90:\scale); \draw[-<-] (i1) arc (180:270:\scale); \draw[->-] (v1) arc (90:0:\scale); \draw[->-] (v2) arc (-90:0:\scale); \draw[->-] (v1) -- (v2); \filldraw (v1) circle (1.3pt); \filldraw (v2) circle (1.3pt); \filldraw (i1) circle (1.3pt); \filldraw (i2) circle (1.3pt); \draw[dashed] (\scale/2,-\scale) -- (\scale/2,\scale); \end{tikzpicture} \\
&& (2b) && (2c) \notag \end{align}
In the depiction above, each row features only one orientation of the graph and each column a possible cut. In this example, there are only two admissible paths that are compatible with a cut. 
The cut diagram $(1a)$ has the following three routes
{
\def\scale{.8}
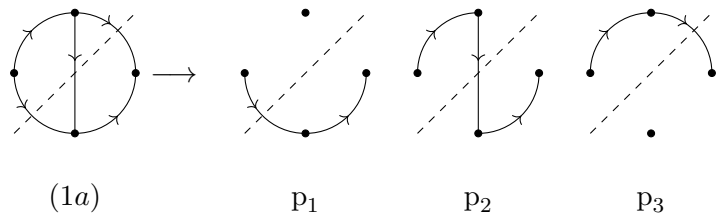
\begin{align*} \begin{array}{ccccc} \begin{tikzpicture}[baseline={([yshift=-0.7ex]0,0)}] \coordinate (i1) at (-\scale,0); \coordinate (i2) at (+\scale,0); \coordinate (v1) at (0,\scale); \coordinate (v2) at (0,-\scale); \draw[->-] (i1) arc (180:90:\scale); \draw[->] (i1) arc (180:270:\scale); \draw[->] (v1) arc (90:0:\scale); \draw[->-] (v2) arc (-90:0:\scale); \draw[->] (v1) -- (v2); \filldraw (v1) circle (1.3pt); \filldraw (v2) circle (1.3pt); \filldraw (i1) circle (1.3pt); \filldraw (i2) circle (1.3pt); \draw[dashed] (-\scale,-\scale) -- (\scale,\scale); \end{tikzpicture} & \longrightarrow &\;\;\;\; \begin{tikzpicture}[baseline={([yshift=-0.7ex]0,0)}] \coordinate (i1) at (-\scale,0); \coordinate (i2) at (+\scale,0); \coordinate (v1) at (0,\scale); \coordinate (v2) at (0,-\scale); \draw[->] (i1) arc (180:270:\scale); \draw[->-] (v2) arc (-90:0:\scale); \filldraw (v1) circle (1.3pt); \filldraw (v2) circle (1.3pt); \filldraw (i1) circle (1.3pt); \filldraw (i2) circle (1.3pt); \draw[dashed] (-\scale,-\scale) -- (\scale,\scale); \end{tikzpicture} &\;\;\;\; \begin{tikzpicture}[baseline={([yshift=-0.7ex]0,0)}] \coordinate (i1) at (-\scale,0); \coordinate (i2) at (+\scale,0); \coordinate (v1) at (0,\scale); \coordinate (v2) at (0,-\scale); \draw[->-] (i1) arc (180:90:\scale); \draw[->-] (v2) arc (-90:0:\scale); \draw[->] (v1) -- (v2); \filldraw (v1) circle (1.3pt); \filldraw (v2) circle (1.3pt); \filldraw (i1) circle (1.3pt); \filldraw (i2) circle (1.3pt); \draw[dashed] (-\scale,-\scale) -- (\scale,\scale); \end{tikzpicture} &\;\;\;\; \begin{tikzpicture}[baseline={([yshift=-0.7ex]0,0)}] \coordinate (i1) at (-\scale,0); \coordinate (i2) at (+\scale,0); \coordinate (v1) at (0,\scale); \coordinate (v2) at (0,-\scale); \draw[->-] (i1) arc (180:90:\scale); \draw[->] (v1) arc (90:0:\scale); \filldraw (v1) circle (1.3pt); \filldraw (v2) circle (1.3pt); \filldraw (i1) circle (1.3pt); \filldraw (i2) circle (1.3pt); \draw[dashed] (-\scale,-\scale) -- (\scale,\scale); \end{tikzpicture} \\
&&&& \\
(1a) & &\;\;\;\;\pp_1 &\;\; \;\;\pp_2 &\;\;\;\; \pp_3 \end{array} \end{align*}
}
Hence, applying the FOPT-cut Feynman rules from above to the cut diagram $(1a)$ 
results in the following expression 
\begin{gather} \begin{gathered} A_{(\boldsymbol{\sigma},\mathfrak{C})_{(1a)}} = -8 \frac{(2\pi)^2 g^{4}}{(8\pi^2)^{5}} \int \frac{\dd^3 \vec y_1 \dd^3 \vec y_2}{|\vec z_1| |\vec z_2| |\vec z_3| |\vec z_4| |\vec z_5| } \\
\times \frac{|\vec z_2|}{(-|\vec z_3|+\tau+i\ep)^2-\vec z_2^{\,2}} \frac{|\vec z_5|}{(|\vec z_1|-|\vec z_3|+\tau+i\ep)^2-\vec z_5^{\,2}} \frac{|\vec z_4|}{(|\vec z_1|+\tau+i\ep)^2-\vec z_4^{\,2}}. \end{gathered} \intertext{ where we accounted for the admissible paths through the cut, $23$, $153$ and $14$ via the appropriate denominators, and $\tau = x_2^0 - x_1^0$. Analogously, applying the Feynman rules to the FOPT-cut graphs $(1b)$ and $(1c)$ results in } \begin{gathered} A_{(\boldsymbol{\sigma},\mathfrak{C})_{(1b)}} = 8 \frac{(2\pi)^2 g^{4}}{(8\pi^2)^{5}} \int \frac{\dd^3 \vec y_1 \dd^3 \vec y_2}{|\vec z_1| |\vec z_2| |\vec z_3| |\vec z_4| |\vec z_5| } \\
\times \frac{|\vec z_1|}{(-|\vec z_4|+\tau+i\ep)^2-\vec z_1^{\,2}} \frac{|\vec z_1|}{(-|\vec z_5|-|\vec z_3|+\tau+i\ep)^2-\vec z_1^{\,2}} \frac{|\vec z_2|}{(-|\vec z_3|+\tau+i\ep)^2-\vec z_2^{\,2}}\,, \end{gathered} \\
\begin{gathered} A_{(\boldsymbol{\sigma},\mathfrak{C})_{(1c)}} (x_1,x_2) = -8 \frac{(2\pi)^2 g^{4}}{(8\pi^2)^{5}} \int \frac{\dd^3 \vec y_1 \dd^3 \vec y_2}{|\vec z_1| |\vec z_2| |\vec z_3| |\vec z_4| |\vec z_5| } \\
\times \frac{|\vec z_4|}{(|\vec z_1|+\tau+i\ep)^2-\vec z_4^{\,2}} \frac{|\vec z_3|}{(|\vec z_1| +|\vec z_5|+\tau+i\ep)^2-\vec z_3^{\,2}} \frac{|\vec z_3|}{(|\vec z_2|+\tau+i\ep)^2-\vec z_3^{\,2}}\,. \end{gathered} \end{gather}
These three FOPT-cut integrals all correspond to the same directed graph with different cuts on it.  Each integrand features three factors in the denominator, which each corresponding to a unique route from the left-most to the right-most vertex.

For the other possible flow orientations of the graph, we only have two routes through the diagram and one closed cyclic energy flow:
\begin{gather} \begin{gathered} A_{(\boldsymbol{\sigma},\mathfrak{C})_{(2b)}} (x_1,x_2) = -4 \frac{(2\pi)^2 g^{4}}{(8\pi^2)^{5}} \int \frac{\dd^3 \vec y_1 \dd^3 \vec y_2}{|\vec z_1| |\vec z_2| |\vec z_3| |\vec z_4| |\vec z_5| } \\
\times \frac{|\vec z_1|}{(-|\vec z_5|-|\vec z_3|+\tau+i\ep)^2-\vec z_1^{\,2}} \frac{|\vec z_2|}{(-|\vec z_3|+\tau+i\ep)^2-\vec z_2^{\,2}} \frac{1}{-|\vec z_5| - |\vec z_3| -|\vec z_4|}\,, \end{gathered} \\
\begin{gathered} A_{(\boldsymbol{\sigma},\mathfrak{C})_{(2c)}} (x_1,x_2) = 4 \frac{(2\pi)^2 g^{4}}{(8\pi^2)^{5}} \int \frac{\dd^3 \vec y_1 \dd^3 \vec y_2}{|\vec z_1| |\vec z_2| |\vec z_3| |\vec z_4| |\vec z_5| } \\
\times \frac{|\vec z_4|}{(|\vec z_1|+\tau+i\ep)^2-\vec z_4^{\,2}} \frac{|\vec z_3|}{(|\vec z_1| +|\vec z_5|+\tau+i\ep)^2-\vec z_3^{\,2}} \frac{1}{|\vec z_2|+|\vec z_5|+|\vec z_1|}\,. \end{gathered} \end{gather}
The energy-flow-oriented cut graph $(2b)$ has the cycle $534$ and the graph $(2c)$ the cycle $152$.

Even though, e.g., $(1a)$ and both $(1b)$ and $(1c)$ have differently sized cuts, the corresponding integrands are of the same dimension. This is a convenient situation from the perspective of the numerical evaluation of these integrals, as we can put all $(1a)$, $(1b)$ and $(1c)$ under the same integral sign. We expect IR singularities to cancel locally in our proposed representation, but we postpone the detailed analysis of this conjecture to a future work.

\providecommand{\href}[2]{#2}\begingroup\raggedright\endgroup

\begin{thebibliography}{10}

\bibitem{Sterman:1993hfp}
G.F.~Sterman, \emph{{An Introduction to quantum field theory}}, Cambridge
  University Press (8, 1993).

\bibitem{tasi_sterman}
G.F.~Sterman, \emph{{Partons, factorization and resummation, TASI 95}},  in
  \emph{{Theoretical Advanced Study Institute in Elementary Particle Physics
  (TASI 95): QCD and Beyond}}, pp.~327--408, 6, 1995
  [\href{https://arxiv.org/abs/hep-ph/9606312}{{\ttfamily hep-ph/9606312}}].

\bibitem{Bierenbaum_2010}
I.~Bierenbaum, S.~Catani, P.~Draggiotis and G.~Rodrigo, \emph{A tree-loop
  duality relation at two loops and beyond},
  \href{https://doi.org/10.1007/JHEP10(2010)073}{\emph{JHEP} {\bfseries 10}
  (2010) 073} [\href{https://arxiv.org/abs/1007.0194}{{\ttfamily 1007.0194}}].

\bibitem{Catani_2008}
S.~Catani, T.~Gleisberg, F.~Krauss, G.~Rodrigo and J.-C.~Winter, \emph{{From
  loops to trees by-passing Feynman's theorem}},
  \href{https://doi.org/10.1088/1126-6708/2008/09/065}{\emph{JHEP} {\bfseries
  09} (2008) 065} [\href{https://arxiv.org/abs/0804.3170}{{\ttfamily
  0804.3170}}].

\bibitem{Capatti_2019}
Z.~Capatti, V.~Hirschi, D.~Kermanschah and B.~Ruijl, \emph{Loop-tree duality
  for multiloop numerical integration},
  \href{https://doi.org/10.1103/PhysRevLett.123.151602}{\emph{Phys. Rev. Lett.}
  {\bfseries 123} (2019) 151602}
  [\href{https://arxiv.org/abs/1906.06138}{{\ttfamily 1906.06138}}].

\bibitem{Runkel_2019}
R.~Runkel, Z.~Sz\H{o}r, J.P.~Vesga and S.~Weinzierl, \emph{{Causality and
  loop-tree duality at higher loops}},
  \href{https://doi.org/10.1103/PhysRevLett.122.111603}{\emph{Phys. Rev. Lett.}
  {\bfseries 122} (2019) 111603}
  [\href{https://arxiv.org/abs/1902.02135}{{\ttfamily 1902.02135}}].

\bibitem{Runkel_2020}
R.~Runkel, Z.~Sz\H{o}r, J.P.~Vesga and S.~Weinzierl, \emph{{Integrands of loop
  amplitudes within loop-tree duality}},
  \href{https://doi.org/10.1103/PhysRevD.101.116014}{\emph{Phys. Rev. D}
  {\bfseries 101} (2020) 116014}
  [\href{https://arxiv.org/abs/1906.02218}{{\ttfamily 1906.02218}}].

\bibitem{Berghoff:2022vah}
M.~Berghoff, \emph{{Schwinger, ltd: loop-tree duality in the parametric
  representation}}, \href{https://doi.org/10.1007/JHEP10(2022)178}{\emph{JHEP}
  {\bfseries 10} (2022) 178}
  [\href{https://arxiv.org/abs/2208.07636}{{\ttfamily 2208.07636}}].

\bibitem{bobadilla2021lotty}
W.J.T.~Bobadilla, \emph{{Lotty \textendash{} the loop-tree duality
  automation}},
  \href{https://doi.org/10.1140/epjc/s10052-021-09235-0}{\emph{Eur. Phys. J. C}
  {\bfseries 81} (2021) 514}
  [\href{https://arxiv.org/abs/2103.09237}{{\ttfamily 2103.09237}}].

\bibitem{capatti2020manifestly}
Z.~Capatti, V.~Hirschi, D.~Kermanschah, A.~Pelloni and B.~Ruijl,
  \emph{Manifestly causal loop-tree duality},
  \href{https://arxiv.org/abs/2009.05509}{{\ttfamily 2009.05509}}.

\bibitem{Sborlini_2021}
G.F.R.~Sborlini, \emph{Geometrical approach to causality in multiloop
  amplitudes},
  \href{https://doi.org/10.1103/physrevd.104.036014}{\emph{Physical Review D}
  {\bfseries 104} (2021) }.

\bibitem{Aguilera_Verdugo_2020}
J.J.~Aguilera-Verdugo, F.~Driencourt-Mangin, R.J.~Hern\'andez-Pinto,
  J.~Plenter, S.~Ramirez-Uribe, A.E.~Renteria~Olivo et~al., \emph{Open loop
  amplitudes and causality to all orders and powers from the loop-tree
  duality}, \href{https://doi.org/10.1103/PhysRevLett.124.211602}{\emph{Phys.
  Rev. Lett.} {\bfseries 124} (2020) 211602}
  [\href{https://arxiv.org/abs/2001.03564}{{\ttfamily 2001.03564}}].

\bibitem{Aguilera_Verdugo_2021}
J.~Jes\'us Aguilera-Verdugo, R.J.~Hern\'andez-Pinto, G.~Rodrigo,
  G.F.R.~Sborlini and W.J.~Torres~Bobadilla, \emph{{Mathematical properties of
  nested residues and their application to multi-loop scattering amplitudes}},
  \href{https://doi.org/10.1007/JHEP02(2021)112}{\emph{JHEP} {\bfseries 02}
  (2021) 112} [\href{https://arxiv.org/abs/2010.12971}{{\ttfamily
  2010.12971}}].

\bibitem{Kromin:2022txz}
S.~Kromin, N.~Schwanemann and S.~Weinzierl, \emph{{Amplitudes within causal
  loop-tree duality}},
  \href{https://doi.org/10.1103/PhysRevD.106.076006}{\emph{Phys. Rev. D}
  {\bfseries 106} (2022) 076006}
  [\href{https://arxiv.org/abs/2208.01060}{{\ttfamily 2208.01060}}].

\bibitem{Capatti_2021}
Z.~Capatti, V.~Hirschi, A.~Pelloni and B.~Ruijl, \emph{Local unitarity: a
  representation of differential cross-sections that is locally free of
  infrared singularities at any order},
  \href{https://doi.org/10.1007/JHEP04(2021)104}{\emph{JHEP} {\bfseries 04}
  (2021) 104} [\href{https://arxiv.org/abs/2010.01068}{{\ttfamily
  2010.01068}}].

\bibitem{Kreimer:2020mwn}
D.~Kreimer, \emph{Outer space as a combinatorial backbone for {Cutkosky} rules
  and coactions},  in \emph{Anti-Differentiation and the Calculation of Feynman
  Amplitudes}, J.~Blümlein and C.~Schneider, eds. (2021),
  \href{https://doi.org/10.1007/978-3-030-80219-6\_12}{DOI}
  [\href{https://arxiv.org/abs/2010.11781}{{\ttfamily 2010.11781}}].

\bibitem{Kreimer:2021jni}
D.~Kreimer and K.~Yeats, \emph{Algebraic interplay between renormalization and
  monodromy},  \href{https://arxiv.org/abs/2105.05948}{{\ttfamily 2105.05948}}.

\bibitem{Dallaway:2022zoz}
W.~Dallaway and K.~Yeats, \emph{{Some results on Landau poles and Feynman
  diagram cut structure by Hopf algebra}},
  \href{https://arxiv.org/abs/2210.01164}{{\ttfamily 2210.01164}}.

\bibitem{Binoth:2000ps}
T.~Binoth and G.~Heinrich, \emph{{An automatized algorithm to compute infrared
  divergent multiloop integrals}},
  \href{https://doi.org/10.1016/S0550-3213(00)00429-6}{\emph{Nucl. Phys. B}
  {\bfseries 585} (2000) 741}
  [\href{https://arxiv.org/abs/hep-ph/0004013}{{\ttfamily hep-ph/0004013}}].

\bibitem{Borowka:2017idc}
S.~Borowka, G.~Heinrich, S.~Jahn, S.P.~Jones, M.~Kerner, J.~Schlenk et~al.,
  \emph{{pySecDec: a toolbox for the numerical evaluation of multi-scale
  integrals}}, \href{https://doi.org/10.1016/j.cpc.2017.09.015}{\emph{Comput.
  Phys. Commun.} {\bfseries 222} (2018) 313}
  [\href{https://arxiv.org/abs/1703.09692}{{\ttfamily 1703.09692}}].

\bibitem{Heinrich:2020ybq}
G.~Heinrich, \emph{{Collider Physics at the Precision Frontier}},
  \href{https://doi.org/10.1016/j.physrep.2021.03.006}{\emph{Phys. Rept.}
  {\bfseries 922} (2021) 1} [\href{https://arxiv.org/abs/2009.00516}{{\ttfamily
  2009.00516}}].

\bibitem{Smirnov:2021rhf}
A.V.~Smirnov, N.D.~Shapurov and L.I.~Vysotsky, \emph{{FIESTA5: Numerical
  high-performance Feynman integral evaluation}},
  \href{https://doi.org/10.1016/j.cpc.2022.108386}{\emph{Comput. Phys. Commun.}
  {\bfseries 277} (2022) 108386}
  [\href{https://arxiv.org/abs/2110.11660}{{\ttfamily 2110.11660}}].

\bibitem{Borinsky:2020rqs}
M.~Borinsky, \emph{Tropical monte carlo quadrature for {Feynman} integrals},
  {\emph{Annales de l'Institut Henri Poincar\' e D} (to appear) }
  [\href{https://arxiv.org/abs/2008.12310}{{\ttfamily 2008.12310}}].

\bibitem{Gong:2008ww}
W.~Gong, Z.~Nagy and D.E.~Soper, \emph{{Direct numerical integration of
  one-loop Feynman diagrams for N-photon amplitudes}},
  \href{https://doi.org/10.1103/PhysRevD.79.033005}{\emph{Phys. Rev. D}
  {\bfseries 79} (2009) 033005}
  [\href{https://arxiv.org/abs/0812.3686}{{\ttfamily 0812.3686}}].

\bibitem{Binoth:2005ff}
T.~Binoth, J.P.~Guillet, G.~Heinrich, E.~Pilon and C.~Schubert, \emph{{An
  algebraic/numerical formalism for one-loop multi-leg amplitudes}},
  \href{https://doi.org/10.1088/1126-6708/2005/10/015}{\emph{JHEP} {\bfseries
  10} (2005) 015} [\href{https://arxiv.org/abs/hep-ph/0504267}{{\ttfamily
  hep-ph/0504267}}].

\bibitem{Nagy:2006xy}
Z.~Nagy and D.E.~Soper, \emph{{Numerical integration of one-loop Feynman
  diagrams for N-photon amplitudes}},
  \href{https://doi.org/10.1103/PhysRevD.74.093006}{\emph{Phys. Rev. D}
  {\bfseries 74} (2006) 093006}
  [\href{https://arxiv.org/abs/hep-ph/0610028}{{\ttfamily hep-ph/0610028}}].

\bibitem{Becker:2010ng}
S.~Becker, C.~Reuschle and S.~Weinzierl, \emph{{Numerical NLO QCD
  calculations}}, \href{https://doi.org/10.1007/JHEP12(2010)013}{\emph{JHEP}
  {\bfseries 12} (2010) 013} [\href{https://arxiv.org/abs/1010.4187}{{\ttfamily
  1010.4187}}].

\bibitem{Becker:2012aqa}
S.~Becker, C.~Reuschle and S.~Weinzierl, \emph{Efficiency improvements for the
  numerical computation of {NLO} corrections},
  \href{https://doi.org/10.1007/JHEP07(2012)090}{\emph{JHEP} {\bfseries 07}
  (2012) 090} [\href{https://arxiv.org/abs/1205.2096}{{\ttfamily 1205.2096}}].

\bibitem{Becker:2012nk}
S.~Becker and S.~Weinzierl, \emph{{Direct contour deformation with arbitrary
  masses in the loop}},
  \href{https://doi.org/10.1103/PhysRevD.86.074009}{\emph{Phys. Rev. D}
  {\bfseries 86} (2012) 074009}
  [\href{https://arxiv.org/abs/1208.4088}{{\ttfamily 1208.4088}}].

\bibitem{Anastasiou:2018rib}
C.~Anastasiou and G.~Sterman, \emph{{Removing infrared divergences from
  two-loop integrals}},
  \href{https://doi.org/10.1007/JHEP07(2019)056}{\emph{JHEP} {\bfseries 07}
  (2019) 056} [\href{https://arxiv.org/abs/1812.03753}{{\ttfamily
  1812.03753}}].

\bibitem{Anastasiou:2020sdt}
C.~Anastasiou, R.~Haindl, G.~Sterman, Z.~Yang and M.~Zeng, \emph{{Locally
  finite two-loop amplitudes for off-shell multi-photon production in
  electron-positron annihilation}},
  \href{https://doi.org/10.1007/JHEP04(2021)222}{\emph{JHEP} {\bfseries 04}
  (2021) 222} [\href{https://arxiv.org/abs/2008.12293}{{\ttfamily
  2008.12293}}].

\bibitem{Soper:1999xk}
D.E.~Soper, \emph{{Techniques for QCD calculations by numerical integration}},
  \href{https://doi.org/10.1103/PhysRevD.62.014009}{\emph{Phys. Rev. D}
  {\bfseries 62} (2000) 014009}
  [\href{https://arxiv.org/abs/hep-ph/9910292}{{\ttfamily hep-ph/9910292}}].

\bibitem{Buchta:2015wna}
S.~Buchta, G.~Chachamis, P.~Draggiotis and G.~Rodrigo, \emph{{Numerical
  implementation of the loop\textendash{}tree duality method}},
  \href{https://doi.org/10.1140/epjc/s10052-017-4833-6}{\emph{Eur. Phys. J. C}
  {\bfseries 77} (2017) 274}
  [\href{https://arxiv.org/abs/1510.00187}{{\ttfamily 1510.00187}}].

\bibitem{Capatti_2020}
Z.~Capatti, V.~Hirschi, D.~Kermanschah, A.~Pelloni and B.~Ruijl,
  \emph{Numerical loop-tree duality: contour deformation and subtraction},
  \href{https://doi.org/10.1007/JHEP04(2020)096}{\emph{JHEP} {\bfseries 04}
  (2020) 096} [\href{https://arxiv.org/abs/1912.09291}{{\ttfamily
  1912.09291}}].

\bibitem{Capatti:2022tit}
Z.~Capatti, V.~Hirschi and B.~Ruijl, \emph{{Local unitarity: cutting raised
  propagators and localising renormalisation}},
  \href{https://doi.org/10.1007/JHEP10(2022)120}{\emph{JHEP} {\bfseries 10}
  (2022) 120} [\href{https://arxiv.org/abs/2203.11038}{{\ttfamily
  2203.11038}}].

\bibitem{Kermanschah:2021wbk}
D.~Kermanschah, \emph{{Numerical integration of loop integrals through local
  cancellation of threshold singularities}},
  \href{https://doi.org/10.1007/JHEP01(2022)151}{\emph{JHEP} {\bfseries 01}
  (2022) 151} [\href{https://arxiv.org/abs/2110.06869}{{\ttfamily
  2110.06869}}].

\bibitem{Veltman2}
M.J.G.~Veltman, \emph{{Unitarity and causality in a renormalizable field theory
  with unstable particles}},
  \href{https://doi.org/10.1016/S0031-8914(63)80277-3}{\emph{Physica}
  {\bfseries 29} (1963) 186}.

\bibitem{Martinus}
M.J.G.~Veltman, \emph{Diagrammatica: The path to {F}eynman rules}, vol.~4,
  Cambridge University Press (5, 2012).

\bibitem{Epstein:1973gw}
H.~Epstein and V.~Glaser, \emph{{The role of locality in perturbation theory}},
  {\emph{Ann. Inst. H. Poincare Phys. Theor. A} {\bfseries 19} (1973) 211}.

\bibitem{Chetyrkin:1980pr}
K.G.~Chetyrkin, A.L.~Kataev and F.V.~Tkachov, \emph{New approach to evaluation
  of multiloop {Feynman} integrals: the {Gegenbauer} polynomial $x$ space
  technique}, \href{https://doi.org/10.1016/0550-3213(80)90289-8}{\emph{Nucl.
  Phys. B} {\bfseries 174} (1980) 345}.

\bibitem{Schnetz:2013hqa}
O.~Schnetz, \emph{{Graphical functions and single-valued multiple
  polylogarithms}},
  \href{https://doi.org/10.4310/CNTP.2014.v8.n4.a1}{\emph{Commun. Number Theory
  Phys.} {\bfseries 08} (2014) 589}
  [\href{https://arxiv.org/abs/1302.6445}{{\ttfamily 1302.6445}}].

\bibitem{Borinsky:2021jdb}
M.~Borinsky, J.A.~Gracey, M.V.~Kompaniets and O.~Schnetz, \emph{{Five-loop
  renormalization of $\phi^3$ theory with applications to the Lee-Yang edge
  singularity and percolation theory}},
  \href{https://doi.org/10.1103/PhysRevD.103.116024}{\emph{Phys. Rev. D}
  {\bfseries 103} (2021) 116024}
  [\href{https://arxiv.org/abs/2103.16224}{{\ttfamily 2103.16224}}].

\bibitem{gfe}
M.~Borinsky and O.~Schnetz, \emph{{Graphical functions in even dimensions}},
  \href{https://doi.org/dx.doi.org/10.4310/CNTP.2022.v16.n3.a3}{\emph{Commun.
  Number Theory Phys.} {\bfseries 16} (2022) 515}
  [\href{https://arxiv.org/abs/2105.05015}{{\ttfamily 2105.05015}}].

\bibitem{Borinsky:2022lds}
M.~Borinsky and O.~Schnetz, \emph{{Recursive computation of Feynman periods}},
  \href{https://doi.org/10.1007/JHEP08(2022)291}{\emph{JHEP} {\bfseries 08}
  (2022) 291} [\href{https://arxiv.org/abs/2206.10460}{{\ttfamily
  2206.10460}}].

\bibitem{Eric1}
E.~Laenen, K.J.~Larsen and R.~Rietkerk, \emph{{Imaginary parts and
  discontinuities of Wilson line correlators}},
  \href{https://doi.org/10.1103/PhysRevLett.114.181602}{\emph{Phys. Rev. Lett.}
  {\bfseries 114} (2015) 181602}
  [\href{https://arxiv.org/abs/1410.5681}{{\ttfamily 1410.5681}}].

\bibitem{Eric2}
E.~Laenen, K.J.~Larsen and R.~Rietkerk, \emph{{Position-space cuts for Wilson
  line correlators}},
  \href{https://doi.org/10.1007/JHEP07(2015)083}{\emph{JHEP} {\bfseries 07}
  (2015) 083} [\href{https://arxiv.org/abs/1505.02555}{{\ttfamily
  1505.02555}}].

\bibitem{Sterman3}
O.~Erdo\u{g}an and G.~Sterman, \emph{{Path description of coordinate-space
  amplitudes}}, \href{https://doi.org/10.1103/PhysRevD.95.116015}{\emph{Phys.
  Rev. D} {\bfseries 95} (2017) 116015}
  [\href{https://arxiv.org/abs/1705.04539}{{\ttfamily 1705.04539}}].

\bibitem{StermanErdogan}
G.~Sterman and O.~Erdo\u{g}an, \emph{{A coordinate description of partonic
  processes}}, \href{https://doi.org/10.22323/1.235.0027}{\emph{PoS} {\bfseries
  RADCOR2015} (2016) 027} [\href{https://arxiv.org/abs/1602.00943}{{\ttfamily
  1602.00943}}].

\bibitem{ErdoganCS}
O.~Erdo\u{g}an, \emph{{Coordinate-space singularities of massless gauge
  theories}}, \href{https://doi.org/10.1103/PhysRevD.89.085016}{\emph{Phys.
  Rev. D} {\bfseries 89} (2014) 085016}
  [\href{https://arxiv.org/abs/1312.0058}{{\ttfamily 1312.0058}}].

\bibitem{Salas-Bernardez:2022cuw}
A.~Salas-Bern\'ardez, \emph{{Explicit computation of jet functions in
  coordinate space}},
  \href{https://doi.org/10.1016/j.nuclphysb.2022.116024}{\emph{Nucl. Phys. B}
  {\bfseries 985} (2022) 116024}
  [\href{https://arxiv.org/abs/2205.05423}{{\ttfamily 2205.05423}}].

\bibitem{gallo1978extreme}
G.~Gallo and C.~Sodini, \emph{Extreme points and adjacency relationship in the
  flow polytope}, \href{https://doi.org/10.1007/BF02575918}{\emph{Calcolo}
  {\bfseries 15} (1978) 277}.

\bibitem{stanley2000acyclic}
R.P.~Stanley, \emph{Acyclic flow polytopes and {K}ostant’s partition
  function},  in \emph{Conference transparencies}, 2000,
  \href{https://math.mit.edu/~rstan/transparencies/kostant.pdf}{https://math.mit.edu/~rstan/transparencies/kostant.pdf}.

\bibitem{baldoni2004counting}
W.~Baldoni-Silva, J.A.~De~Loera and M.~Vergne, \emph{Counting integer flows in
  networks}, \href{https://doi.org/10.1007/s10208-003-0088-8}{\emph{Found.
  Comput. Math.} {\bfseries 4} (2004) 277}.

\bibitem{meszaros2015flow}
K.~M\'{e}sz\'{a}ros and A.H.~Morales, \emph{Flow polytopes of signed graphs and
  the {K}ostant partition function},
  \href{https://doi.org/10.1093/imrn/rnt212}{\emph{Int. Math. Res. Not. IMRN}
  (2015) 830}.

\bibitem{meszaros2017polytope}
K.~M\'{e}sz\'{a}ros, A.H.~Morales and B.~Rhoades, \emph{The polytope of
  {T}esler matrices},
  \href{https://doi.org/10.1007/s00029-016-0241-2}{\emph{Selecta Math. (N.S.)}
  {\bfseries 23} (2017) 425}.

\bibitem{benedetti2019combinatorial}
C.~Benedetti, R.S.~Gonz\'{a}lez~D'Le\'{o}n, C.R.H.~Hanusa, P.E.~Harris,
  A.~Khare, A.H.~Morales et~al., \emph{A combinatorial model for computing
  volumes of flow polytopes},
  \href{https://doi.org/10.1090/tran/7743}{\emph{Trans. Amer. Math. Soc.}
  {\bfseries 372} (2019) 3369}.

\bibitem{Kaneko:2009qx}
T.~Kaneko and T.~Ueda, \emph{{A geometric method of sector decomposition}},
  \href{https://doi.org/10.1016/j.cpc.2010.04.001}{\emph{Comput. Phys. Commun.}
  {\bfseries 181} (2010) 1352}
  [\href{https://arxiv.org/abs/0908.2897}{{\ttfamily 0908.2897}}].

\bibitem{Arkani-Hamed:2013jha}
N.~Arkani-Hamed and J.~Trnka, \emph{{The amplituhedron}},
  \href{https://doi.org/10.1007/JHEP10(2014)030}{\emph{JHEP} {\bfseries 10}
  (2014) 030} [\href{https://arxiv.org/abs/1312.2007}{{\ttfamily 1312.2007}}].

\bibitem{Brown:2015fyf}
F.~Brown, \emph{{Feynman amplitudes, coaction principle, and cosmic Galois
  group}}, \href{https://doi.org/10.4310/CNTP.2017.v11.n3.a1}{\emph{Commun.
  Num. Theor. Phys.} {\bfseries 11} (2017) 453}
  [\href{https://arxiv.org/abs/1512.06409}{{\ttfamily 1512.06409}}].

\bibitem{Panzer:2019yxl}
E.~Panzer, \emph{{Hepp's bound for Feynman graphs and matroids}},
  \href{https://arxiv.org/abs/1908.09820}{{\ttfamily 1908.09820}}.

\bibitem{Arkani-Hamed:2017tmz}
N.~Arkani-Hamed, Y.~Bai and T.~Lam, \emph{Positive geometries and canonical
  forms}, \href{https://doi.org/10.1007/JHEP11(2017)039}{\emph{JHEP} {\bfseries
  11} (2017) 039} [\href{https://arxiv.org/abs/1703.04541}{{\ttfamily
  1703.04541}}].

\bibitem{Schultka:2018nrs}
K.~Schultka, \emph{{Toric geometry and regularization of Feynman integrals}},
  \href{https://arxiv.org/abs/1806.01086}{{\ttfamily 1806.01086}}.

\bibitem{Arkani-Hamed:2019mrd}
N.~Arkani-Hamed, S.~He and T.~Lam, \emph{{Stringy canonical forms}},
  \href{https://doi.org/10.1007/JHEP02(2021)069}{\emph{JHEP} {\bfseries 02}
  (2021) 069} [\href{https://arxiv.org/abs/1912.08707}{{\ttfamily
  1912.08707}}].

\bibitem{Ananthanarayan:2020fhl}
B.~Ananthanarayan, S.~Banik, S.~Friot and S.~Ghosh, \emph{Multiple series
  representations of {N-fold Mellin-Barnes} integrals},
  \href{https://doi.org/10.1103/PhysRevLett.127.151601}{\emph{Phys. Rev. Lett.}
  {\bfseries 127} (2021) 151601}
  [\href{https://arxiv.org/abs/2012.15108}{{\ttfamily 2012.15108}}].

\bibitem{Arkani-Hamed:2022cqe}
N.~Arkani-Hamed, A.~Hillman and S.~Mizera, \emph{{Feynman polytopes and the
  tropical geometry of UV and IR divergences}},
  \href{https://doi.org/10.1103/PhysRevD.105.125013}{\emph{Phys. Rev. D}
  {\bfseries 105} (2022) 125013}
  [\href{https://arxiv.org/abs/2202.12296}{{\ttfamily 2202.12296}}].

\bibitem{Collins:1989gx}
J.C.~Collins, D.E.~Soper and G.F.~Sterman, \emph{Factorization of hard
  processes in {QCD}},
  \href{https://doi.org/10.1142/9789814503266_0001}{\emph{Adv. Ser. Direct.
  High Energy Phys.} {\bfseries 5} (1989) 1}
  [\href{https://arxiv.org/abs/hep-ph/0409313}{{\ttfamily hep-ph/0409313}}].

\bibitem{Cutkosky:1960sp}
R.E.~Cutkosky, \emph{{Singularities and discontinuities of Feynman
  amplitudes}}, \href{https://doi.org/10.1063/1.1703676}{\emph{J. Math. Phys.}
  {\bfseries 1} (1960) 429}.

\bibitem{le1991quantum}
M.~Le~Bellac et~al., \emph{Quantum and statistical field theory}, Oxford
  University Press (1991).

\bibitem{itzykson2012quantum}
C.~Itzykson and J.-B.~Zuber, \emph{Quantum field theory}, Courier Corporation
  (2012).

\bibitem{bondy1976graph}
J.A.~Bondy and U.S.R.~Murty, \emph{Graph theory with applications}, Macmillan
  London (1976).

\bibitem{Weinberg:1959nj}
S.~Weinberg, \emph{{High-energy behavior in quantum field theory}},
  \href{https://doi.org/10.1103/PhysRev.118.838}{\emph{Phys. Rev.} {\bfseries
  118} (1960) 838}.

\bibitem{Collins:2020euz}
J.~Collins, \emph{{A new and complete proof of the Landau condition for pinch
  singularities of Feynman graphs and other integrals}},
  \href{https://arxiv.org/abs/2007.04085}{{\ttfamily 2007.04085}}.

\bibitem{baldoni2008kostant}
W.~Baldoni and M.~Vergne, \emph{Kostant partitions functions and flow
  polytopes}, \href{https://doi.org/10.1007/s00031-008-9019-8}{\emph{Transform.
  Groups} {\bfseries 13} (2008) 447}.

\bibitem{polytope_fourier}
S.~Robins, \emph{{A friendly introduction to Fourier analysis on polytopes}},
  \href{https://arxiv.org/abs/2104.06407}{{\ttfamily 2104.06407}}.

\bibitem{Kinoshita:1962ur}
T.~Kinoshita, \emph{{Mass singularities of Feynman amplitudes}},
  \href{https://doi.org/10.1063/1.1724268}{\emph{J. Math. Phys.} {\bfseries 3}
  (1962) 650}.

\bibitem{Lee:1964is}
T.D.~Lee and M.~Nauenberg, \emph{Degenerate systems and mass singularities},
  \href{https://doi.org/10.1103/PhysRev.133.B1549}{\emph{Phys. Rev.} {\bfseries
  133} (1964) B1549}.

\bibitem{Hannesdottir:2022bmo}
H.S.~Hannesdottir and S.~Mizera, \emph{{What is the $i\varepsilon$ for the
  S-matrix?}},  \href{https://arxiv.org/abs/2204.02988}{{\ttfamily
  2204.02988}}.

\bibitem{Frye:2018xjj}
C.~Frye, H.~Hannesdottir, N.~Paul, M.D.~Schwartz and K.~Yan, \emph{Infrared
  finiteness and forward scattering},
  \href{https://doi.org/10.1103/PhysRevD.99.056015}{\emph{Phys. Rev. D}
  {\bfseries 99} (2019) 056015}
  [\href{https://arxiv.org/abs/1810.10022}{{\ttfamily 1810.10022}}].

\bibitem{Arkani-Hamed:2012zlh}
N.~Arkani-Hamed, J.L.~Bourjaily, F.~Cachazo, A.B.~Goncharov, A.~Postnikov and
  J.~Trnka, \emph{Grassmannian Geometry of Scattering Amplitudes}, Cambridge
  University Press (4, 2016),
  \href{https://doi.org/10.1017/CBO9781316091548}{10.1017/CBO9781316091548},
  [\href{https://arxiv.org/abs/1212.5605}{{\ttfamily 1212.5605}}].

\bibitem{Arkani-Hamed:2014dca}
N.~Arkani-Hamed, A.~Hodges and J.~Trnka, \emph{Positive amplitudes in the
  amplituhedron}, \href{https://doi.org/10.1007/JHEP08(2015)030}{\emph{JHEP}
  {\bfseries 08} (2015) 030} [\href{https://arxiv.org/abs/1412.8478}{{\ttfamily
  1412.8478}}].

\bibitem{Hodges:2009hk}
A.~Hodges, \emph{{Eliminating spurious poles from gauge-theoretic amplitudes}},
  \href{https://doi.org/10.1007/JHEP05(2013)135}{\emph{JHEP} {\bfseries 05}
  (2013) 135} [\href{https://arxiv.org/abs/0905.1473}{{\ttfamily 0905.1473}}].

\bibitem{Hodges:2011wm}
A.~Hodges, \emph{{New expressions for gravitational scattering amplitudes}},
  \href{https://doi.org/10.1007/JHEP07(2013)075}{\emph{JHEP} {\bfseries 07}
  (2013) 075} [\href{https://arxiv.org/abs/1108.2227}{{\ttfamily 1108.2227}}].

\bibitem{Britto:2004ap}
R.~Britto, F.~Cachazo and B.~Feng, \emph{{New recursion relations for tree
  amplitudes of gluons}},
  \href{https://doi.org/10.1016/j.nuclphysb.2005.02.030}{\emph{Nucl. Phys. B}
  {\bfseries 715} (2005) 499}
  [\href{https://arxiv.org/abs/hep-th/0412308}{{\ttfamily hep-th/0412308}}].

\bibitem{Britto:2005fq}
R.~Britto, F.~Cachazo, B.~Feng and E.~Witten, \emph{{Direct proof of tree-level
  recursion relation in Yang-Mills theory}},
  \href{https://doi.org/10.1103/PhysRevLett.94.181602}{\emph{Phys. Rev. Lett.}
  {\bfseries 94} (2005) 181602}
  [\href{https://arxiv.org/abs/hep-th/0501052}{{\ttfamily hep-th/0501052}}].

\bibitem{abramowitz1964handbook}
M.~Abramowitz and I.A.~Stegun, \emph{Handbook of mathematical functions with
  formulas, graphs, and mathematical tables}, vol.~55, US Government printing
  office (1964).

\bibitem{Connes:1999yr}
A.~Connes and D.~Kreimer, \emph{{Renormalization in quantum field theory and
  the Riemann-Hilbert problem. 1. The Hopf algebra structure of graphs and the
  main theorem}}, \href{https://doi.org/10.1007/s002200050779}{\emph{Commun.
  Math. Phys.} {\bfseries 210} (2000) 249}
  [\href{https://arxiv.org/abs/hep-th/9912092}{{\ttfamily hep-th/9912092}}].

\bibitem{Figueroa:2004hb}
H.~Figueroa and J.M.~Gracia-Bondia, \emph{{Combinatorial Hopf algebras in
  quantum field theory. I}},
  \href{https://doi.org/10.1142/S0129055X05002467}{\emph{Rev. Math. Phys.}
  {\bfseries 17} (2005) 881}
  [\href{https://arxiv.org/abs/hep-th/0408145}{{\ttfamily hep-th/0408145}}].

\bibitem{Borinsky:2014xwa}
M.~Borinsky, \emph{{Feynman graph generation and calculations in the Hopf
  algebra of Feynman graphs}},
  \href{https://doi.org/10.1016/j.cpc.2014.07.023}{\emph{Comput. Phys. Commun.}
  {\bfseries 185} (2014) 3317}
  [\href{https://arxiv.org/abs/1402.2613}{{\ttfamily 1402.2613}}].

\bibitem{Beekveldt:2020kzk}
R.~Beekveldt, M.~Borinsky and F.~Herzog, \emph{{The Hopf algebra structure of
  the $R^{*}$-operation}},
  \href{https://doi.org/10.1007/JHEP07(2020)061}{\emph{JHEP} {\bfseries 07}
  (2020) 061} [\href{https://arxiv.org/abs/2003.04301}{{\ttfamily
  2003.04301}}].

\bibitem{Borinsky:2015mga}
M.~Borinsky, \emph{{Algebraic lattices in QFT renormalization}},
  \href{https://doi.org/10.1007/s11005-016-0843-9}{\emph{Lett. Math. Phys.}
  {\bfseries 106} (2016) 879}
  [\href{https://arxiv.org/abs/1509.01862}{{\ttfamily 1509.01862}}].

\bibitem{Herzog:2018ily}
F.~Herzog, \emph{{Geometric IR subtraction for final state real radiation}},
  \href{https://doi.org/10.1007/JHEP08(2018)006}{\emph{JHEP} {\bfseries 08}
  (2018) 006} [\href{https://arxiv.org/abs/1804.07949}{{\ttfamily
  1804.07949}}].

\bibitem{Ma:2019hjq}
Y.~Ma, \emph{A forest formula to subtract infrared singularities in amplitudes
  for wide-angle scattering},
  \href{https://doi.org/10.1007/JHEP05(2020)012}{\emph{JHEP} {\bfseries 05}
  (2020) 012} [\href{https://arxiv.org/abs/1910.11304}{{\ttfamily
  1910.11304}}].

\bibitem{Bloch:2015efx}
S.~Bloch and D.~Kreimer, \emph{Cutkosky rules and outer space},
  \href{https://arxiv.org/abs/1512.01705}{{\ttfamily 1512.01705}}.

\bibitem{Berghoff:2020bug}
M.~Berghoff and D.~Kreimer, \emph{{Graph complexes and Feynman rules}},
  \href{https://arxiv.org/abs/2008.09540}{{\ttfamily 2008.09540}}.

\end{thebibliography}
\end{document}